\documentclass[pdflatex,sn-nature]{sn-jnl}

\usepackage{graphicx}
\usepackage{amsmath,amssymb}
\usepackage{xcolor}
\usepackage{colortbl}
\definecolor{bcrtblue}{HTML}{0072B2} %
\usepackage{subcaption}
\usepackage{rotating}
\usepackage{framed}
\usepackage[nameinlink,noabbrev]{cleveref}

\usepackage{booktabs}
\usepackage{multirow}
\usepackage{tabularx}
\usepackage{makecell}
\usepackage{array}
\usepackage{ragged2e}

\newcolumntype{Z}{>{\raggedleft\arraybackslash}p{30mm}}
\newcolumntype{Y}{>{\raggedleft\arraybackslash}X}

\usepackage{algorithm}
\usepackage{algorithmic}

\usepackage{pifont}
\usepackage{adjustbox}
\usepackage[switch]{lineno}
\usepackage{comment}
\usepackage{enumitem} %
\usepackage{placeins}

\begin{document}
\title[Behavioural Reflection Test]{The Behavioural Reflection Test: A time-efficient measure of reflective reasoning in morally and epistemically charged decisions}
\author*[1]{\fnm{Sion} \sur{Weatherhead}}\email{s.weatherhead@unsw.edu.au}
\author[1]{\fnm{Flora} \sur{Salim}}\email{flora.salim@unsw.edu.au}
\author[2]{\fnm{Aaron} \sur{Belbasis}}\email{aaron.belbasis@aurecongroup.com.au}
\author*[1]{\fnm{Ben R.} \sur{Newell}}\email{ben.newell@unsw.edu.au}
\affil[1]{\orgname{University of New South Wales, Sydney}}
\affil[2]{\orgname{Aurecon Group, Melbourne}}

\abstract{
Cognitive reflection, the tendency to override an intuitive conclusion by reflecting on it, is linked to how people navigate information from reliable and misleading sources. The Cognitive Reflection Test (CRT) can predict how reflection plays out in decisions and language when it is measured alongside relevant behaviour, such as social-media activity. However, CRT items may need periodic replacement: as they become widely known, people who report having seen them score higher. The Behavioural Reflection Test (BRT) was designed to record construct-relevant behaviour directly: in two open-ended, morally and epistemically charged scenarios, participants say what they would do and why. Its responses are not scored as right or wrong, so there is no correct answer to find out after a first encounter. We also provide a new four-item bespoke CRT (bCRT) as a low-exposure anchor against which to validate the BRT. Among 473 online adults, higher bCRT predicted how people sought information when a salient claim conflicted with institutional evidence (contacting the source directly and consulting institutional and official records), but not the decisions they reached. It also predicted expressed worry and more emotionally engaged, risk-attentive language. The widely used four-item CRT2, adjusted for prior exposure, predicted none of these search, worry or language outcomes. The bCRT showed convergent validity and was more informative than CRT2 from about one standard deviation above mean ability. Though open-ended, the BRT took a median 11.8 minutes.
}
\maketitle

Reasoning research often distinguishes between relatively intuitive and more reflective forms of thought, frequently discussed under dual-process theories such as Kahneman's `System 1' and `System 2' \cite{kahneman2012Thinkingfastslow}. However, contemporary critiques argue that treating these as mechanistically discrete systems with fixed attribute clusters \cite{schul2009TwoNotAlways} -- for example, intuitive thought as necessarily fast, emotional, and automatic, and reflective thought as necessarily slow, unemotional, and deliberative -- is not empirically well justified \cite{newell2023OpenMindedSearching,evans2013DualProcessTheoriesHigher}. A more defensible use of this distinction is therefore descriptive rather than strongly architectural: intuitive and reflective thought can be treated as modes of processing tendencies without assuming rigidly separable systems \cite{evans2013DualProcessTheoriesHigher}.

This distinction remains consequential in the contemporary information environment. Social media platforms, algorithmic feeds, and now ubiquitous large language models (LLMs) such as ChatGPT present people with a dense mixture of high-quality evidence, misinformation, and disinformation \cite{chen2024Combatingmisinformationage, meyrowitsch2023AIchatbotsmisinformation}. Understanding when people engage in more reflective, evidence-sensitive reasoning, and how this varies across individuals, is therefore important for explaining how people navigate morally and epistemically loaded decisions in everyday life.

\subsection*{Cognitive Reflection Test: significance and criticisms}
The Cognitive Reflection Test (CRT) \cite{frederick2005} operationalises individual differences in this reflective style. The original three-item CRT consists of `trick' questions that elicit an immediately compelling but incorrect `intuitive' response and a correct answer that typically requires inhibiting that response and reasoning further. Despite its brevity, the CRT has shown striking predictive power: higher scores are associated with greater tolerance for beneficial risk~\cite{cokely2009Cognitiveabilitiessuperior,frederick2005}, more willingness to delay gratification \cite{frederick2005}, avoidance of biases \cite{oechssler2009Cognitiveabilitiesbehavioral}, actively open-minded thinking applied across task domains \cite{oechssler2009Cognitiveabilitiesbehavioral}, and tendency toward utilitarian moral judgements that rely on deliberate cost--benefit analysis \cite{greene2009Dualprocessmoralitypersonal,frederick2005}. Crucially, CRT performance predicts unique variance in reasoning and decision tasks beyond standard measures of intelligence \cite{toplak2011CognitiveReflectionTest}, suggesting that it taps a relatively stable tendency to override default responses rather than raw cognitive capacity alone.

At the same time, the CRT has attracted sustained psychometric criticism. Several studies argue that much of its variance reflects numerical or other cognitive ability rather than reflective thinking per se \cite{sinayev2015Cognitivereflectionvsa,campitelli2014Doescognitivereflection,welsh2022WhatCRTIntelligence}. However, alternative versions such as the four-item CRT2 \cite{thomson2016} and the ten-item CRT-V \cite{sirota2021Measuringcognitivereflection} were developed to minimise mathematical demands while preserving the core `intuitive lure versus correct answer' structure. The CRT and CRT2 retain consistent correlations with constructs such as Need for Cognition (NFC)~\cite{frederick2005,thomson2016,pennycook2016} and continue to predict performance on tasks that require overriding initial impressions~\cite{alos-ferrer2016CognitiveReflectionDecision}, supporting the view that cognitive reflection is a meaningful trait-level construct. Their correlations with Conscientiousness are smaller and less consistent~\cite{knezevic2024Lifetimeprevalencequestionable,juanchich2016CognitiveReflectionPredicts,welsh2022WhatCRTIntelligence}.

Importantly, the influence of cognitive reflection is not confined to laboratory puzzles. Mosleh et al. \cite{mosleh2021} linked CRT scores to large-scale `in-the-wild' behaviour on Twitter. Using Linguistic Inquiry and Word Count (LIWC), they showed that higher CRT scores predicted greater use of \textit{inhibition}, \textit{insight} (cognitive complexity), \textit{negative emotion}, \textit{moral} (prescriptive reasoning), and \textit{political} word categories in tweets. More reflective individuals thus used language that was more analytically complex, normatively loaded, and politically engaged, extending CRT's reach from problem-solving performance to everyday linguistic style.

In the same dataset, higher CRT scores predicted better informational hygiene: more links shared to high-quality outlets and fewer to misleading sources, even after accounting for ideology and demographics \cite{mosleh2021}. Political orientation mattered independently: misinformation sharing trended with both lower reflection and conservatism \cite{deppe2015Reflectiveliberalsintuitive}. Reflection, however, is not necessarily reducible to it. For example, Pennycook et al. showed that low CRT scores among 2016 Trump voters were largely driven by party-crossing Democrats, tying an intuitive style of engagement to concrete voting \textit{behaviour} rather than stated ideology \cite{pennycook2019CognitiveReflection2016}.

Together, these findings position the CRT as a compact but useful tool for detecting a cognitive style that is expressed in how people reason, talk, and choose in ecologically rich settings. However, the very success of classic CRT items has created a methodological problem. Thomson and Oppenheimer \cite{thomson2016} in their 2016 paper warned that the original items were widely known and shared, and found that participants who reported having seen them scored higher, as later studies did~\cite{haigh2016,bialek2018}. Whether this weakens the test's predictive validity is disputed: large studies of repeated exposure find that it survives~\cite{bialek2018,meyer2018}, whereas a version with altered numbers correlated more strongly with validation measures than the original items~\cite{woike2019}. As a precaution, Thomson and Oppenheimer called for expanded item pools and alternative CRT variants~\cite{thomson2016}, and variants such as the CRT-V have followed~\cite{sirota2021Measuringcognitivereflection}, but such solutions also erode one of the CRT's main practical strengths: brevity.

Longer CRTs demand more time and may prompt respondents to become vigilant for `trick questions', and disguising items among decoys extends administration further. More fundamentally, decontextualised brainteasers may not capture the reflective tendencies that matter when people navigate messier, real-life decisions. Richer behavioural measures such as Raven's Matrices \cite{maran2020Intelligencepredictschoice} and IQ tests \cite{richardson2002WhatIQTests} can give rich, predictive data, but trade off time cost (45 minutes \cite{langener2022shortenedversionRavens} and 72 minutes \cite{ryan2007Administrationtimeestimates} respectively) and construct specificity, while brief self-reports such as NFC index motivation rather than how reflective thinking is enacted in judgement~\cite{cacioppo1984EfficientAssessmentNeed}.

This leaves two related gaps. First, despite what we now know about individual differences in concrete decisional behaviour and linguistic expression among more reflective individuals, these findings have not been leveraged to create an efficient, but rich behavioural measure of reflection outside of trick-questions. Second, there is a need for a brief, low-exposure measure of cognitive reflection that preserves the core lure-versus-correct structure while avoiding the saturation of classic items. Addressing these gaps together is useful: a fresher reflection measure can help anchor the construct validity of a new behavioural task, while the behavioural task can show whether that measure and the legacy CRT items predict the same theoretically relevant signatures of reflection.

\subsection*{Aims and hypotheses}
The present study had two aims. First, we introduced a compact Behavioural Reflection Test (BRT): a pair of open-ended scenarios in which participants explain and justify decisions in morally and epistemically charged situations, designed to capture construct-relevant behaviour linked to cognitive reflection---online informational hygiene, linguistic expression, and moral judgement \cite{mosleh2021,pennycook2014roleanalyticthinking,pennycook2016}---without Mosleh et al.'s reliance on massive personal-data collection. Because BRT responses are not scored as right or wrong, they record how people reason and what they prefer. Unlike a trick question, a scenario seen before has no correct answer to find out, so prior exposure alone should not change the response. Second, we developed and validated a brief bespoke CRT (bCRT) that reduces item exposure while preserving the classic lure-versus-correct structure, serving both as a lower-exposure reflection measure and as a contemporaneous construct anchor for the BRT. Comparing bCRT with CRT2 and their associations with NFC and BRT outcomes also shows, descriptively, whether a low-exposure measure and saturated legacy items relate to behaviour in the same way.

\subsubsection*{BRT: decisional hypotheses}
The BRT comprises two scenario types that vary in how individuals must arbitrate between higher-quality institutional evidence and salient but less reliable alternative sources, under conditions of social and/or personal stakes. Across both scenarios, participants must integrate evidential quality, assess risk, and navigate norm-relevant considerations when forming and justifying a decision. The scenarios differ in domain (social commitment versus health decision-making) but share this underlying evidential conflict structure. The scenarios are described in \hyperref[par:brt]{Methods} and given in full in the \hyperref[app:brt_full_items]{Supplementary Information}.

Prior work linking CRT to normatively appropriate, evidence-sensitive behaviour and specific patterns of moral evaluation \cite[e.g.,][]{pennycook2014roleanalyticthinking,pennycook2016,mosleh2021} motivates the view that more reflective individuals are more likely to seek out high-quality information, better at integrating evidential quality and fairness considerations, and more affectively engaged with the hazards a decision carries.

Accordingly, we predicted that higher bCRT scores would be associated with the following decisional patterns, listed in the order the results are reported:

\begin{enumerate}[label=\textbf{H\arabic*.}]
  \item \textbf{Epistemic search.} When a salient claim conflicts with institutional evidence, more reflective respondents are more likely to plan to check it with expert or official sources, or with the party the claim concerns (H1a), rather than through vague or no search.
  \item \textbf{Evidence-weighted decisions.} More reflective respondents are more likely to reach the decision the higher-quality evidence supports and to ground it in a plan to check, and less likely to act on the low-quality claim without scrutiny.
  \item \textbf{Affective engagement.} More reflective respondents express more negative affect toward the scenario's hazard and its source~\cite{mosleh2021}.
  \item \textbf{Fairness and norm-sensitive reasoning.} More reflective respondents appeal more spontaneously to fairness, loyalty and related moral considerations, especially where norms or social obligations conflict.
\end{enumerate}

Each prediction was evaluated with one primary family of coded outcomes (\hyperref[par:decisional]{Methods}), built from directly coded reasoning features or from deterministic composites that combine a decision with its justification.

\subsubsection*{BRT: linguistic hypotheses}
Complementing the decisional outcomes, we used LIWC to test whether bCRT predicts systematic differences in the linguistic expression of reasoning in the BRT scenarios. Drawing on Mosleh et al. \cite{mosleh2021} and related work, we expected higher bCRT to be positively associated with markers of analytic engagement, \textit{negative emotion} and normatively loaded reasoning. Because both scenarios turn on evaluating a potential hazard, we also expected greater explicit consideration of risk.

\subsubsection*{Psychometric hypotheses for bCRT}
At the psychometric level, we expected that:
\begin{enumerate}
  \item bCRT would show adequate internal consistency.
  \item bCRT scores would correlate positively with CRT2, NFC and Conscientiousness, and with relevant BRT outcomes, supporting convergent validity while leaving room for non-redundant prediction relative to CRT2.
\end{enumerate}

\section*{Methods}

\subsection*{Participants}
The main validation sample comprised $n=473$ adults recruited through Prolific. Eligibility criteria required participants to be at least 18 years old, fluent in English, and resident in the UK or USA. The realised sample was UK-majority (341 UK and 132 US residents). Participants were paid at Prolific's recommended ethical hourly rate. This sample completed CRT2, the final bCRT, the BRT, the NFC (18-item), the Big Five Inventory--10 (BFI-10), and demographic questions. Two participants had an incomplete CRT2 total and no NFC or BFI-10 responses, so analyses involving CRT2 or the trait scales use $n=471$, and bCRT and BRT analyses use the full $n=473$. The $n$ is reported with each analysis. Exclusions and participant flow are given in the \hyperref[app:cleaning]{Supplementary Information}. A pilot study with approximately 240 undergraduate students preceded the main study and is described in the \hyperref[app:pilot]{Supplementary Information}.

\subsection*{Measures}

\paragraph{CRT2.}\phantomsection\label{par:crt2}
The four-item CRT2 \cite{thomson2016} was administered as a benchmark measure of cognitive reflection. For each item, participants provided an open numeric or short-text response. Following Thomson and Oppenheimer, responses were coded into three categories: \textit{correct}, \textit{intuitive incorrect}, or \textit{other}. Each answer was matched against item-specific answer keys, and answers the keys did not resolve were classified by three large language models from different providers (GPT-5.6 Sol, Gemini 3.5 Flash and DeepSeek V4-Flash), keeping the majority label (\hyperref[app:crt_scoring]{Supplementary Information}). We computed a total score (number of correct responses). After each CRT2 item, participants indicated whether they had seen the question before. These familiarity judgements were summed into a familiarity count. A familiarity-adjusted CRT2 total subtracted 0.167 points for each CRT2 item seen before: the familiarity effect, estimated controlling for age and held fixed across analyses (\hyperref[app:familiarity]{Supplementary Information}). The LIWC comparator models instead entered the familiarity count as a covariate.

\paragraph{Bespoke Cognitive Reflection Test (bCRT).}
To obtain a fresher measure that balances numeracy and logic-based reasoning, we assembled candidate CRT-style items from existing `trick question' sources available online, following precedent \cite{frederick2005,thomson2016}. Items were chosen to follow the classic CRT structure (a salient but wrong intuitive answer, a correct answer, and a clear scoring rubric) while varying in surface content. Four items that showed acceptable difficulty and discrimination and elicited clear intuitive lures in the pilot study were selected to form the final bCRT (\hyperref[app:pilot]{Supplementary Information}). Example items include a profit calculation riddle and a non-numerical logic item in which a married person is looking at an unmarried person.\footnote{The full text of the four bCRT items, scoring keys, and intuitive lures is provided in the \hyperref[app:bcrt_items]{Supplementary Information}.}

In the Prolific sample, responses to the four selected items were coded into \textit{correct}, \textit{intuitive}, or \textit{other} by the same procedure as CRT2 (keys and lures in Supplementary Table~\ref{tab:bcrt_keys}), and a total score was computed. After each bCRT item, participants indicated whether they had previously seen a question of that form. Familiarity rates for these items were low, so bCRT familiarity counts were summarised descriptively but not used as primary covariates.

\paragraph{Behavioural Reflection Test.}\phantomsection\label{par:brt}
The BRT was developed as a brief, open-ended behavioural measure of reflective thinking, motivated by Mosleh et al.'s finding that cognitive reflection predicts systematic variation in spontaneous language use on social media \cite{mosleh2021}. The task comprised two short hypothetical scenarios designed to cue real-time reasoning about evidence, trust, and social obligations. In each, participants were asked to write at least 150 words, or four or more sentences, describing what they would do and why, `as if thinking aloud', and were prompted to cover their first reaction, what they would say, which information they trusted, any checks they would make and their final choice (\hyperref[app:brt_full_items]{Supplementary Information}).

In the \emph{restaurant scenario}, participants organise a birthday dinner with a non-refundable deposit when an unhelpful second friend forwards alarming customer reviews that contradict professional critics under time pressure. In the \emph{medication scenario}, a treating doctor's standard-of-care prescription is challenged by a popular health influencer emphasising rare or anecdotal harms. Both scenarios pit institutional sources (professional critics; medical professionals and clinical guidelines) against salient but lower-quality alternatives (crowd reviews; influencers) under mild time pressure and clear social stakes. Full scenario texts are provided in the \hyperref[app:brt_full_items]{Supplementary Information}.

\emph{Linguistic coding.}
For linguistic analyses, each participant's two responses were concatenated and analysed with LIWC-22 \cite{boyd2022DevelopmentPsychometricProperties}. In line with prior work linking function words and content categories to psychological processes and thinking style \cite{tausczik2010PsychologicalMeaningWords}, we grouped LIWC categories into a small set of theory-driven families relevant to reflective reasoning and epistemic vigilance: structural and quantity markers, cognitive and motivational language, auxiliary verbs, risk-focused language, affective content, and normative content, plus an exploratory time-focus family, added after the pilot study. The exact mapping between LIWC categories and these families is provided in the \hyperref[app:liwc_families]{Supplementary Information}.

LIWC was selected because the language analyses were designed to test psychologically interpretable categories previously linked to cognitive reflection. Although alternative tools such as SEANCE \cite{crossley2017SentimentAnalysisSocial} or Empath \cite{fast2016EmpathUnderstandingTopic} provide useful social-cognitive, affective, or data-driven lexical features, substituting them would alter the target constructs and reduce comparability with the motivating LIWC-based findings.

\begin{figure}[tb]
  \centering
  \makebox[\textwidth][c]{\includegraphics[width=170mm]{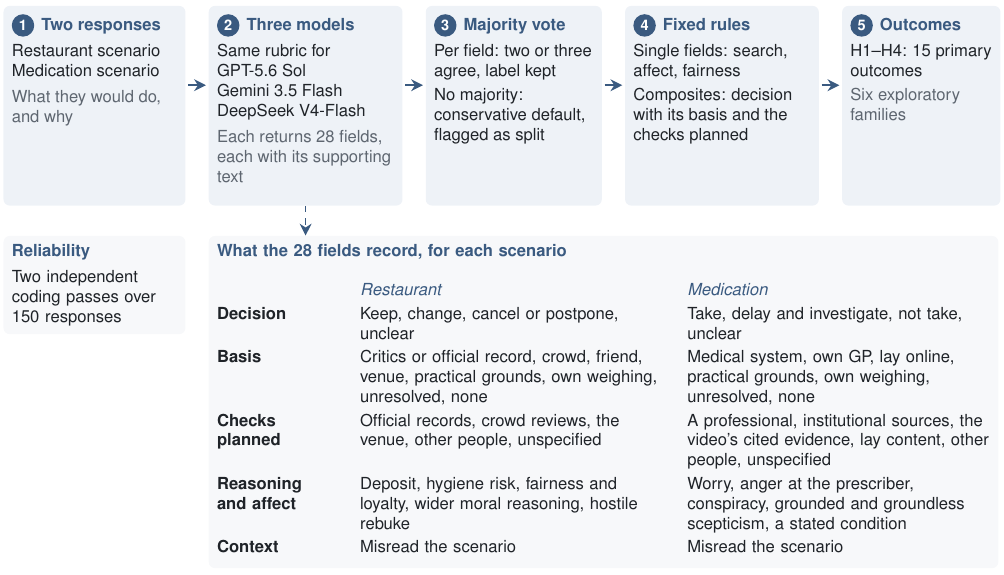}}
  \makebox[\linewidth][c]{\begin{minipage}{170mm}\captionsetup{position=below}\caption{\textbf{How does a written BRT response become a coded outcome?} The numbered boxes are the coding steps, read left to right; the lower panel lists what the 28 coded fields record in each scenario (Prolific sample, $N=473$).}\label{fig:coding_workflow}\end{minipage}}
\end{figure}

\emph{Decisional coding.}\phantomsection\label{par:decisional}
Open-ended responses were coded by the same three language models under a fixed, pre-specified rubric. Figure~\ref{fig:coding_workflow} shows how each response became the coded outcomes. Higher-quality sources were defined in advance as institutional or expert evidence (e.g., doctors, regulators, clinical guidance, professional reviews) and contrasted with crowd reviews, social media, anecdote, or unspecified reliance, and ambiguous cases were assigned the more conservative category. The full rubric and coding reliability are reported in the \hyperref[app:brt_coding_scheme]{Supplementary Information}.

From these coded dimensions, each decisional hypothesis was tested by one primary family of binary outcomes, fifteen in total (Table~\ref{tab:hypothesis_outcomes}). For example, the composite keeping the booking on evidence grounds requires the decision to keep together with hygiene-risk reasoning, an official-records check, or reliance on the critics or official record. Two overlaps that follow from these definitions are noted with the H2 estimates (Supplementary Table~\ref{tab:brt_primary}). Additional coded and derived outcomes were analysed as six exploratory families and are reported in the \hyperref[app:brt_exploratory_methods]{Supplementary Information}.

\begin{table}[tb]
\centering
\footnotesize
\caption{\textbf{Which outcomes test each decisional hypothesis?} Each row is one binary outcome. Prediction: \textcolor{bcrtblue}{$\uparrow$} more frequent with higher bCRT, $\downarrow$ less frequent, \textcolor{gray}{--} no directional prediction. R, restaurant scenario; M, medication scenario. One field, coded directly; composite, derived from several coded fields by fixed rules.}
\label{tab:hypothesis_outcomes}
\begin{tabularx}{\textwidth}{@{}>{\raggedright\arraybackslash\leftskip=1em}Xccl@{}}
\toprule
\multicolumn{1}{@{}l}{Outcome} & Scenario & Prediction & Coded as \\
\midrule
\multicolumn{4}{@{}l}{\textbf{H1 Epistemic search}\quad\textcolor{gray}{(7 outcomes)}} \\
Phones the restaurant (H1a) & R & \textcolor{bcrtblue}{$\uparrow$} & one field \\
Consults hygiene or official records & R & \textcolor{bcrtblue}{$\uparrow$} & one field \\
Asks friends or family & R & \textcolor{gray}{--} & one field \\
Consults regulators, journals or official guidance & M & \textcolor{bcrtblue}{$\uparrow$} & one field \\
Consults a qualified professional & M & \textcolor{bcrtblue}{$\uparrow$} & one field \\
Verifies the video's cited sources & M & \textcolor{bcrtblue}{$\uparrow$} & one field \\
Asks friends or family & M & \textcolor{gray}{--} & one field \\
\addlinespace
\multicolumn{4}{@{}l}{\textbf{H2 Evidence-weighted decisions}\quad\textcolor{gray}{(5 outcomes)}} \\
Keeps the booking on evidence grounds & R & \textcolor{bcrtblue}{$\uparrow$} & composite \\
Accepts on institutional grounds with a verification plan & M & \textcolor{bcrtblue}{$\uparrow$} & composite \\
Delays with a verification plan and a stated condition & M & \textcolor{bcrtblue}{$\uparrow$} & composite \\
Acts on the claim with no doubt and no rule & M & $\downarrow$ & composite \\
Refuses outright, worry-marked & M & $\downarrow$ & composite \\
\addlinespace
\multicolumn{4}{@{}l}{\textbf{H3 Affective engagement}\quad\textcolor{gray}{(2 outcomes)}} \\
Hostile rebuke of the unhelpful friend & R & \textcolor{bcrtblue}{$\uparrow$} & one field \\
Voices worry about the medication & M & \textcolor{bcrtblue}{$\uparrow$} & one field \\
\addlinespace
\multicolumn{4}{@{}l}{\textbf{H4 Fairness and norm-sensitive reasoning}\quad\textcolor{gray}{(1 outcome)}} \\
Appeals to fairness or loyalty toward Friend B & R & \textcolor{bcrtblue}{$\uparrow$} & one field \\
\bottomrule
\end{tabularx}
\end{table}

\paragraph{Need for Cognition.}
The 18-item NFC scale was administered as a dispositional measure of enjoyment of effortful thinking \cite{cacioppo1984EfficientAssessmentNeed}. NFC was chosen as a convergent-validity target because the same scale was used to validate the original CRT and CRT2~\cite{frederick2005,thomson2016}. Items are rated on a Likert-type scale and summed (after reverse scoring where appropriate) to yield a total NFC score. The NFC was presented at the end of the survey to minimise any priming of reflective self-concept before the CRT sections.

\paragraph{Big Five Inventory-10.}
Personality traits were assessed using the BFI-10, a brief measure of the Big Five dimensions (Openness, Conscientiousness, Extraversion, Agreeableness, Neuroticism) \cite{rammstedt2007Measuringpersonalityone}. Each dimension is assessed with two items rated on a Likert scale. Items are combined (with appropriate reversals) to yield five trait scores. In the present study, Conscientiousness was the personality trait of interest for convergent validity, given theorised links to deliberation and self-control. In a nationally representative sample of 1,003 adults, CRT scores correlated modestly with Conscientiousness ($r=.09$), and a need-for-cognition scale correlated with both ($r=.28$ with CRT scores, $.33$ with Conscientiousness)~\cite{knezevic2024Lifetimeprevalencequestionable}.

\paragraph{Demographics.}
Participants reported age, gender, coarse-grained ethnicity (Caucasian, black, Asian, mixed or other), highest level of education, country of residence, and language background.

\subsection*{Procedure}
All data were collected online via Qualtrics. After providing informed consent, participants completed the two BRT scenarios, a `reasoning block' (a randomised collection of final four-item bCRT, four decoy CRT questions, and CRT2; 12 items in total), the NFC scale, the BFI-10, and demographics in a single session. Overall survey completion typically fell within the expected $\sim$20 minute range. Completion-time distributions for the BRT and CRT blocks are shown in Supplementary Figure~\ref{fig:block_time_distributions}.

\subsection*{Data Analysis}

\paragraph{Reliability and item response model.}\phantomsection\label{par:irt}
Internal consistency (Cronbach's $\alpha$) was computed for CRT2, bCRT, and their combined total, and the eight CRT2 and bCRT items were calibrated together in one two-parameter logistic (2PL) item response theory (IRT) model by marginal maximum likelihood, so that the information curves of CRT2, bCRT and the combined test share one ability scale. Confidence intervals and bands come from a person bootstrap (1,000 resamples). The difference between bCRT and CRT2 test information was computed in each resample, giving a pointwise 95\% interval at each level of $\theta$.
\paragraph{Convergent validity.}
To assess convergent validity, we examined associations between bCRT, CRT2, NFC, and Big Five traits in the Prolific sample. We computed Pearson correlations between CRT totals and NFC and BFI scores, with 95\% confidence intervals. The familiarity-adjusted CRT2 total was correlated with the other measures.
\paragraph{Behavioural validity via BRT language.}\phantomsection\label{par:liwc}
For BRT language we ran two separate model families: \emph{primary} ordinary least squares (OLS) models regressing each LIWC family outcome on raw bCRT total (the direct association between the low-exposure anchor and BRT language), and \emph{secondary} CRT2 comparator models (raw CRT2 total, with CRT2 familiarity count as a covariate).

We report unstandardised coefficients, $p$-values, and $R^2$. LIWC categories were split into those for which prior work yields a directional prediction---\textit{insight}, \textit{negative emotion}, and \textit{moral}, following Mosleh et al.~\cite{mosleh2021}, and \textit{risk}, given scenarios built around hazard and evidential conflict---and the remaining categories, examined as exploratory. Each predicted category was tested singly and is reported with its per-test $p$, whereas exploratory categories were Benjamini--Hochberg corrected within LIWC families ($p_{\mathrm{FDR}}$). Of Mosleh et al.'s remaining predictive categories, \textit{inhibition} is unavailable in the current LIWC dictionary and \textit{political} language was not modelled, as the BRT scenarios contain no political content. Summary LIWC variables were treated as descriptive rather than focal (\hyperref[app:liwc_families]{Supplementary Information}). A sensitivity analysis excluding the eight attention-check failures is reported in the \hyperref[app:sensitivity]{Supplementary Information}.

\paragraph{Behavioural validity via BRT decisions.}
To test the decisional hypotheses, we used the four primary outcome families defined above.

For each binary outcome, we fit a separate logistic regression model using raw bCRT total as the focal predictor, so that odds ratios (ORs) are per additional bCRT item answered correctly. We report Wald 95\% confidence intervals and two-sided likelihood-ratio $p$-values. Where an outcome had fewer than 20 events, or the fit showed separation, Firth-penalised logistic regression with profile-likelihood intervals was used. No demographic covariates were included in the primary models. Benjamini--Hochberg correction at $q=.05$ was applied within each family separately, never across families. Beside every odds ratio we report the average marginal effect (percentage points per bCRT item) and, for every primary outcome, the minimum detectable odds ratio (MDOR) at 80\% power, estimated from 1,000 datasets simulated at the observed sample size, outcome prevalence and bCRT distribution.

Each model was refitted with raw CRT2 and with the familiarity-adjusted CRT2 total (\hyperref[par:crt2]{Measures}) as the focal predictor. A joint model and further sensitivity analyses are described in the \hyperref[app:brt_exploratory_methods]{Supplementary Information}.

\subsection*{Writing assistance}
A large language model (Claude, Anthropic) was used to assist with editing and restructuring the manuscript text. All changes were reviewed by the authors, who take full responsibility for the content.

\section*{Results}
\subsection*{Descriptive performance and inter-correlations}
Of the 473 Prolific participants who passed data-quality checks,
mean performance on the CRT2 was reasonably high (2.88 of 4 items correct),
whereas performance on the new four-item bCRT was lower (1.44 of 4 items correct), consistent with its items being harder (Supplementary Table~\ref{tab:irt_params}).
Internal consistency was modest and similar for the two measures: CRT2 Cronbach's $\alpha = .530$, close to CRT2 at its inception ($\alpha = .511$~\cite{thomson2016}),
and bCRT $\alpha = .545$, with similar mean split-half estimates.
The combined eight-item scale (CRT2 + bCRT) showed $\alpha = .649$.

Reliability therefore does not distinguish the two four-item measures in this sample.

Prior-exposure rates confirmed the low-exposure rationale directly: bCRT items were seen far less often (5--32\% across items) than CRT2 items (41--75\%; Table~\ref{tab:familiarity}).

\begin{table}[htbp]
\centering
\footnotesize
\caption{\textbf{How many participants had seen each CRT2 and bCRT item before?} Per-item prior-exposure (``seen before'') rates for CRT2 and bCRT items (Prolific sample; CRT2 rates $n=471$, bCRT rates $n=473$, reflecting two participants with an incomplete CRT2 total). bCRT items were markedly less familiar, consistent with the low-exposure rationale.}
\label{tab:familiarity}
\begin{tabular*}{\textwidth}{@{\extracolsep{\fill}}lcccc}
\toprule
 & Item 1 & Item 2 & Item 3 & Item 4 \\
\midrule
CRT2 seen (\%) & 75 & 59 & 74 & 41 \\
bCRT seen (\%) & 32 & 11 & \phantom{0}5 & \phantom{0}5 \\
\bottomrule
\end{tabular*}
\end{table}

Table~\ref{tab:crt_trait_correlations} summarises the correlations between CRT scores, NFC, and Conscientiousness. CRT2 and bCRT totals were moderately correlated, $r=.364$, 95\% CI $[.283,.440]$, indicating that the new items tap a related but not redundant construct.
Adjusting CRT2 scores for item familiarity (subtracting the effect of each CRT2 item seen before, estimated controlling for age; \hyperref[par:crt2]{Methods}) had a negligible impact on this association ($r=.368$, 95\% CI $[.287,.444]$).

\begin{figure}[htbp]
  \centering
  \includegraphics[width=\textwidth]{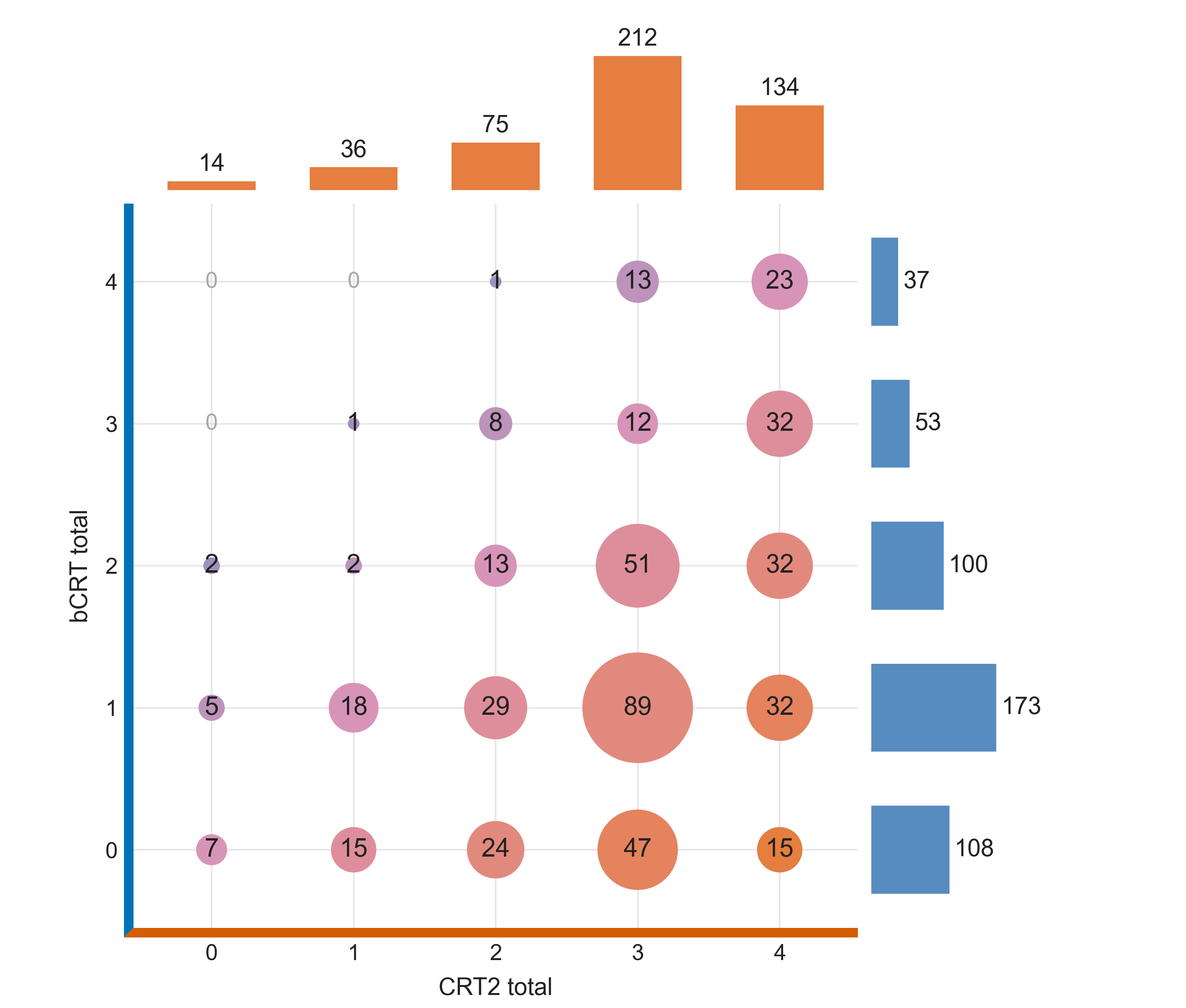}
  \caption{\textbf{How do total scores on CRT2 and the bespoke CRT (bCRT) relate?} Each circle's area is proportional to the number of participants with that CRT2 total ($x$) and bCRT total ($y$); the count is printed in the cell. Circle colour runs from blue (bCRT total above CRT2 total) through reddish purple (equal totals) to vermillion (CRT2 total above), and the axis bars carry each test's colour. Bars along the top and right give the number of participants at each CRT2 and bCRT total. $n=471$.}
  \label{fig:crt_scatter}
\end{figure}

Both CRT measures showed small positive associations with NFC (Table~\ref{tab:crt_trait_correlations}).
bCRT was modestly but significantly correlated with Conscientiousness, whereas CRT2 was not, its confidence interval spanning zero.
NFC itself correlated moderately with Conscientiousness. Together with the count grid in Figure~\ref{fig:crt_scatter}, this pattern supports the bCRT's convergent validity while indicating that the two four-item measures are related but not interchangeable.

Completion times also supported the practical feasibility of the BRT.
Median time for the BRT block was 11.83 minutes, with an interquartile range of 7.99--16.21 minutes. The distribution was right-skewed (mean = 13.45 minutes), but most participants completed the task well under 20 minutes, with longer completion times confined to a sparse tail.

\begin{table}[htbp]
\centering
\footnotesize
\caption{\textbf{How do the CRT scores relate to each other, to Need for Cognition and to Conscientiousness?} Correlations among CRT scores, Need for Cognition, and Conscientiousness (Prolific sample, $N=471$). Values are Pearson $r$ with 95\% confidence intervals.}
\label{tab:crt_trait_correlations}
\begin{tabular*}{\textwidth}{@{\extracolsep{\fill}}llrcr}
\toprule
Predictor & Outcome & $r$ & 95\% CI & $n$ \\
\midrule
CRT2 total         & bCRT total          & 0.364 & [0.283, 0.440] & 471 \\
CRT2 total         & Need for Cognition  & 0.177 & [0.088, 0.263] & 471 \\
CRT2 total         & Conscientiousness   & 0.067 & [-0.023, 0.156] & 471 \\
bCRT total         & Need for Cognition  & 0.105 & [0.015, 0.193] & 471 \\
bCRT total         & Conscientiousness   & 0.098 & [0.007, 0.186] & 471 \\
Need for Cognition & Conscientiousness   & 0.369 & [0.289, 0.445] & 471 \\
\bottomrule
\end{tabular*}
\end{table}

\subsection*{Item characteristics and information functions}
\begin{figure}[htbp]
  \centering
  \includegraphics[width=\textwidth]{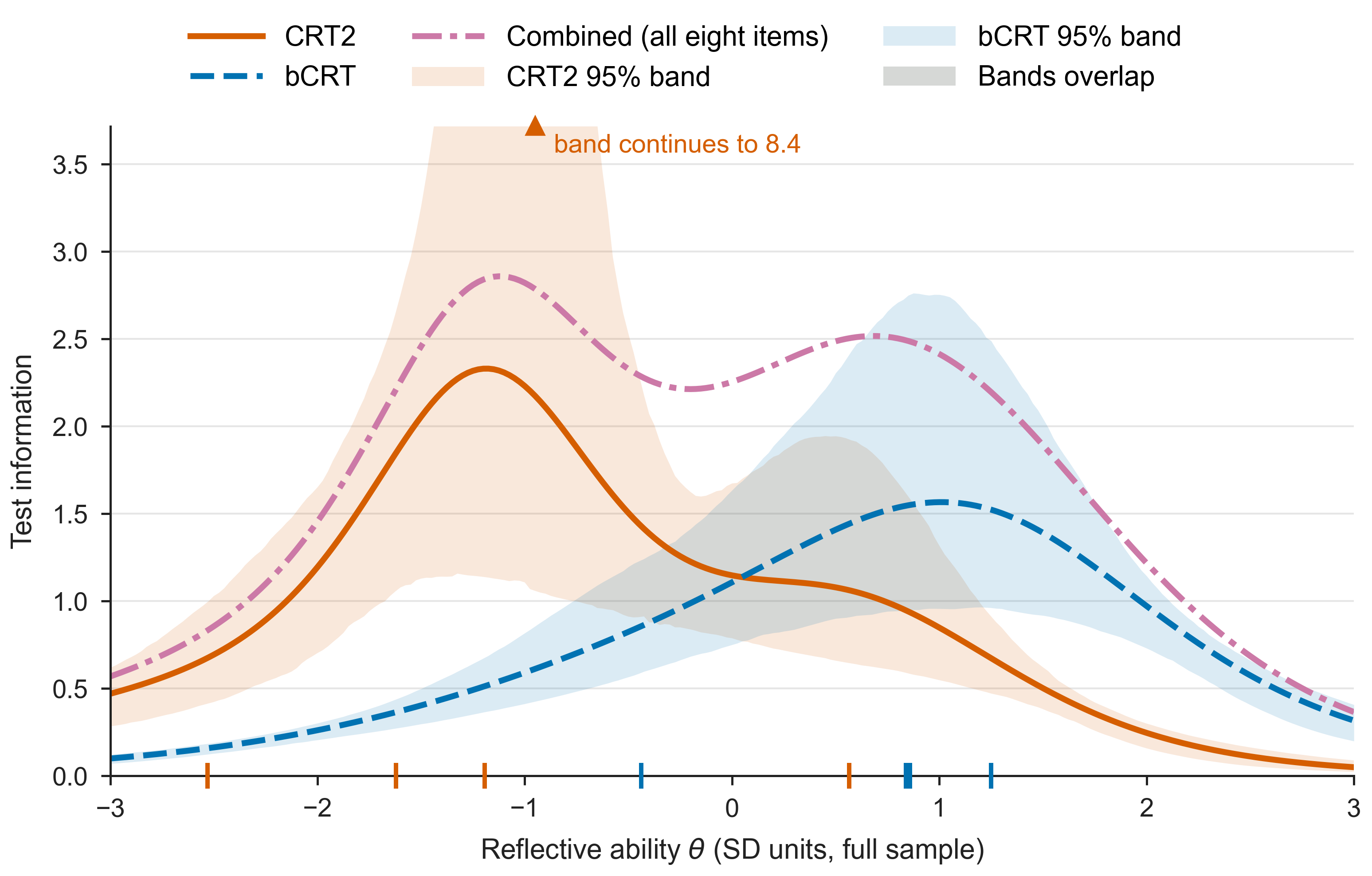}
  \caption{\textbf{Where on the ability scale does each test measure best?} Test information curves (TIC) for CRT2, bCRT and the combined eight-item test, from one two-parameter logistic IRT calibration of all eight items, over reflective ability $\theta$, with person-bootstrap 95\% bands for CRT2 and bCRT. The CRT2 band runs above the axis because the discrimination of one CRT2 item, Emily's father's three daughters, varied widely across bootstrap resamples (Supplementary Table~\ref{tab:irt_params}), and an item's information rises with the square of its discrimination. Ticks on the $\theta$ axis: item difficulties, in each test's colour. $N=473$ (CRT2 items, $n=471$).}
  \label{fig:irt_tic}
\end{figure}

We next examined item-level psychometrics for the bCRT relative to CRT2 using the two-parameter logistic item response model fitted to all eight items (\hyperref[par:irt]{Methods}; full parameter estimates are given in Supplementary Table~\ref{tab:irt_params}).
Test information curves (TIC) in Figure~\ref{fig:irt_tic} show that CRT2 was most informative about one standard deviation below mean ability and the bCRT about one standard deviation above it. The bCRT was more informative than CRT2 only from about one standard deviation above the mean, where the bootstrap interval for the difference between the two curves excluded zero. Nearer the mean, the interval included zero. The combined eight-item test (CRT2 + bCRT) delivered substantially greater information across a broad band of $\theta$, indicating that the new items improve measurement precision without sacrificing coverage of easier items.

\subsection*{BRT: search, decision and affect outcomes}
We then tested the decisional hypotheses by asking whether bCRT scores predicted how participants sought information, what they decided, and how they reacted in the two BRT scenarios: a restaurant booking disrupted by an unhelpful friend (`Friend~A') ahead of a birthday dinner for `Friend~B', and a doctor-prescribed medication (`Medex') challenged by an online influencer's warning (outcome definitions in Table~\ref{tab:hypothesis_outcomes}, full scenarios in the \hyperref[app:brt_full_items]{Supplementary Information}). Behaviourally, most participants kept the restaurant booking (70.8\% keep, 4.9\% change, 1.9\% cancel or postpone, 22.4\% unclear/conditional; $N=473$).
Responses were more dispersed in the medication scenario (33.4\% take, 50.1\% delay and investigate, 12.3\% not take, 4.2\% unclear/conditional).
What participants said their medication decision rested on was similarly spread: their own weighing of the evidence (37.2\%), the medical system (doctors, clinical guidance and regulators as an accountable system; 35.9\%), no stated basis (8.5\%), lay sources such as the video, other online content or friends and family (8.2\%), practical grounds (5.9\%), or a personal relationship with their own doctor (4.0\%).

\paragraph{Epistemic search (H1).}
The clearest behavioural signal was in which sources participants said they would consult to check the contested claim (Figure~\ref{fig:dissociation}a; Supplementary Table~\ref{tab:brt_primary}). Higher bCRT predicted four of the seven search outcomes after Benjamini--Hochberg correction within the family.
The largest effect was verification at the primary source: each additional bCRT item raised the odds of planning to phone the restaurant (H1a; OR per item~$=1.52$, 95\% CI $[1.21, 1.91]$, $q=.002$).
Higher scorers were also more likely to plan to consult regulators, journals or official guidance about the medication (OR per item~$=1.33$, $[1.13, 1.57]$, $q=.002$) and hygiene or official records about the restaurant (OR per item~$=1.28$, $[1.04, 1.57]$, $q=.038$),
and to ask friends or family about the medication (OR per item~$=1.24$, $[1.05, 1.47]$, $q=.024$), an outcome for which we held no directional prediction.
Consulting a qualified professional, asking friends or family about the restaurant and verifying the video's cited sources were not significantly associated with bCRT after correction.
At the other end of the search dimension, vague-only search---stating an intention to check without naming any source---is reported with the exploratory outcome families (Supplementary Table~\ref{tab:brt_exploratory}).

\paragraph{Evidence-weighted decisions (H2).}
In contrast, bCRT did not predict the decisions participants reached (Figure~\ref{fig:dissociation}a; Supplementary Table~\ref{tab:brt_primary}). None of the five decision outcomes was significant after correction (all $q\geq.65$), at either pole:
on the reflective side, keeping the booking on evidence grounds, accepting the medication on institutional grounds with a verification plan, and delaying with a verification plan and a stated condition;
on the intuitive side, acting on the claim without doubt or a decision rule and outright, worry-marked refusal.
These nulls are bounded: at the observed prevalences the sample had 80\% power to detect odds ratios of 1.25--1.35 per bCRT item in the predicted direction for the reflective outcomes, and of 0.60--0.73 for the intuitive outcomes (Supplementary Table~\ref{tab:brt_primary}, MDOR column).
Stricter variants of the acceptance and delay composites, restricted to verification plans without any lay check, were not positively associated with bCRT either (rows beneath H2 in Supplementary Table~\ref{tab:brt_primary}).

\paragraph{Affective engagement (H3).}
Higher bCRT predicted voicing worry about the medication (OR per item~$=1.44$, 95\% CI $[1.22, 1.70]$, $q<.001$; Figure~\ref{fig:dissociation}a; Supplementary Table~\ref{tab:brt_primary}),
about 8.6 percentage points per additional bCRT item.
Hostile rebuke of the unhelpful friend was not associated with bCRT.

\paragraph{Fairness and norm-sensitive reasoning (H4).}
Appeals to fairness or loyalty toward Friend~B were not significantly associated with bCRT ($p=.099$; Figure~\ref{fig:dissociation}a; Supplementary Table~\ref{tab:brt_primary}).
With a single outcome, this family required no correction.

Across the four families, five of the fifteen primary outcomes reached $q<.05$ within their family: four of the seven search outcomes and worry.
None of the sixteen exploratory outcomes did (Supplementary Table~\ref{tab:brt_exploratory}).
All five kept a per-test $p<.05$ with age, gender and education entered as covariates, and no response had a split coder vote on the fields these five outcomes are built from, so the refit excluding split responses leaves them unchanged (Supplementary Table~\ref{tab:brt_cov_nosplit}).
Estimates with bCRT entered as an ordered factor are given in Supplementary Table~\ref{tab:brt_ordered_factor}.

\paragraph{Response length.}\phantomsection\label{par:resplength}
Higher bCRT scorers wrote longer responses: bCRT correlated with the total word count across both scenarios, $r=.179$, 95\% CI $[.090, .265]$, $p<.001$.
Longer responses also named more sources: word count predicted every search outcome (Supplementary Table~\ref{tab:brt_response_length}).

\subsection*{BRT: linguistic markers}
We next tested whether bCRT scores predicted linguistic markers of analytic engagement, risk consideration, and normatively salient reasoning in the BRT scenarios (\hyperref[par:liwc]{Methods}).

The results gave partial support for the linguistic hypothesis. Among the predicted categories, higher bCRT was associated with greater use of \textit{negative emotion} and \textit{risk} words (Table~\ref{tab:liwc_bcrt}),
whereas \textit{insight} and \textit{moral} language were unrelated ($p=.61$ and $.60$).
Among the exploratory categories, higher bCRT predicted greater \textit{present-focus} and lower \textit{discrepancy}, \textit{differentiation}, and \textit{auxiliary verb} use (Table~\ref{tab:liwc_bcrt}). The expectation of more overtly normative wording was therefore not borne out: \textit{moral} language, tested directly, showed no association. Word count was also positively associated with bCRT, whereas words per sentence was not after correction. Both were treated as descriptive rather than focal (\hyperref[par:resplength]{Response length}).

The corresponding CRT2 comparator models, using raw CRT2 total with the CRT2 familiarity count entered as a covariate, are reported in Supplementary Table~\ref{tab:liwc_crt2}. No CRT2 association was significant for the same focal LIWC categories. Visual slope diagnostics for the primary bCRT LIWC models and the bCRT--CRT2 \textit{risk} category prediction comparison are provided in Supplementary Figures~\ref{fig:liwc_grid}--\ref{fig:risk_slope_overlay_supp}.

\begin{figure}[htbp]
  \centering
  \makebox[\textwidth][c]{\includegraphics[width=170mm]{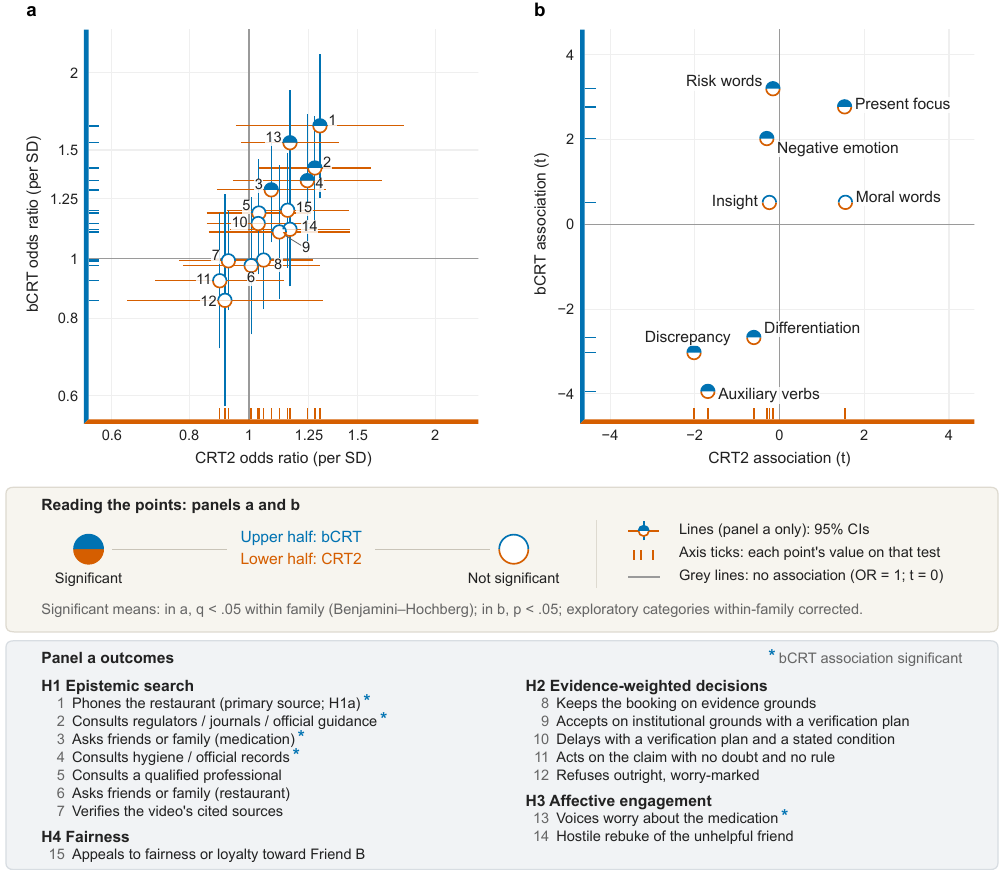}}
  \makebox[\linewidth][c]{\begin{minipage}{170mm}\captionsetup{position=below,skip=8pt}\caption{\textbf{Which coded outcomes and language categories do bCRT and CRT2 predict?} \textbf{a}, Coded outcomes: odds ratio per SD of bCRT ($y$) against odds ratio per SD of CRT2 ($x$), with 95\% CIs, from separate logistic models for the fifteen primary coded outcomes, numbered by family in the list below the panels. CRT2 is familiarity-adjusted (\hyperref[par:crt2]{Methods}). \textbf{b}, Language: $t$ value of bCRT ($y$) against $t$ value of CRT2 ($x$) from separate OLS models for each focal LIWC category; CRT2 enters with its count of items seen before as a covariate. \textit{Insight}, \textit{negative emotion}, \textit{moral} and \textit{risk} words were predicted a priori; for the exploratory categories, significance is the within-family corrected $p$ printed in Table~\ref{tab:liwc_bcrt}. Unstandardised odds ratios in Supplementary Tables~\ref{tab:brt_primary} and~\ref{tab:brt_block_crt2_adj}; language in Supplementary Table~\ref{tab:liwc_crt2}. bCRT $n=473$; CRT2 $n=471$.}\label{fig:dissociation}\end{minipage}}
\end{figure}

Figure~\ref{fig:dissociation} brings this language contrast together with the coded outcomes. Where bCRT was associated with an outcome---the focal language categories and, among the coded outcomes, the four search outcomes and worry---CRT2 was not:
the familiarity-adjusted CRT2 reached $q<.05$ on none of the fifteen primary coded outcomes, and neither did raw CRT2 (Supplementary Tables~\ref{tab:brt_block_crt2_raw} and~\ref{tab:brt_block_crt2_adj}).
The comparison is descriptive, not a formal test of the difference between instruments.

\begin{table}[!htbp]
\centering
\footnotesize
\caption{\textbf{Which language categories does bCRT predict?} Behavioural Reflection Test (BRT): psychologically interpretable LIWC categories significantly predicted by raw bCRT total in primary bCRT LIWC models (Prolific sample, $N=473$). Each row is a separate OLS model of the form LIWC category $\sim$ bCRT total; $\beta$ is the unstandardised coefficient in original LIWC-score units. Predicted categories ($\dagger$: \textit{negative emotion}, following Mosleh et al.; \textit{risk}, given the scenario design) were each tested singly and are reported with their per-test $p$; exploratory categories are Benjamini--Hochberg corrected within LIWC families ($p_{\mathrm{FDR}}$), except \textit{present-focus}, which is the sole exploratory member of its family and is therefore reported with its per-test $p$.}
\label{tab:liwc_bcrt}

\begin{tabular*}{\textwidth}{@{\extracolsep{\fill}}lrrrrr@{}}
\toprule
LIWC category      & $\beta$ & SE    & $t$   & $p$  & $R^{2}$ \\
\midrule
Negative emotion$^{\dagger}$ & 0.055 & 0.027 & 2.03 & .043 & 0.009 \\
Discrepancy                  & -0.219 & 0.072 & -3.03 & .013 & 0.019 \\
Differentiation              & -0.176 & 0.066 & -2.67 & .020 & 0.015 \\
Auxiliary verbs              & -0.328 & 0.083 & -3.94 & $<$.001 & 0.032 \\
Risk words$^{\dagger}$       & 0.077 & 0.024 & 3.20 & .001 & 0.021 \\
Present-focus                & 0.205 & 0.074 & 2.78 & .006 & 0.016 \\
\bottomrule
\end{tabular*}
\end{table}

\FloatBarrier
\section*{Discussion}

The present study makes two connected contributions. The primary contribution is the BRT: a compact behavioural test that reveals how reflective thinking is expressed in the language people use and in how they seek information when weighing high-quality evidence against noisier social information. The second contribution is the bCRT, a fresher CRT-style anchor against which to validate the BRT. Together, the two measures offer a pragmatic way to study reflection both as a latent trait and as an enacted pattern of reasoning in ecologically plausible scenarios, while remaining brief enough for online administration.

Turning to the BRT, the linguistic results provide convergent evidence that the task is sensitive to reflective reasoning, while complicating any simple `more reflection = more complexity words' account. The clearest functional-word signal comes from auxiliary verb use. As shown in Table~\ref{tab:liwc_bcrt}, higher bCRT scores were associated with reduced auxiliary verb use. Auxiliary verbs (e.g., `is', `have', `will') are typically associated with a more informal, narrative style and have been linked to poorer academic performance \cite{pennebaker2014WhenSmallWords}. Their reduced use among higher-scoring participants is therefore consistent with a more analytic, less narrative register---not a briefer one, since higher scorers wrote longer responses overall.

The bCRT was also associated with greater use of \textit{risk}-related language. The \textit{risk} category in LIWC appears to capture \textit{attention} to hazards rather than a generic appetite or aversion for risk \cite{thorpehuerta2021Exploringdiscussionshealth}. Here, higher bCRT predicted both this heightened attention and more targeted verification (checking the claim at its source and against institutional and official records), but not the final decision. This suggests that higher-bCRT participants more explicitly identified the risk-bearing features of the situation and planned to check them.
In Frederick's account, more reflective individuals are willing to accept small or controlled risks when the expected payoff is favourable \cite{frederick2005}. That account predicts a decisional difference---keeping the booking or following medical advice more often---which we did not detect.

Higher bCRT also predicted greater \textit{present-focus} (Table~\ref{tab:liwc_bcrt}). Unlike the categories above, this was not a theoretically motivated prediction: \textit{present-focus} emerged in the pilot and was carried into the main (validation) run as an exploratory time-focus check, where it replicated. Its dispositional correlates are mixed---elsewhere \textit{present-focus} tracks impulsivity and lower well-being \cite{park2017LivingPresentFuture}---so we read the signal cautiously, most plausibly as engagement with the \textit{immediacy} of the decision rather than a stable temporal orientation.

Consistent with Mosleh et al. \cite{mosleh2021}, higher bCRT scores were also associated with greater use of \textit{negative emotion} words. At the linguistic level we read this as a marker of affective engagement with the adverse, risky, or socially consequential features of the decision context, in line with Baumeister et al.'s account of emotion as a feedback system that makes decision-relevant signals more salient and supports counterfactual reasoning and guides behaviour \cite{baumeister2007HowEmotionShapes}. The coded outcomes---the behaviours each hypothesis predicted, judged from what each response says rather than counted from its words---point the same way: higher bCRT also predicted \textit{voicing worry about the medication} (H3), though not a \textit{hostile rebuke of the unhelpful friend}.

The negative associations with \textit{discrepancy} and \textit{differentiation} markers complicate the interpretation. In LIWC, \textit{discrepancy} captures words such as `should', `would', and `could', which mark conditional, counterfactual, or obligation-oriented framing \cite{tausczik2010PsychologicalMeaningWords}. \textit{Differentiation} markers support contrastive distinctions and have often been treated as indicators of cognitive complexity \cite{tausczik2010PsychologicalMeaningWords,pennebaker2005LinguisticMarkersPsychological}. On face value, lower use of these categories among higher-bCRT participants is not what a simple reflection-as-more-counterfactual-processing account would predict.

One plausible interpretation is that higher-scoring participants resolved uncertainty through verification rather than verbal deliberation. Higher bCRT predicted naming concrete checks (H1), yet did not predict which option participants chose or which source their decision rested on. A participant who has settled how they would check a claim has less need to rehearse competing possibilities in words (perhaps the institution is wrong; perhaps the anecdote should be taken seriously). The uncertainty is handed to an action rather than argued through. On this reading, the lower use of \textit{discrepancy}, \textit{differentiation} and \textit{auxiliary verbs} among higher-bCRT participants reflects a verification-oriented stance: emotionally engaged and risk-attentive, relying on planned checks set out in a more analytic, less narrative register rather than on counterfactual or contrastive scaffolding. The scenarios also give participants nothing further to reason over, so a verification plan defers the reasoning to evidence they do not yet have. Fewer counterfactual words may therefore partly reflect what the task affords. This interpretation should, however, be treated cautiously. A LIWC count records how often a participant used a category's words, not the stance they took. The \textit{discrepancy} word `should', for example, can state an obligation, voice a doubt or set out a plan. The associations show which categories tracked reflection in these decisions, and they become more interpretable when paired with the coded outcomes.

The coded outcomes show more specifically that what moved with reflection was information-seeking, not the decision that followed. Higher bCRT predicted plans to check the contested claim at its source or against institutional records (H1), such as \textit{phoning the restaurant} or \textit{consulting regulators, journals or official guidance} about the medication (Figure~\ref{fig:dissociation}a). However, it predicted none of the five evidence-weighted decision outcomes (H2). These tested whether more reflective participants would reach the decision the stronger evidence supports, as in \textit{accepting the medication on institutional grounds with a verification plan}, and would less often act on the weaker claim without scrutiny, as in \textit{acting on the claim without doubt or a decision rule} (Table~\ref{tab:hypothesis_outcomes}).
The BRT therefore separates two things that are often run together: how people seek to verify a contested claim, and what they decide. What participants said their decision rested on was close to unrelated to the checks they planned (Supplementary Table~\ref{tab:anchor_check_overlap}).

The decision null, however, is bounded: for four of the five decision outcomes, the sample could detect effects smaller than the primary-source verification effect (Supplementary Table~\ref{tab:brt_primary}, MDOR column). \textit{Outright, worry-marked refusal}, the rarest, is the exception.
One reading of the null is that these scenarios allow more than one reflective decision. By design, the Medex counter-source is more credible than the restaurant's---an apparent GP citing medical journals, rather than crowd reviews---so accepting the prescription on the balance of evidence and delaying to verify it are both defensible, evidence-sensitive choices. The present outcomes credit both, as separate primary outcomes, and neither was associated with bCRT. Having two defensible choices therefore cannot by itself explain the null. In the restaurant scenario, changing or cancelling the booking was costly: a non-refundable deposit, three hours' notice and few alternative venues. Most participants kept the booking.
The practical frame may thus have crowded out the epistemic one at the credulous end of the decision, and a future version seeking variation there should relax the sunk cost or the time pressure.

Ultimately, the broader pattern shows convergence with Mosleh et al. \cite{mosleh2021}. As in their data, higher reflection here coincided with better informational hygiene---in the BRT, plans to verify a contested claim (H1). However, the convergence is not complete. Unlike Mosleh et al.'s study, neither LIWC \textit{insight} nor \textit{moral} wording was associated with bCRT in our models, and neither was the rubric-coded \textit{appeal to fairness or loyalty} (H4).

One plausible explanation for the absent moral vocabulary is contextual. Social media environments such as Twitter may more strongly incentivise moralised, \textit{judgement-laden} language \cite{brady2020MADModelMoral}, whereas our scenarios asked for a private decision. The BRT therefore reproduces the informational-hygiene part of the language-behaviour signature that motivated its design, but not, at the effect sizes this sample can detect, its moral part.

Taken together, the language and the coded outcomes show that more reflective participants voiced more worry, used more \textit{negative emotion} words and planned more checks. The key interpretive question is what this \textit{negative emotion} signal reflects.

One reading is that \textit{negative emotion} terms index \emph{evaluative focus} rather than felt negative affect, which the words alone do not establish. On this account, expressing concern about a medication's reported harms reflects sharper tracking of the decision-relevant features of the problem. This is broadly compatible with prior findings that higher CRT is associated with more utilitarian, deliberative responding in moral dilemmas \cite{baron2015WhydoesCognitivea}, and with neurobiological work interpreted as showing \textit{reduced} engagement of affect-related circuitry such as the vmPFC \cite{greene2009Dualprocessmoralitypersonal,greene2007WhyareVMPFCa}. 

An alternative reading is also possible. More reflective participants may genuinely experience stronger negative affect while reasoning in morally charged tasks. Hong et al. found that people with high metacognitive monitoring ability, measured on a separate memory task, outscored others on the CRT after a positive mood induction but not after a negative one~\cite{hong2023Negativeemotioncan}. While this suggests that negative affect can disrupt reflective performance in otherwise stronger performers, \textit{negative emotion} words in a written response are not an induced mood, so the two results are not directly comparable. If the words do mark felt affect, it came alongside more verification, not less: higher bCRT predicted both more \textit{negative emotion} language and more planned checks.

Under both readings, these results suggest that more reflective participants took the video's warning seriously enough to worry about the medication and to plan to check it. Taking the warning seriously was not the same as acting on it: bCRT did not predict \textit{outright, worry-marked refusal} of the medication, or \textit{acting on the claim without doubt or a decision rule} (Supplementary Table~\ref{tab:brt_primary}). This fits the expectation that more reflective people verify a contested claim (H1). In a compact, scenario-based setting, the BRT reproduces the link Mosleh et al. found between reflection and \textit{negative emotion} language \cite{mosleh2021}. Future work could pair the BRT with psychophysiological measures or other direct indices of arousal to better characterise how \textit{negative emotion}, moral and epistemic reasoning, and reflective traits interact.

A final strand concerns the two reflection measures themselves. CRT2 predicted none of the focal LIWC categories or coded outcomes that bCRT predicted (Figure~\ref{fig:dissociation}).
This contrast is clearest for \textit{risk}-category words: the bCRT pattern is consistent with heightened attention to hazards, whereas the corresponding CRT2 model is effectively flat after familiarity is controlled (Table~\ref{tab:liwc_bcrt} and Supplementary Table~\ref{tab:liwc_crt2}). CRT2 also did not predict \textit{negative emotion}, \textit{discrepancy}, \textit{differentiation}, \textit{auxiliary verbs} or \textit{present-focus}, each of which bCRT predicted.

Participants who reported having seen more CRT2 items scored higher on it (about 0.17 points per previously seen item; \hyperref[par:crt2]{Methods}). This is an association, not a measured cost of exposure: people who remember the items may also be better able to solve them~\cite{meyer2018}, and whether exposure weakens the CRT's predictive validity is disputed~\cite{bialek2018,meyer2018,woike2019}. bCRT's convergence with CRT2, NFC and Conscientiousness (Table~\ref{tab:crt_trait_correlations}) is consistent with the two tests tapping a shared reflective disposition. Why only bCRT predicted how participants searched and what they worried about is not established: exposure is one candidate and a difference in what the two item sets capture is another, and the 75 participants who had seen none of the CRT2 items (\hyperref[app:irt]{Supplementary Information}) are too few to separate them. A practical implication is that researchers using the CRT construct in contemporary samples can add newer items such as ours, or less familiar CRT variants, alongside legacy items as a precaution.

Several limitations qualify these conclusions. First, the BRT scenarios are necessarily stylised, compressing key features of everyday reflection into short vignettes, and they record what participants say they would do. This improves control and feasibility but may obscure domain-specific nuances, particularly in moral and political contexts where identities and group loyalties are more entrenched. Extending the BRT to other domains and higher-stakes settings would test the generality of the observed patterns. Future BRT iterations could also add a second, locked phase that presents brief facts from several sources and elicits how the participant weighs them, so that the reasoning a verification plan defers can be observed.

Second, although LLM-based coding allowed scalable operationalisation of decision variables, it remains conditional on the rubric's quality and the models' alignment with human judgement. We reduced subjectivity by pre-defining evidence-source quality, instructing conservative defaults and taking the majority label of three models from different providers. The panel's agreement across two independent coding passes is reported in the \hyperref[app:coding_reliability]{Supplementary Information}. Future work should triangulate these outputs against independent human coding.

Third, whether repeated exposure changes BRT responses was not tested. Although responses are not scored as right or wrong, the coding scheme rewards particular kinds of response, which a returning participant could learn. A retest study would show whether scores shift.

In sum, reflective thinking leaves measurable traces in how people write and in how they seek information. The primary contribution is the BRT, a compact behavioural task that captures how reflection is enacted in emotionally charged, socially embedded decisions. The short bCRT complements it as a new low-exposure measure of reflective ability that links in the expected direction to NFC and Conscientiousness and anchors the BRT's interpretation. More reflective participants articulated risks, voiced worry, and sought to check contested claims at their source and against institutional records. Reflection did not, however, predict the decisions they made. By pairing a scalable behavioural-linguistic task with refreshed psychometrics, the work offers a practical framework for studying when and how reflection is expressed in real-world-style reasoning, and a foundation for work on interventions, individual differences, and computational modelling of reflective judgement.

\section*{Ethics approval and consent to participate}
This study was approved by the University of New South Wales Human Research Ethics Committee (approval reference iRECS7125). All participants were presented with a Participant Information Statement and Consent Form and provided informed consent through the Qualtrics survey platform before taking part.

\section*{Data availability}
The de-identified person-level data ($N=473$) will be made public in a Hugging Face dataset of the Cruise Research Group (\url{https://huggingface.co/datasets/CRUISEResearchGroup/behavioural-reflection-test}) no later than journal publication. Prolific identifiers and other potentially identifying metadata were removed prior to deposition. Responses from the same participants also underpin a companion paper, accepted at NeurIPS 2026, examining large language model performance on these scenarios. They are reused here with no additional data collection.

\section*{Code availability}
The analysis code that reproduces every table, figure and reported number from these data, together with the rubric and prompts used for the language-model panel coding, will be made public in a GitHub repository of the Cruise Research Group (\url{https://github.com/cruiseresearchgroup/behavioural-reflection-test}) no later than journal publication.

\section*{Competing interests}
A.B. is employed by Aurecon Group. The PhD scholarship supporting S.W. was jointly funded by Aurecon Group and the CSIRO, and Aurecon Group additionally funded participant reimbursement for the online study. These organisations had no role in the study design, data collection, analysis, interpretation, or the decision to publish, and this work does not evaluate any product or service of either organisation. The authors declare no other competing interests.

\section*{Funding}
S.W. was supported by a PhD scholarship jointly funded by Aurecon Group and the Commonwealth Scientific and Industrial Research Organisation (CSIRO). Participant reimbursement for the online (Prolific) study was funded by Aurecon Group. The funders had no role in study design, data collection and analysis, the decision to publish, or preparation of the manuscript.

\section*{Author contributions}
S.W. conceived the study; designed the scenarios, scope, and hypotheses; obtained ethics approval; implemented the survey and collected the data; performed all analyses; and wrote the manuscript. B.R.N. provided supervision and cognitive-science guidance on study design, and contributed to the manuscript through critical review and editing. F.S. provided supervision and contributed to the manuscript through critical review and editing. A.B. contributed to funding acquisition and critically reviewed and commented on the manuscript. All authors approved the final version.

\bibliography{sn-bibliography} 
\section*{Supplementary Information}
\setcounter{table}{0}\setcounter{figure}{0}\renewcommand{\tablename}{Supplementary Table}\renewcommand{\figurename}{Supplementary Fig.}\def\theHtable{S\arabic{table}}\def\theHfigure{S\arabic{figure}}
\FloatBarrier
\subsection*{Supplementary Methods: Data Cleaning and Participant Flow}\phantomsection\label{app:cleaning}

\subsubsection*{Exclusions and participant flow}
Exclusions were applied in a fixed order and logged per case:
\begin{enumerate}
  \item \textbf{Blank identifier.} One response with a blank Prolific identifier
  (3\% progress) was removed.
  \item \textbf{Incomplete responses.} Responses that did not complete the reasoning block of the standard four-item Cognitive Reflection Test (CRT2) and the bespoke CRT (bCRT)---which follows the Behavioural Reflection Test (BRT) writing tasks---were removed
  under a single generic criterion---``did not complete enough of the survey for
  inclusion''. This set comprised 139 unfinished rows in total, including the blank-identifier response above and the one duplicated identifier (both unfinished); the majority (112 of 139) terminated immediately after the writing tasks and before the reasoning block, leaving no reflection score with which to model their BRT responses.

  \item \textbf{Reinstatement of near-complete records.} Four unfinished records were
  reinstated on inspection: two were functionally complete (97\% progress; full CRT2, bCRT,
  Need for Cognition (NFC) and Big Five Inventory (BFI) data) and were marked unfinished only because the terminal page-submit timer
  did not fire; two further records held complete bCRT and BRT data but an incomplete CRT2
  total (three of four CRT2 items) and no NFC/BFI responses. These two were reinstated for the
  bCRT and BRT analyses, with their CRT2 answers and the trait scales left as missing
  (no imputation).
  \item \textbf{Invalid response.} One finished response was removed because BRT1 response suggested a misunderstanding of the task, and BRT2 response reproduced the scenario prompt verbatim rather than answering it.
\end{enumerate}

After these steps the analytic sample comprised $N = 473$ participants, or $N = 471$ for analyses requiring a complete CRT2 total or the trait scales.

\subsubsection*{Content review and the attention-check item}
Several records flagged by automated quality signals were retained after inspection, each on its own grounds. One record's fast completion time and another's low reCAPTCHA score reflect response speed and browser/network fingerprinting rather than answer quality; both records carried coherent, on-task, human-authored responses. The two shared-IP pairs each comprised distinct Prolific identifiers---which Prolific's own controls prevent from being the same account twice---with distinct, non-overlapping responses, consistent with separate respondents behind a shared connection (a household network or a common VPN exit); these records were retained on the basis of their distinct content. One participant wrote three words in the first writing task and their answer to that scenario in the second task's box, leaving the second scenario unanswered; the record was retained, with the first scenario coded from that answer and the second taking the rubric's default labels.

The BFI block included a directed-response catch item (correct response: ``Extremely
Uncharacteristic''). Of the 41 responses flagged on this item, 8 were genuine failures (an
off-key substantive selection) and 33 were \emph{blank}. Given the BFI block was
configured for forced response, the presence of blank catch-item responses suggests
the force-response constraint did not apply to this item for a subset of respondents; blank
responses were therefore treated as item-level missingness rather than attention failures. 
After manual inspection, the 8 genuine failures were retained---all wrote substantive, on-task BRT responses and
had complete, coherent CRT data---and attention-check status was instead encoded as a covariate
(pass, fail, or missing). Because this item indexes attentiveness during the trait questionnaire block rather than the earlier, directly inspected reasoning responses, it bears on the trait-scale analyses only and was not used as an exclusion criterion for the BRT, bCRT, or CRT2 models. As a sensitivity analysis for the Linguistic Inquiry and Word Count (LIWC) models, the 8 genuine failures were dropped and the 33 blank responses retained ($n=465$; Supplementary Results: Sensitivity analyses and response length).

\FloatBarrier
\subsection*{Supplementary Methods: LIWC families and category mapping}\phantomsection\label{app:liwc_families}

LIWC scores were treated as the proportion of words in each response belonging to each category. For the main analyses, categories were grouped into families motivated by theory and prior LIWC work linking function words and psychological processes \cite{tausczik2010PsychologicalMeaningWords}:

\begin{itemize}
  \item \textbf{Quantity/structure}: \textit{word count} and \textit{words per sentence}.
  \item \textbf{Function words}: \textit{articles}, \textit{prepositions}, \textit{first-person singular pronouns}, \textit{other pronouns}, \textit{auxiliary verbs}, \textit{conjunctions}, and \textit{adjectives}.
  \item \textbf{Cognitive processes}: \textit{insight}, \textit{causation}, \textit{discrepancy}, \textit{tentativeness}, \textit{differentiation}, and \textit{certainty}.
  \item \textbf{Affect}: \textit{negative emotion}.
  \item \textbf{Time-focus}: \textit{present-focus}.
  \item \textbf{Motivational/risk language}: \textit{risk-related} terms.
  \item \textbf{Normative content}: \textit{moral}.
\end{itemize}

The models and the split between predicted and exploratory categories are described in Methods. Single-member exploratory families involve no multiple comparison and are reported with the uncorrected $p$. Broad LIWC summary variables (Analytic, Clout, Authentic, and Tone) were retained in the analysis outputs for exploratory/descriptive use but were not included in the focal manuscript tables.

\FloatBarrier
\subsection*{Supplementary Methods: Full Behavioural Reflection Test scenarios}
\label{app:brt_full_items}

Participants completed two open-ended BRT scenarios designed to elicit reasoning about evidence quality, trust, risk, and social obligation. Participants were instructed to respond in free text as if thinking aloud while deciding what to do in each situation.

\subsubsection*{BRT1: Restaurant booking scenario}

\begin{quote}
Please aim for at least 150 words, or 4+ sentences. There are no upper limits.

\vspace{0.5em}

One month ago, you and Friend A were supposed to find and book a restaurant for Friend B's birthday.

However, when you asked Friend A to help, they stopped responding to your texts. As a result, you found a restaurant on your own and paid a non-refundable deposit for the booking.

Now, three hours before the dinner, Friend A sends you a message with screenshots of Google Reviews for the restaurant you chose --- claiming some diners got food poisoning --- and says, ``maybe we shouldn't go...''

You remember the restaurant has great reviews from food critics. Dinner is tonight; most nearby venues are already full and you need to make a decision quickly.

Write your thoughts as if thinking aloud --- from the moment you read Friend A's message until you decide what to do (keep or change reservation). Include anything that crosses your mind --- and consider:

\begin{enumerate}[label=\Alph*.]
    \item Your immediate gut or emotional reaction
    \item What (if anything) you'd say to Friend A right away
    \item How you decide which information is trustworthy
    \item Any extra checks or actions you'd take before choosing
    \item Finish by stating your final choice (keep the booking or change plans) and why.
\end{enumerate}
\end{quote}

\subsubsection*{BRT2: Medication decision scenario}

\begin{quote}
Aim for at least 150 words, or 4+ sentences. There are no upper limits.

\vspace{0.5em}

A doctor prescribed you ``Medex'' for recurring headaches --- informing you that clinical trials indicate it works well and has minimal side effects, which are mild.

That evening, while on social media, you come across a video by Dr George Miles, a practising GP with a large online following. Dr Miles warns that Medex poses serious long-term risk to liver health that patients are often not informed about. He shows images of medical journal articles supporting his claim.

The post is very popular and has mixed comments. Some thank Dr Miles for the warning; others say the data are being misrepresented.

Write your thoughts as if thinking aloud, from the moment you watch the video until you decide whether to start taking Medex. Include anything that occurs to you --- but please consider:

\begin{enumerate}[label=\Alph*.]
    \item Your immediate emotional or gut reaction
    \item Anything you would say or message to your clinician, friends, or family
    \item How you decide which information is trustworthy
    \item Any checks or actions you would take before choosing
    \item Finish by stating your final decision and why.
\end{enumerate}
\end{quote}

\FloatBarrier
\subsection*{Supplementary Methods: BRT coding rubric, panel and derived outcomes}
\label{app:brt_coding_scheme}

This section gives the rubric, panel and rules behind the coding workflow in Methods (Figure~\ref{fig:coding_workflow}). The rubric defined every allowable category in advance, with worked examples. Each model received both scenario texts and the participant's two responses and returned the 28 fields in Supplementary Figure~\ref{fig:brt_rubric} as one structured object, each code with the text span supporting it; a non-default code without a supporting span was invalid.

\begin{figure}[!tbp]
  \centering
  \makebox[\textwidth][c]{\begin{subfigure}{170mm}\centering\includegraphics[width=\linewidth]{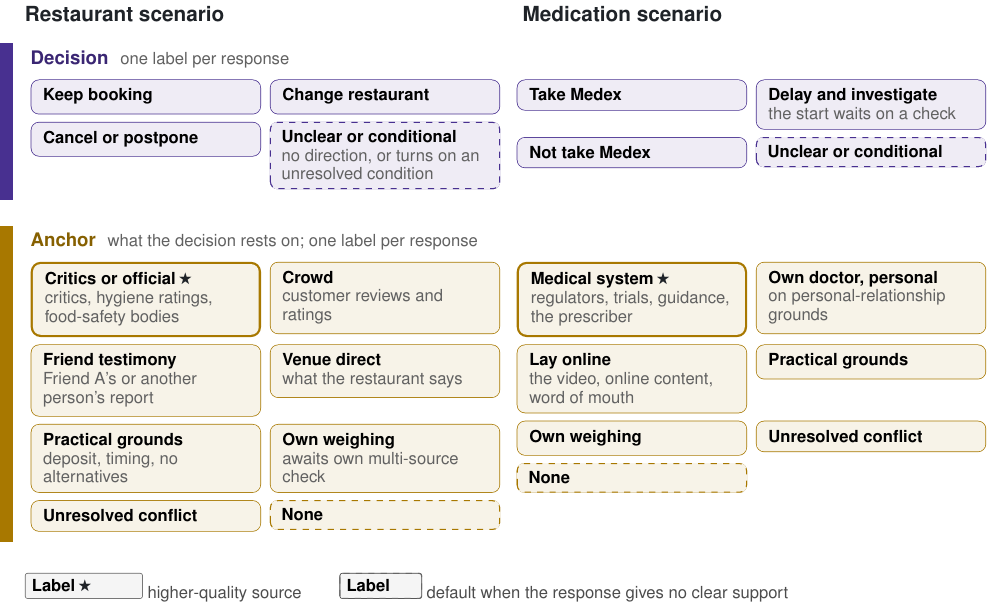}\caption{Decision and anchor}\end{subfigure}}
  \makebox[\linewidth][c]{\begin{minipage}{170mm}\captionsetup{position=below}\caption{\textbf{What does each of the 28 coded fields record?} The coding rubric, by scenario: one box per label or flag, coloured by field type, with a short meaning in grey. Decision and anchor take one label per response; planned checks, reasoning and affect, and scenario misread are each coded present or absent. Bold frame and star: higher-quality source, as defined in Methods. Dashed frame: the default label, assigned when the response gives no clear support. \textbf{a}, decision and anchor; \textbf{b}, planned checks, reasoning and affect, and scenario misread.}\label{fig:brt_rubric}\end{minipage}}
\end{figure}

\begin{figure}[!tbp]
  \ContinuedFloat
  \centering
  \makebox[\textwidth][c]{\begin{subfigure}{170mm}\centering\includegraphics[width=\linewidth]{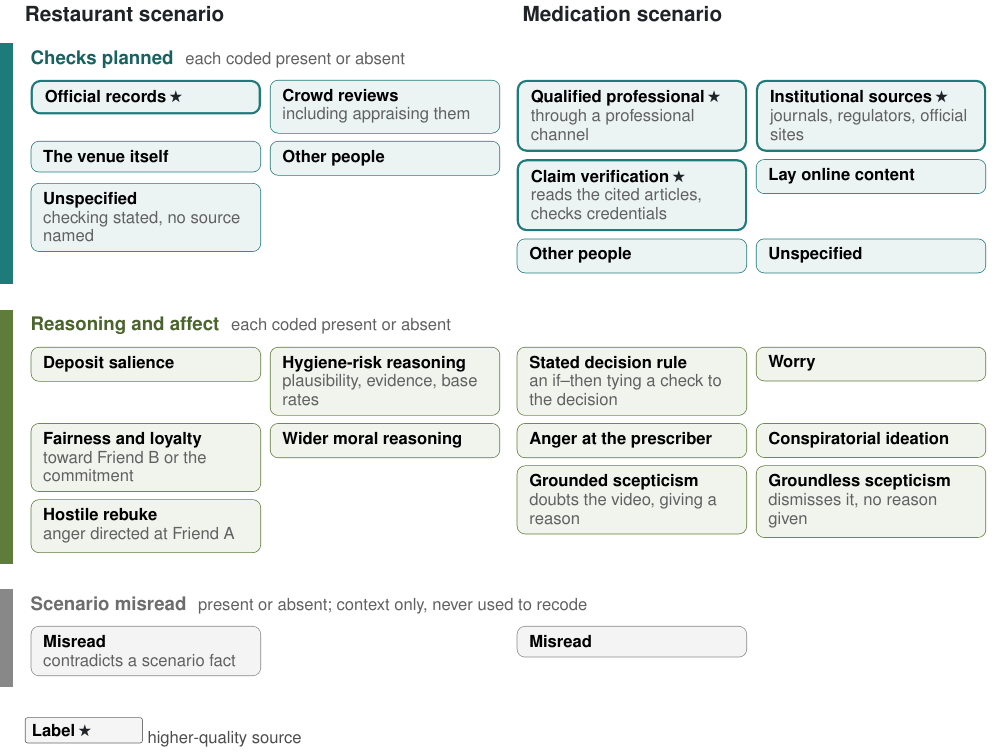}\caption{Planned checks, reasoning and affect, and scenario misread}\end{subfigure}}
  \makebox[\linewidth][c]{\begin{minipage}{170mm}\captionsetup{position=below}\caption{(continued)}\end{minipage}}
\end{figure}

\paragraph{Panel.}
The three models were GPT-5.6 Sol (reasoning effort low), Gemini 3.5 Flash (minimal thinking) and DeepSeek V4-Flash (thinking disabled, served via Fireworks). Each field took a label from a closed set, and outputs were validated against a fixed schema. The consensus label for each field was the majority across the three models; a field with no majority took the rubric's conservative default and was flagged as split. Sampling was left at each provider's default: one model does not accept a temperature setting, a temperature of zero does not guarantee identical outputs across providers, and residual ambiguity in the rubric is more useful expressed as measurable disagreement between two coding passes, which the reliability gates evaluate, than hidden by forced agreement. Labels were frozen once minted.

\paragraph{Derived outcomes.}
Derived outcomes were computed from the consensus codes by fixed rules in one implementation (Supplementary Table~\ref{tab:brt_derived}); a missing ingredient made the derived value missing.

\begin{table}[!htbp]
\centering
\footnotesize
\caption{\textbf{How is each derived outcome built from the coded fields?} Rules applied to the consensus codes (Supplementary Figure~\ref{fig:brt_rubric}). Upper four tiers: evidence exclusive, evidence mixed, consult only or consult mixed.}
\label{tab:brt_derived}
\begin{tabularx}{\textwidth}{@{}>{\raggedright\arraybackslash}p{48mm}>{\raggedright\arraybackslash}X@{}}
\toprule
Derived outcome & Rule \\
\midrule
\rowcolor{black!6}\multicolumn{2}{@{}l}{\textbf{Profiles and analysed forms}} \\
\rowcolor{black!6}Verification-quality profile, medication & Evidence-grade checking: institutional sources or claim verification; consult-grade: a qualified professional; lay: lay online content or other people. Evidence exclusive (evidence-grade, no lay check); evidence mixed (evidence-grade with a lay check); consult only (professional, no evidence-grade or lay check); consult mixed (professional with a lay check, no evidence-grade); lay only; unspecified only; none. The search-plan profile merges the last three into no plan \\
\rowcolor{black!6}Verification-quality profile, restaurant & Official exclusive (official check, no crowd or personal check; contacting the venue is ignored); official mixed; non-institutional only; unspecified only; none \\
\rowcolor{black!6}Unspecified search, analysed form & Unspecified coded and no specific check coded in the same answer, per scenario \\
\multicolumn{2}{@{}l}{\textbf{Primary composites (H2)}} \\
Keeps the booking on evidence grounds & Keep booking, and hygiene-risk reasoning, an official check or the critics-or-official anchor \\
Accepts on institutional grounds with a verification plan & Take Medex, medical-system anchor, a medication profile in the upper four tiers, and no conspiratorial ideation \\
Delays with a verification plan and a stated condition & Delay and investigate, a medication profile in the upper four tiers, and a stated decision rule \\
Acts on the claim with no doubt and no rule & Delay and investigate or not take Medex, no grounded or groundless scepticism, and no stated decision rule \\
Refuses outright, worry-marked & Not take Medex, worry, and no stated decision rule \\
\rowcolor{black!6}\multicolumn{2}{@{}l}{\textbf{Sensitivity composites}} \\
\rowcolor{black!6}Accepts on institutional grounds, exclusive-tier plan or no plan & As the acceptance composite, with an evidence-exclusive, consult-only or no-check profile \\
\rowcolor{black!6}Delays with an exclusive-tier plan and a stated condition & As the delay composite, with an evidence-exclusive or consult-only profile \\
\multicolumn{2}{@{}l}{\textbf{Exploratory derived outcomes}} \\
Keeps the booking & Keep booking \\
Takes or delays the medication & Take Medex, or delay and investigate \\
Rests on the critics' or official record & Critics-or-official anchor \\
Rests on the medical system & Medical-system anchor \\
Takes the medication with any check plan & Take Medex with at least one planned check \\
\bottomrule
\end{tabularx}
\end{table}

\paragraph{Anchor and planned checks.}
Because the anchor and the planned checks are coded separately, a participant can rest a decision on one source and plan to check another. The two were close to statistically unrelated: in each scenario, the number of participants who both rested the decision on the higher-quality source and planned a check of that grade was close to the number expected if anchor and check were independent (Supplementary Table~\ref{tab:anchor_check_overlap}). Supplementary Tables~\ref{tab:si_crosstab_brt1} and~\ref{tab:si_crosstab_brt2} give the full counts by anchor, decision and planned-check profile.

\begin{table}[!htbp]
\centering
\footnotesize
\caption{\textbf{Do participants who rest the decision on a higher-quality source also plan to check one?} Anchor: rests on the critics or official record (restaurant) or the medical system (medication). Check: plans an official check (restaurant: official exclusive or official mixed) or an evidence-grade check (medication: evidence exclusive or evidence mixed). Expected: participants in both if anchor and check were unrelated, with the corresponding Jaccard index. $N=473$.}
\label{tab:anchor_check_overlap}
\begin{tabular*}{\textwidth}{@{\extracolsep{\fill}}lrrrrrrrr@{}}
\toprule
Scenario & $n$ & Anchor & Check & Both & Expected & Jaccard & \makecell[r]{Jaccard\\expected} & $\phi$ \\
\midrule
Restaurant & 473 & 85 & 68 & 11 & 12.2 & 0.077 & 0.087 & -0.019 \\
Medication & 473 & 170 & 258 & 98 & 92.7 & 0.297 & 0.277 & 0.047 \\
\bottomrule
\end{tabular*}
\end{table}

\begin{table}[!htbp]
\centering
\footnotesize
\caption{\textbf{Restaurant: what did the decision rest on, what was decided, and which checks were planned?} Counts by anchor (rows) against the decision and against the planned-check profile (columns); each column group sums to $N=473$.}
\label{tab:si_crosstab_brt1}
\begin{tabular*}{\textwidth}{@{\extracolsep{\fill}}lrrrrrrrrrr@{}}
\toprule
& \multicolumn{4}{c}{Decision} & \multicolumn{5}{c}{Planned checks} & \\ \cmidrule(lr){2-5}\cmidrule(lr){6-10}
Anchor & \rotatebox{90}{Cancel or postpone} & \rotatebox{90}{Change restaurant} & \rotatebox{90}{Keep booking} & \rotatebox{90}{Unclear or conditional} & \rotatebox{90}{No check} & \rotatebox{90}{Non-institutional only} & \rotatebox{90}{Official only} & \rotatebox{90}{Official and other} & \rotatebox{90}{Unspecified only} & Total \\
\midrule
Critics or official record & 0 & 0 & 79 & 6 & 39 & 33 & 4 & 7 & 2 & 85 \\
Crowd reviews & 2 & 8 & 12 & 4 & 5 & 20 & 1 & 0 & 0 & 26 \\
Friend's testimony & 1 & 1 & 0 & 2 & 2 & 2 & 0 & 0 & 0 & 4 \\
The restaurant itself & 0 & 0 & 2 & 6 & 0 & 8 & 0 & 0 & 0 & 8 \\
Practical grounds & 6 & 5 & 206 & 7 & 97 & 97 & 8 & 19 & 3 & 224 \\
Own weighing & 0 & 2 & 7 & 51 & 1 & 39 & 0 & 18 & 2 & 60 \\
Unresolved conflict & 0 & 0 & 0 & 1 & 1 & 0 & 0 & 0 & 0 & 1 \\
None & 0 & 7 & 29 & 29 & 16 & 37 & 1 & 10 & 1 & 65 \\
\midrule
Total & 9 & 23 & 335 & 106 & 161 & 236 & 14 & 54 & 8 & 473 \\
\bottomrule
\end{tabular*}
\end{table}

\begin{table}[!htbp]
\centering
\footnotesize
\caption{\textbf{Medication: what did the decision rest on, what was decided, and which checks were planned?} Counts by anchor (rows) against the decision and against the planned-check profile (columns), as in Supplementary Table~\ref{tab:si_crosstab_brt1}; each column group sums to $N=473$.}
\label{tab:si_crosstab_brt2}
\begin{tabular*}{\textwidth}{@{\extracolsep{\fill}}lrrrrrrrrrrrr@{}}
\toprule
& \multicolumn{4}{c}{Decision} & \multicolumn{7}{c}{Planned checks} & \\ \cmidrule(lr){2-5}\cmidrule(lr){6-12}
Anchor & \rotatebox{90}{Delay and investigate} & \rotatebox{90}{Not take Medex} & \rotatebox{90}{Take Medex} & \rotatebox{90}{Unclear or conditional} & \rotatebox{90}{Professional and lay} & \rotatebox{90}{Professional only} & \rotatebox{90}{Evidence only} & \rotatebox{90}{Evidence and lay} & \rotatebox{90}{Lay only} & \rotatebox{90}{No check} & \rotatebox{90}{Unspecified only} & Total \\
\midrule
Medical system & 69 & 3 & 92 & 6 & 15 & 33 & 53 & 45 & 4 & 15 & 5 & 170 \\
Own GP, personal grounds & 5 & 0 & 14 & 0 & 1 & 5 & 1 & 7 & 0 & 5 & 0 & 19 \\
Lay online sources & 4 & 29 & 4 & 2 & 6 & 6 & 4 & 3 & 6 & 7 & 7 & 39 \\
Practical grounds & 0 & 11 & 17 & 0 & 4 & 5 & 6 & 3 & 3 & 6 & 1 & 28 \\
Own weighing & 145 & 7 & 18 & 6 & 24 & 24 & 54 & 58 & 9 & 0 & 7 & 176 \\
Unresolved conflict & 0 & 0 & 0 & 1 & 0 & 0 & 1 & 0 & 0 & 0 & 0 & 1 \\
None & 14 & 8 & 13 & 5 & 7 & 5 & 15 & 8 & 0 & 4 & 1 & 40 \\
\midrule
Total & 237 & 58 & 158 & 20 & 57 & 78 & 134 & 124 & 22 & 37 & 21 & 473 \\
\bottomrule
\end{tabular*}
\end{table}

\subsubsection*{Coding reliability}
\label{app:coding_reliability}
Panel reliability was assessed by coding a fixed panel of 150 responses (random seed 42) twice, independently, and comparing the consensus labels across the two passes (Supplementary Table~\ref{tab:coding_reliability}). For each field we report percentage agreement and Fleiss' $\kappa$, together with the prevalence-adjusted bias-adjusted kappa (PABAK) and Gwet's AC1, which remain interpretable where a flag is rare and $\kappa$ becomes unstable. Every field with a gate met it except three primary fields, the restaurant anchor, the medication decision and the medication anchor, which are reported with their reliability, and the raw medication unspecified-search flag, whose analysed form met it. Each model row in Supplementary Tables~\ref{tab:brt_primary} and~\ref{tab:brt_exploratory} reports the minimum agreement and $\kappa$ across the fields its outcome is built from.

\begin{table}[!htbp]
\centering
\footnotesize
\caption{\textbf{How consistently did the panel code each field?} Agreement between the two coding passes over the 150-response panel. Agr.: \% of responses receiving the same consensus label in both passes. $\kappa$: Fleiss' kappa; PABAK: prevalence-adjusted bias-adjusted kappa; AC1: Gwet's AC1. Prev.: panel prevalence of the flag in the 150 (binary fields). Gate: $\geq$95\% agreement for the two decision and two anchor fields, $\geq$90\% for contributing fields; ---, no gate assigned. $^{\S}$Prevalence below 5\% in the full sample: $\kappa$ is unstable and PABAK and AC1 are the interpretable statistics. $^{\ddagger}$Below its gate; reported with its reliability. $^{\P}$Raw flag, retained as an ingredient of the verification-quality profiles; its analysed form (\texttt{\_sup}; Supplementary Table~\ref{tab:brt_derived}) is in the last two rows.}
\label{tab:coding_reliability}
\begin{tabular*}{\textwidth}{@{\extracolsep{\fill}}lrrrrrl@{}}
\toprule
Field & Agr.\ (\%) & $\kappa$ & PABAK & AC1 & Prev.\ (\%) & Gate \\
\midrule
\texttt{BRT1\_decision} & 98.0 & 0.96 & 0.96 & 0.98 & --- & pass \\
\texttt{BRT1\_anchor} & 85.3 & 0.79 & 0.71 & 0.83 & --- & below$^{\ddagger}$ \\
\texttt{BRT2\_decision} & 90.0 & 0.85 & 0.80 & 0.87 & --- & below$^{\ddagger}$ \\
\texttt{BRT2\_anchor} & 79.3 & 0.73 & 0.59 & 0.76 & --- & below$^{\ddagger}$ \\
\texttt{BRT1\_check\_official} & 99.3 & 0.97 & 0.99 & 0.99 & 10.7 & pass \\
\texttt{BRT1\_check\_crowd} & 96.7 & 0.93 & 0.93 & 0.93 & 45.3 & pass \\
\texttt{BRT1\_check\_venue} & 100.0 & 1.00 & 1.00 & 1.00 & 12.7 & pass \\
\texttt{BRT1\_check\_persons} & 97.3 & 0.89 & 0.95 & 0.96 & 14.7 & pass \\
\texttt{BRT1\_check\_unspecified} & 94.0 & 0.61 & 0.88 & 0.93 & 10.7 & pass \\
\texttt{BRT1\_deposit\_salience} & 99.3 & 0.99 & 0.99 & 0.99 & 48.0 & --- \\
\texttt{BRT1\_hygiene\_risk\_reasoning} & 90.7 & 0.77 & 0.81 & 0.84 & 70.0 & pass \\
\texttt{BRT1\_moral\_fairness\_loyalty} & 93.3 & 0.80 & 0.87 & 0.90 & 21.3 & pass \\
\texttt{BRT1\_moral\_extended} & 77.3 & 0.54 & 0.55 & 0.55 & 48.7 & --- \\
\texttt{BRT1\_emotional\_outburst} & 91.3 & 0.78 & 0.83 & 0.86 & 26.7 & --- \\
\texttt{BRT1\_stem\_misread} & 98.0 & 0.79$^{\S}$ & 0.96 & 0.98 & 5.3 & --- \\
\texttt{BRT2\_check\_professional} & 92.0 & 0.83 & 0.84 & 0.85 & 62.7 & pass \\
\texttt{BRT2\_check\_institutional\_sources} & 96.7 & 0.87 & 0.93 & 0.96 & 14.0 & pass \\
\texttt{BRT2\_check\_claim\_verification} & 95.3 & 0.88 & 0.91 & 0.92 & 28.0 & pass \\
\texttt{BRT2\_check\_lay\_content} & 96.0 & 0.89 & 0.92 & 0.94 & 25.3 & pass \\
\texttt{BRT2\_check\_persons} & 98.7 & 0.96 & 0.97 & 0.98 & 21.3 & pass \\
\texttt{BRT2\_check\_unspecified} & 89.3 & 0.77 & 0.79 & 0.80 & 34.0 & below$^{\P}$ \\
\texttt{BRT2\_contingency\_stated} & 92.7 & 0.84 & 0.85 & 0.86 & 40.0 & pass \\
\texttt{BRT2\_emo\_worry} & 96.0 & 0.92 & 0.92 & 0.92 & 46.7 & pass \\
\texttt{BRT2\_emo\_prescriber\_reactance} & 97.3 & 0.83 & 0.95 & 0.97 & 10.0 & --- \\
\texttt{BRT2\_conspiracy\_ideation} & 100.0 & 1.00$^{\S}$ & 1.00 & 1.00 & 5.3 & pass \\
\texttt{BRT2\_sceptic\_grounded} & 96.7 & 0.93 & 0.93 & 0.93 & 51.3 & pass \\
\texttt{BRT2\_sceptic\_groundless} & 98.7 & 0.49$^{\S}$ & 0.97 & 0.99 & 1.3 & pass \\
\texttt{BRT2\_stem\_misread} & 95.3 & 0.44$^{\S}$ & 0.91 & 0.95 & 4.7 & --- \\
\texttt{BRT1\_check\_unspecified\_sup} & 98.7 & 0.74$^{\S}$ & 0.97 & 0.99 & 3.3 & pass \\
\texttt{BRT2\_check\_unspecified\_sup} & 99.3 & 0.95$^{\S}$ & 0.99 & 0.99 & 6.7 & pass \\
\bottomrule
\end{tabular*}
\end{table}

\FloatBarrier
\subsection*{Supplementary Methods: Exploratory and sensitivity models}
\label{app:brt_exploratory_methods}

Beyond the primary outcomes, coded and derived outcomes that bear on the same hypotheses without testing them directly were analysed as six exploratory families, each tagged with the hypothesis it extends: \emph{appraises the claim's evidence} (doubts the video with a stated reason; evaluates the plausibility, evidence or base rates of the claim; extends H2); \emph{what the decision rested on} (anchor on the critics or official record; anchor on the medical system; extends H2); \emph{which choice they made} (keeps the booking; takes or delays the medication---refusal being the only choice that sides with the low-quality source; extends H2); \emph{low-scrutiny responses} (vague-only search in each scenario, conspiratorial ideation, anger at the prescriber; the negative poles of H1 and H2); \emph{miscellaneous} (reads the reviews or crowd sources, consults lay online content, states an if-then rule, wider moral reasoning, sunk-cost salience, takes the medication with any check plan; no directional prediction); and the \emph{medication search-plan profile}.

Each binary outcome was analysed with the primary-family models and within-family correction (Methods). All 15 primary models were fitted on all $N=473$ participants with no missing outcome or predictor value; each was an ordinary maximum-likelihood logistic regression with no separation, and none had fewer than the 20 events below which a Firth-penalised fit is used (smallest event count 29, outright worry-marked refusal). In total the analysis comprises 15 primary tests across four families and 16 exploratory binary tests across five families, plus one multinomial model as the sixth exploratory family; no battery-wide correction was computed.
The search-plan profile was analysed with one multinomial logistic regression over its five unordered categories, with no plan (no evidence-grade or professional plan) as the reference, one likelihood-ratio test as the family test, and a relative-risk ratio per category; no ordering among the categories is assumed. The restaurant profile was not modelled: its only informative category (an official source named) is the official-records check already tested in H1. Four further outcomes were reported descriptively only, as counts at each bCRT score with no model fitted: credulous switching in the restaurant scenario (changing or cancelling the booking with no hygiene-risk reasoning and no planned check), groundless scepticism of the video, and a misread of each scenario (Supplementary Table~\ref{tab:si_counts_by_bcrt}).

Sensitivity analyses comprised stricter variants of the two H2 plan-based composites, admitting only verification plans without any lay check (rows beneath H2 in Supplementary Table~\ref{tab:brt_primary}); the predictor comparison described in Methods, extended by a joint model entering bCRT, familiarity-adjusted CRT2 and NFC as standardised predictors, each block re-running its within-family corrections (Supplementary Tables~\ref{tab:brt_block_crt2_raw}--\ref{tab:brt_block_joint}); models adding age, gender and education, and a refit excluding responses on which the panel split on any field an outcome is built from (Supplementary Table~\ref{tab:brt_cov_nosplit}); and bCRT entered as an ordered factor (Supplementary Table~\ref{tab:brt_ordered_factor}).

\FloatBarrier
\subsection*{Supplementary Methods: CRT item pools and participant procedures}
\subsubsection*{Response scoring}
\label{app:crt_scoring}

This section details the CRT2 and bCRT scoring summarised in Methods. Before matching, each answer was normalised; the keys list the correct answer, the intuitive lure and explicit other answers for each item (Supplementary Table~\ref{tab:bcrt_keys} for the bCRT). Each language model was shown the item and its keys, and an answer with no majority label was coded \textit{other}. Blank answers were coded as missing. Every vote is released. Against a blinded annotation of 125 human answers by one author, the first version of this scoring agreed on 86.4\% of answers ($\kappa = .79$); the released version agrees with the same annotation, after the annotation's own errors were corrected, on 99.2\%, a consistency check rather than an independent validation.

\subsubsection*{Candidate bCRT item pool and final retained items}\phantomsection\label{app:bcrt_items}

The four-item bCRT used in the Prolific validation sample was selected from a broader pool of candidate items evaluated during development. The final retained items and additional tested candidates are listed below.

\paragraph{Final bCRT items.}

\begin{enumerate}
\item A man buys a cow for \$800. He sells it for \$1000. Then he buys the cow back for \$1100 and sells it again for \$1300. How much money did he make in total?

\item A snail climbs up a 10-foot wall. Each day it climbs 3 feet but slides down 2 feet each night. How many days will it take to reach the top?

\item A rope ladder hangs over the side of a ship. The bottom rung touches the water. The distance between rungs is 20 cm and the tide rises at 15 cm per hour. How long until three rungs are covered?

\item Jack is looking at Anne, but Anne is looking at George. Jack is married, while George is not. Is a married person looking at an unmarried person?
\end{enumerate}

\begin{table}[!htbp]
\centering
\footnotesize
\caption{\textbf{Which answer is correct, and which is the intuitive lure, on each bCRT item?} Items in the order listed above. Answers matching neither were coded \textit{other} (Response scoring); the full answer patterns, including explicit other answers, are released with the analysis code.}
\label{tab:bcrt_keys}
\begin{tabular*}{\textwidth}{@{\extracolsep{\fill}}>{\raggedright\arraybackslash}p{48mm}>{\raggedright\arraybackslash}p{38mm}>{\raggedright\arraybackslash}p{32mm}@{}}
\toprule
Item & Correct answer & Intuitive lure \\
\midrule
Cow bought and sold twice & 400 (dollars) & 300 (dollars) \\
Snail on a 10-foot wall & eight days & ten days \\
Rope ladder and rising tide & never --- the ladder rises with the tide & 4 hours (or 2 hours 40 minutes) \\
Married person looking at an unmarried person & yes & cannot be determined \\
\bottomrule
\end{tabular*}
\end{table}

\paragraph{Additional candidate items evaluated during development.}

\begin{enumerate}
\item If it takes 5 minutes to boil one egg, how long would it take to boil 4 eggs?

\item How many months have 28 days?

\item How many 2-cent stamps are in a dozen?

\item You are in a dark room with a candle, a wood stove, and a gas lamp. You only have one match. What do you light first?

\item What do you get when you divide 10 by half and add 10?

\item A plane crashes on the border of the United States and Canada. Where do they bury the survivors?

\item A man is looking at a portrait. Someone asks him whose portrait he is looking at. He replies, ``Brothers and sisters I have none, but that man's father is my father's son.'' Whose portrait is the man looking at?

\item A doctor gives you three pills and tells you to take one every half hour. How long will the pills last?
\end{enumerate}

\subsubsection*{Pool of decoy questions}

\begin{enumerate}
\item If a train leaves Melbourne at 3pm and travels at 100 km/h, how long does it take to reach Sydney 800 km away?

\item A warehouse had 400 boxes of toys. At the first delivery, 50 boxes were shipped out. At the second delivery, another 150 boxes were shipped out. How many boxes are left?

\item A fleet of trucks is carrying 1200 boxes. Each truck can carry 300 boxes. How many trucks are needed to carry all the boxes?

\item John, Lisa, and Mark go on a camping trip. Each person brings one item: a tent, a bag of marshmallows, and a flashlight. John brought the flashlight, and Lisa did not bring any food. What did Mark bring?

\item A recipe calls for 2 cups of flour to make 12 cookies. How many cups of flour are needed to make 24 cookies?

\item A factory produces 150 cars per day. How many cars will it produce in 4 days?

\item If a clock shows 3:45pm, how many minutes are left until 5:00pm?

\item A bookshelf has 5 shelves, and each shelf can hold 8 books. If you have 34 books, how many empty spaces are left on the bookshelf?

\item In a class of 30 students, 18 are girls. What percentage of the class is boys?

\item If you have a 5-litre jug and a 3-litre jug, and you fill the 5-litre jug completely, how many times can you fill the 3-litre jug from it?
\end{enumerate}

\subsubsection*{Pilot study}\phantomsection\label{app:pilot}

The pilot sample comprised approximately 240 undergraduate psychology students recruited via the UNSW SONA system in exchange for course credit. After providing informed consent, they completed CRT2, candidate bCRT items (each participant was randomly assigned to one of two sets), an earlier extended version of the BRT (8 items for the same 2 scenarios), and demographic questions. Aside from checking basic descriptives and data quality, these data were used mainly for bCRT item selection. We computed standard item statistics (proportion correct, proportion intuitive errors, item--total correlations) for all candidate bCRT items and CRT2 items, and fitted two-parameter logistic item response models separately for CRT2, the bCRT pool, and the combined item set to estimate item difficulty and discrimination. Items with poor discrimination, extreme difficulty, or ambiguous lure patterns were discarded. The four final bCRT items were chosen to span a reasonable difficulty range and to reliably elicit the intended intuitive response. Exploratory text analyses of the pilot responses did not inform item selection, but led us to add time-focus language as an exploratory family in the main study.

Responses in both studies were collected in real time, anonymised, and stored securely on the UNSW Qualtrics platform. Participants in both studies were provided with study information at recruitment, informed of their right to withdraw, and debriefed after participation.

\FloatBarrier
\subsection*{Supplementary Results: primary outcome families}
\begin{figure}[htbp]
  \centering
  \begin{subfigure}{\linewidth}\centering\includegraphics[width=\linewidth]{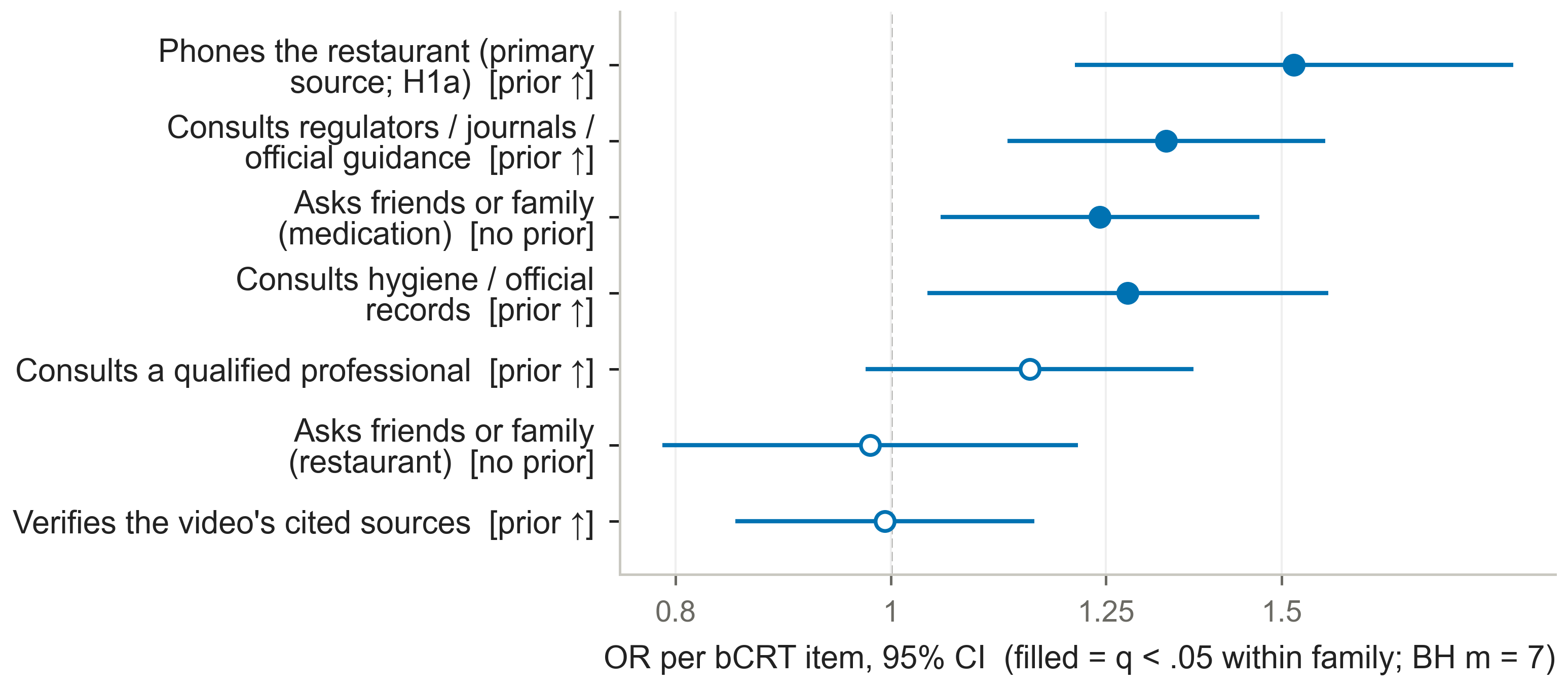}\caption{Family H1, epistemic search}\end{subfigure}\\
  \begin{subfigure}{\linewidth}\centering\includegraphics[width=\linewidth]{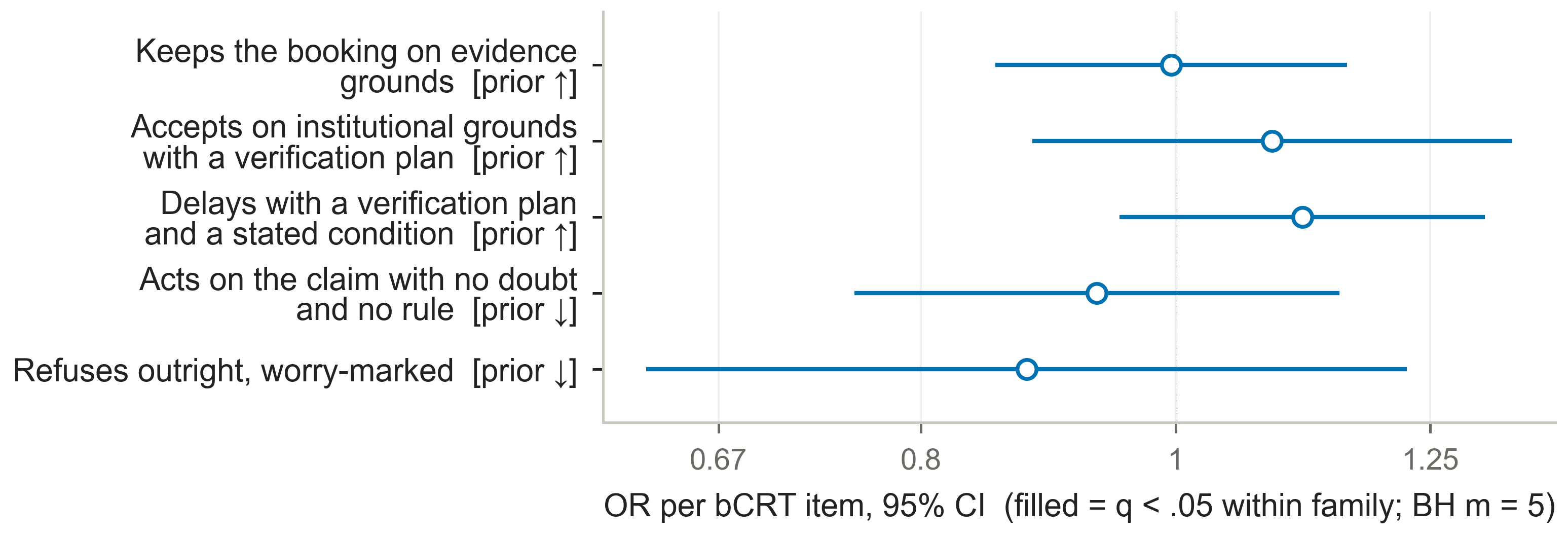}\caption{Family H2, evidence-weighted decisions}\end{subfigure}\\
  \begin{subfigure}{\linewidth}\centering\includegraphics[width=\linewidth]{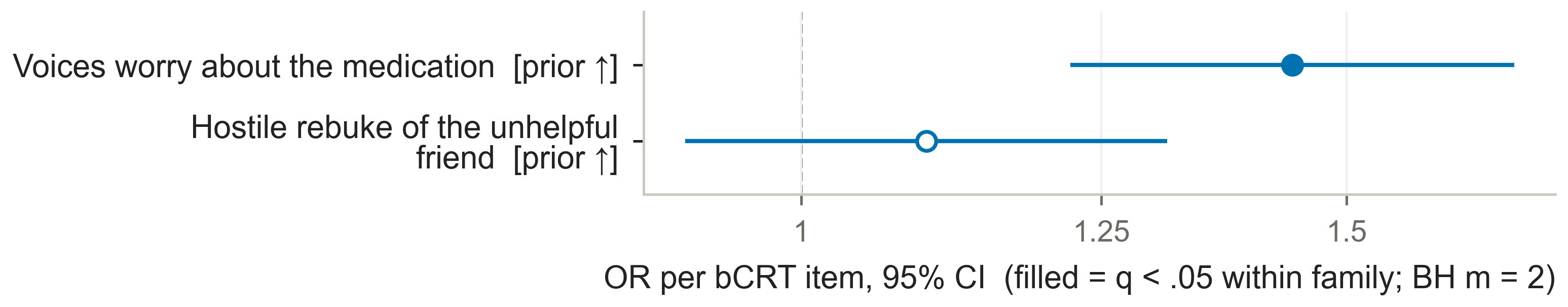}\caption{Family H3, affective engagement}\end{subfigure}\\
  \begin{subfigure}{\linewidth}\centering\includegraphics[width=\linewidth]{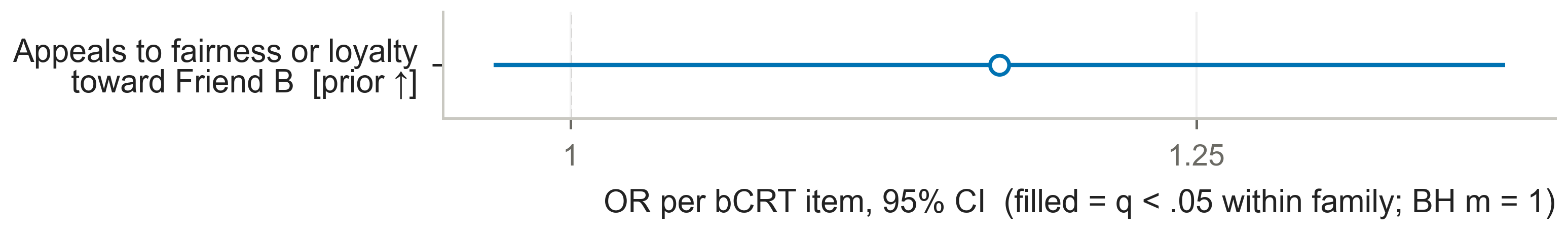}\caption{Family H4, fairness and norm-sensitive reasoning}\end{subfigure}
  \caption{\textbf{Which primary coded outcomes does bCRT predict, per item?} Odds ratios per additional bCRT item (95\% confidence intervals) from separate logistic models for each outcome in the four primary families. The dashed line marks an odds ratio of 1; the predicted direction is shown beside each outcome. $N=473$. Full estimates in Supplementary Table~\ref{tab:brt_primary}.}\label{fig:brt_families}
\end{figure}

Supplementary Table~\ref{tab:brt_primary} gives the full estimates for the fifteen primary coded outcomes reported in the main text, by hypothesis family; Supplementary Figure~\ref{fig:brt_families} plots their odds ratios per bCRT item. Supplementary Table~\ref{tab:si_prevalence} gives how often each coded outcome occurred.

\begin{table}[!htbp]
\centering
\footnotesize
\caption{\textbf{How common is each coded outcome?} Prevalence among all participants ($N=473$) with Wilson 95\% intervals. \textbf{a}, the fifteen primary outcomes; \textbf{b}, the exploratory outcomes, the two H2 sensitivity composites and the four descriptive-only outcomes.}
\label{tab:si_prevalence}
\begin{tabularx}{\textwidth}{@{}>{\raggedright\arraybackslash}Xrrrr@{}}
\toprule
Outcome & $N$ & Events & Prevalence (\%) & 95\% CI \\
\midrule
\multicolumn{5}{@{}l}{\textbf{a. Primary outcomes}} \\
\multicolumn{5}{@{}l}{\textit{H1 Epistemic search}} \\
\quad Phones the restaurant (primary source; H1a) & 473 & 54 & 11.4 & [8.9, 14.6] \\
\quad Consults regulators / journals / official guidance & 473 & 142 & 30.0 & [26.1, 34.3] \\
\quad Asks friends or family (medication) & 473 & 138 & 29.2 & [25.3, 33.4] \\
\quad Consults hygiene / official records & 473 & 68 & 14.4 & [11.5, 17.8] \\
\quad Consults a qualified professional & 473 & 328 & 69.3 & [65.0, 73.3] \\
\quad Asks friends or family (restaurant) & 473 & 70 & 14.8 & [11.9, 18.3] \\
\quad Verifies the video's cited sources & 473 & 191 & 40.4 & [36.1, 44.9] \\
\addlinespace
\multicolumn{5}{@{}l}{\textit{H2 Evidence-weighted decisions}} \\
\quad Keeps the booking on evidence grounds & 473 & 272 & 57.5 & [53.0, 61.9] \\
\quad Accepts on institutional grounds with a verification plan & 473 & 70 & 14.8 & [11.9, 18.3] \\
\quad Delays with a verification plan and a stated condition & 473 & 158 & 33.4 & [29.3, 37.8] \\
\quad Acts on the claim with no doubt and no rule & 473 & 75 & 15.9 & [12.8, 19.4] \\
\quad Refuses outright, worry-marked & 473 & 29 & 6.1 & [4.3, 8.7] \\
\addlinespace
\multicolumn{5}{@{}l}{\textit{H3 Affective engagement}} \\
\quad Voices worry about the medication & 473 & 265 & 56.0 & [51.5, 60.4] \\
\quad Hostile rebuke of the unhelpful friend & 473 & 107 & 22.6 & [19.1, 26.6] \\
\addlinespace
\multicolumn{5}{@{}l}{\textit{H4 Fairness}} \\
\quad Appeals to fairness or loyalty toward Friend B & 473 & 103 & 21.8 & [18.3, 25.7] \\
\bottomrule
\end{tabularx}
\end{table}

\begin{table}[!htbp]
\ContinuedFloat
\centering
\footnotesize
\caption{(continued)}
\begin{tabularx}{\textwidth}{@{}>{\raggedright\arraybackslash}Xrrrr@{}}
\toprule
Outcome & $N$ & Events & Prevalence (\%) & 95\% CI \\
\midrule
\multicolumn{5}{@{}l}{\textbf{b. Exploratory, sensitivity and descriptive-only outcomes}} \\
\multicolumn{5}{@{}l}{\textit{Appraises the claim's evidence}} \\
\quad Evaluates the plausibility, evidence or base rates of the claim & 473 & 352 & 74.4 & [70.3, 78.1] \\
\quad Doubts the video with a stated reason & 473 & 265 & 56.0 & [51.5, 60.4] \\
\addlinespace
\multicolumn{5}{@{}l}{\textit{What the decision rested on}} \\
\quad Rests on the critics' / official record & 473 & 85 & 18.0 & [14.8, 21.7] \\
\quad Rests on the medical system & 473 & 170 & 35.9 & [31.7, 40.4] \\
\addlinespace
\multicolumn{5}{@{}l}{\textit{Which choice they made}} \\
\quad Keeps the booking & 473 & 335 & 70.8 & [66.6, 74.7] \\
\quad Takes or delays the medication & 473 & 395 & 83.5 & [79.9, 86.6] \\
\addlinespace
\multicolumn{5}{@{}l}{\textit{Low-scrutiny responses}} \\
\quad Vague-only search, restaurant & 473 & 8 & 1.7 & [0.9, 3.3] \\
\quad Vague-only search, medication & 473 & 21 & 4.4 & [2.9, 6.7] \\
\quad Anger at the prescriber & 473 & 44 & 9.3 & [7.0, 12.3] \\
\quad Conspiratorial ideation & 473 & 19 & 4.0 & [2.6, 6.2] \\
\addlinespace
\multicolumn{5}{@{}l}{\textit{Miscellaneous}} \\
\quad Reads the reviews / crowd sources & 473 & 258 & 54.5 & [50.0, 59.0] \\
\quad Sunk-cost salience & 473 & 246 & 52.0 & [47.5, 56.5] \\
\quad Wider moral reasoning & 473 & 226 & 47.8 & [43.3, 52.3] \\
\quad Consults lay online content & 473 & 92 & 19.5 & [16.1, 23.3] \\
\quad States an if-then rule & 473 & 219 & 46.3 & [41.9, 50.8] \\
\quad Takes the medication with any check plan & 473 & 134 & 28.3 & [24.5, 32.6] \\
\addlinespace
\multicolumn{5}{@{}l}{\textit{Sensitivity composites}} \\
\quad Accepts on institutional grounds, exclusive-tier plan or no plan & 473 & 58 & 12.3 & [9.6, 15.5] \\
\quad Delays with an exclusive-tier plan and a stated condition & 473 & 84 & 17.8 & [14.6, 21.5] \\
\addlinespace
\multicolumn{5}{@{}l}{\textit{Descriptive only}} \\
\quad Misread the scenario, restaurant & 473 & 19 & 4.0 & [2.6, 6.2] \\
\quad Groundless scepticism, medication & 473 & 6 & 1.3 & [0.6, 2.7] \\
\quad Misread the scenario, medication & 473 & 14 & 3.0 & [1.8, 4.9] \\
\quad Credulous switching, restaurant & 473 & 9 & 1.9 & [1.0, 3.6] \\
\bottomrule
\end{tabularx}
\end{table}

\begin{sidewaystable}
\centering
\footnotesize
\caption{\textbf{Which primary coded outcomes does bCRT predict?} The fifteen primary coded outcomes, by hypothesis family (Prolific sample, $N=473$; $m$: outcomes in the family). Each row is a separate logistic regression on raw bCRT total; OR per additional bCRT item answered correctly, Wald 95\% CI, likelihood-ratio $p$; $q$ is Benjamini--Hochberg corrected within the family only ($^{*}$: $q<.05$; with $m=1$ the correction is the identity and the per-test $p$ is the reported test). Prior: the predicted direction, printed for interpretation (all tests two-sided). AME: average marginal effect, percentage points per bCRT item. MDOR: minimum detectable OR at 80\% power in the predicted direction (both directions where no prior is held; Methods). Reliability: minimum dual-pass \% agreement / Fleiss' $\kappa$ over the coded fields the outcome is built from. H1 members do not share one predicted direction: the two personal-network members (asking friends or family) carry no directional prior. In H2 the reflective pole (three members, prior $\uparrow$) and the intuitive pole (two members, prior $\downarrow$) read one construct from both ends. By construction, outright, worry-marked refusal is a subset of voicing worry (H3), and 24 of its 29 cases also meet the definition of acting on the claim with no doubt and no rule. The two H2 sensitivity rows restrict the acceptance and delay composites to exclusive verification plans (evidence-grade or professional checks with no lay check alongside; the acceptance variant also admits no plan). The H3 rebuke field codes anger words and expressive acts directed at Friend A (affect), not a normative judgement.}
\label{tab:brt_primary}
\begin{tabularx}{\linewidth}{@{}>{\raggedright\arraybackslash}Xcrlrrrrr@{}}
\toprule
Outcome & Prior & Events & OR [95\% CI] & $p$ & $q$ & AME & MDOR & Reliability \\
\midrule
\multicolumn{9}{@{}l}{\textit{H1 Epistemic search: which sources respondents plan to consult ($m=7$)}} \\
\quad Phones the restaurant (primary source; H1a)$^{*}$ & $\uparrow$ & 54 & 1.52 [1.21, 1.91] & $<$.001 & .002 & +4.1 & 1.39 & 100.0/1.00 \\
\quad Consults regulators / journals / official guidance$^{*}$ & $\uparrow$ & 142 & 1.33 [1.13, 1.57] & $<$.001 & .002 & +5.9 & 1.26 & 96.7/0.87 \\
\quad Asks friends or family (medication)$^{*}$ & --- & 138 & 1.24 [1.05, 1.47] & .010 & .024 & +4.4 & 1.26/0.78 & 98.7/0.96 \\
\quad Consults hygiene / official records$^{*}$ & $\uparrow$ & 68 & 1.28 [1.04, 1.57] & .022 & .038 & +3.0 & 1.35 & 99.3/0.97 \\
\quad Consults a qualified professional & $\uparrow$ & 328 & 1.16 [0.97, 1.37] & .093 & .130 & +3.0 & 1.27 & 92.0/0.83 \\
\quad Asks friends or family (restaurant) & --- & 70 & 0.98 [0.79, 1.21] & .845 & .937 & -0.3 & 1.35/0.72 & 97.3/0.89 \\
\quad Verifies the video's cited sources & $\uparrow$ & 191 & 0.99 [0.85, 1.16] & .937 & .937 & -0.1 & 1.25 & 95.3/0.88 \\
\addlinespace
\multicolumn{9}{@{}l}{\textit{H2 Evidence-weighted decisions ($m=5$)}} \\
\quad Keeps the booking on evidence grounds & $\uparrow$ & 272 & 1.00 [0.85, 1.16] & .962 & .962 & -0.1 & 1.25 & 85.3/0.77 \\
\quad Accepts on institutional grounds with a verification plan & $\uparrow$ & 70 & 1.09 [0.88, 1.34] & .433 & .653 & +1.1 & 1.35 & 79.3/0.73 \\
\quad Delays with a verification plan and a stated condition & $\uparrow$ & 158 & 1.12 [0.95, 1.31] & .175 & .653 & +2.5 & 1.26 & 89.3/0.77 \\
\quad Acts on the claim with no doubt and no rule & $\downarrow$ & 75 & 0.93 [0.75, 1.15] & .522 & .653 & -0.9 & 0.73 & 90.0/0.49 \\
\quad Refuses outright, worry-marked & $\downarrow$ & 29 & 0.88 [0.63, 1.22] & .435 & .653 & -0.8 & 0.60 & 90.0/0.84 \\
\addlinespace
\multicolumn{9}{@{}l}{\textit{Composite sensitivities (per-test $p$ only; not part of the family correction)}} \\
\quad Accepts on institutional grounds, exclusive-tier plan or no plan & $\uparrow$ & 58 & 0.79 [0.61, 1.02] & .060 & --- & -2.5 & --- & 79.3/0.73 \\
\quad Delays with an exclusive-tier plan and a stated condition & $\uparrow$ & 84 & 1.06 [0.87, 1.29] & .540 & --- & +0.9 & --- & 89.3/0.77 \\
\addlinespace
\multicolumn{9}{@{}l}{\textit{H3 Affective engagement ($m=2$)}} \\
\quad Voices worry about the medication$^{*}$ & $\uparrow$ & 265 & 1.44 [1.22, 1.70] & $<$.001 & $<$.001 & +8.6 & 1.25 & 96.0/0.92 \\
\quad Hostile rebuke of the unhelpful friend & $\uparrow$ & 107 & 1.10 [0.92, 1.31] & .312 & .312 & +1.6 & 1.29 & 91.3/0.78 \\
\addlinespace
\multicolumn{9}{@{}l}{\textit{H4 Fairness and norm-sensitive reasoning ($m=1$)}} \\
\quad Appeals to fairness or loyalty toward Friend B & $\uparrow$ & 103 & 1.17 [0.97, 1.40] & .099 & .099 & +2.6 & 1.29 & 93.3/0.80 \\
\bottomrule
\end{tabularx}
\end{sidewaystable}

\FloatBarrier

\FloatBarrier
\subsection*{Supplementary Results: Item response parameters}\phantomsection\label{app:irt}

\begin{table}[!htbp]
\centering
\footnotesize
\caption{\textbf{How discriminating and how difficult is each CRT2 and bCRT item?} Two-parameter logistic (2PL) item parameters with person-bootstrap 95\% CIs, from one calibration of all eight items by marginal maximum likelihood (Prolific sample, $N=473$; CRT2 items, $n=471$; binary correct scoring). $a$: discrimination; $b$: difficulty, the location on the latent reflection scale $\theta$.}
\label{tab:irt_params}
\begin{tabular*}{\textwidth}{@{\extracolsep{\fill}}lrr@{}}
\toprule
Item & $a$ [95\% CI] & $b$ [95\% CI] \\
\midrule
\multicolumn{3}{@{}l}{\textit{CRT2}} \\
\quad Race: passing the person in second place & 1.05 [0.63, 1.62] & -1.62 [-2.50, -1.20] \\
\quad Farmer: 15 sheep, all but 8 die & 1.00 [0.54, 1.66] & -2.53 [-4.29, -1.77] \\
\quad Emily's father's three daughters & 2.67 [1.77, 5.79] & -1.19 [-1.46, -0.97] \\
\quad Dirt in a 3$\times$3$\times$3 ft hole & 1.86 [1.37, 2.63] & 0.57 [0.39, 0.76] \\
\multicolumn{3}{@{}l}{\textit{bCRT}} \\
\quad Cow bought and sold twice & 1.19 [0.87, 1.63] & -0.44 [-0.68, -0.25] \\
\quad Snail on a 10-foot wall & 1.48 [1.04, 2.07] & 0.86 [0.63, 1.17] \\
\quad Rope ladder and rising tide & 1.72 [1.15, 2.66] & 1.25 [0.99, 1.66] \\
\quad Married person looking at an unmarried person & 0.73 [0.43, 1.11] & 0.84 [0.51, 1.41] \\
\bottomrule
\end{tabular*}
\end{table}

\begin{figure}[htbp]
  \centering
  \includegraphics[width=\textwidth]{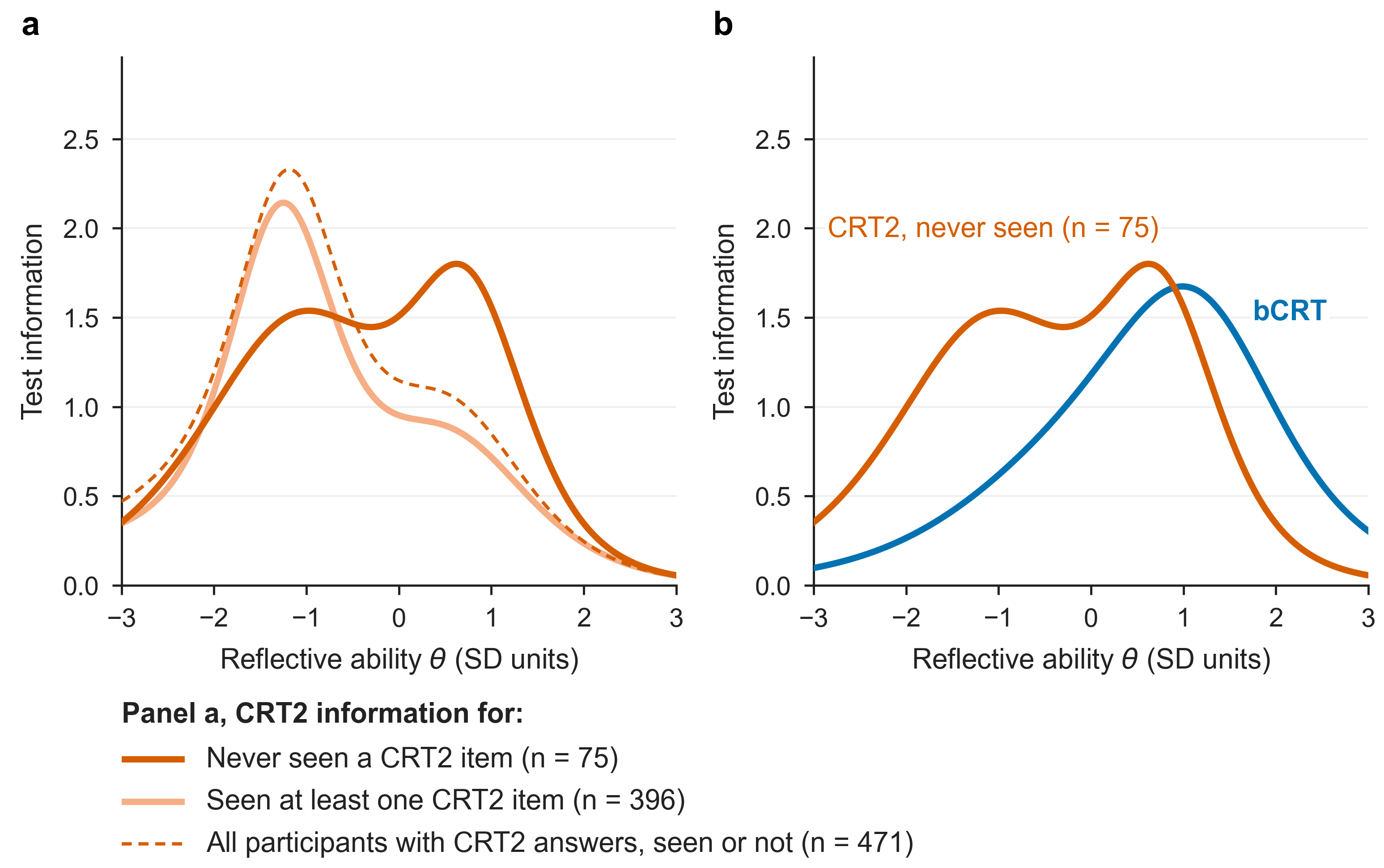}
  \caption{\textbf{Did prior exposure change how the CRT2 items measure?} Test information over reflective ability $\theta$ from the two-group calibration (bCRT item parameters equal across groups, CRT2 item parameters free), on the scale of Figure~\ref{fig:irt_tic}. \textbf{a}, CRT2 information for participants who had never seen a CRT2 item (dark vermillion, $n=75$) and for those who had seen at least one (light vermillion, $n=396$); dashed: the CRT2 curve of Figure~\ref{fig:irt_tic} ($n=471$). \textbf{b}, bCRT information (blue) against the never-seen CRT2 curve.}
  \label{fig:irt_exposure}
\end{figure}

To check whether prior exposure to CRT2 items changed how they measure, we fitted a two-group model to participants who had never seen a CRT2 item ($n=75$) and those who had seen at least one ($n=396$; the two participants whose partial CRT2 answers were set to missing are not included), with the bCRT item parameters held equal across groups and the CRT2 item parameters free (Supplementary Figure~\ref{fig:irt_exposure}). Letting the CRT2 parameters differ improved fit (likelihood-ratio $\chi^2(8)=17.56$, $p=.025$); the Race and Farmer items were easier for the exposed group (difficulty $-0.51$ to $-1.63$ and $-1.07$ to $-3.18$). The never-seen group is small ($n=75$), so we treat this as a sensitivity check and do not interpret it further.

\FloatBarrier
\subsection*{Supplementary Results: Full LIWC models for bCRT}

Supplementary Table~\ref{tab:liwc_bcrt_full} reports the primary bCRT model for every category in the LIWC families listed in the Supplementary Methods.

\begin{table}[htbp]
\centering
\footnotesize
\caption{\textbf{Does bCRT predict each LIWC category?} Behavioural Reflection Test (BRT): primary bCRT LIWC models for non-summary LIWC families (Prolific sample, $N=473$). Each row comes from a separate OLS model of the form LIWC category $\sim$ bCRT total. $\beta$ is the unstandardised coefficient; 95\% confidence intervals (CI) are $t$-based. Predicted categories ($\dagger$) are reported with their per-test $p$; all other categories are Benjamini--Hochberg corrected within each LIWC family ($p_{\mathrm{FDR}}$), except single-member families, which are reported with the uncorrected $p$. Intervals are not corrected for multiple comparisons.}
\label{tab:liwc_bcrt_full}
\begin{tabular*}{\textwidth}{@{\extracolsep{\fill}}llrrrrr}
\toprule
Family & LIWC category & $\beta$ & 95\% CI & $p$ & $R^{2}$ & $n$ \\
\midrule
QTY    & Word count & 15.797 & [7.938, 23.657] & $<$.001 & 0.032 & 473 \\
       & Words per sentence & 1.068 & [-0.079, 2.214] & .068 & 0.007 & 473 \\
\midrule
FUNC   & Articles & -0.036 & [-0.165, 0.093] & .598 & 0.001 & 473 \\
       & Prepositions & -0.162 & [-0.319, -0.004] & .103 & 0.009 & 473 \\
       & First-person singular & 0.050 & [-0.107, 0.207] & .598 & 0.001 & 473 \\
       & Other pronouns & -0.116 & [-0.252, 0.019] & .161 & 0.006 & 473 \\
       & Auxiliary verbs & -0.328 & [-0.492, -0.164] & $<$.001 & 0.032 & 473 \\
       & Conjunctions & 0.038 & [-0.103, 0.178] & .598 & 0.001 & 473 \\
       & Adjectives & 0.145 & [0.016, 0.273] & .096 & 0.010 & 473 \\
\midrule
COG    & Insight$^{\dagger}$ & 0.030 & [-0.085, 0.145] & .605 & 0.001 & 473 \\
       & Causation & 0.026 & [-0.050, 0.102] & .530 & 0.001 & 473 \\
       & Discrepancy & -0.219 & [-0.361, -0.077] & .013 & 0.019 & 473 \\
       & Tentativeness & -0.071 & [-0.204, 0.063] & .499 & 0.002 & 473 \\
       & Differentiation & -0.176 & [-0.306, -0.046] & .020 & 0.015 & 473 \\
       & Certainty & 0.015 & [-0.032, 0.061] & .530 & 0.001 & 473 \\
\midrule
AFFECT & Negative emotion$^{\dagger}$ & 0.055 & [0.002, 0.109] & .043 & 0.009 & 473 \\
\midrule
TIME   & Present-focus & 0.205 & [0.060, 0.351] & .006 & 0.016 & 473 \\
\midrule
MOT    & Risk words$^{\dagger}$ & 0.077 & [0.030, 0.124] & .001 & 0.021 & 473 \\
\midrule
MOS    & Moral language$^{\dagger}$ & 0.008 & [-0.021, 0.037] & .603 & 0.001 & 473 \\
\bottomrule
\end{tabular*}
\end{table}

\FloatBarrier
\subsection*{Supplementary Results: CRT2 LIWC comparator models and slope diagnostics}
\label{app:liwc_comparator_diagnostics}

Supplementary Table~\ref{tab:liwc_crt2} reports the CRT2 comparator models for the six focal LIWC categories of the main text (Table~\ref{tab:liwc_bcrt}).

\begin{table}[!htbp]
\centering
\footnotesize
\caption{\textbf{Does CRT2 predict the six LIWC categories bCRT predicts?} BRT: CRT2 comparator models for the six focal bCRT-associated LIWC categories (Prolific sample, $N=471$). Each row comes from a separate OLS model of the form LIWC category $\sim$ CRT2 total + CRT2 familiarity count. $\beta$ is the unstandardised coefficient for CRT2 total in original LIWC-score units, with a $t$-based 95\% CI. Predicted categories ($\dagger$: \textit{negative emotion}, following Mosleh et al.; \textit{risk}, given the scenario design) are reported with their per-test $p$; exploratory categories are Benjamini--Hochberg corrected within LIWC families, except \textit{present-focus} (sole exploratory member of its family), reported with its per-test $p$. Intervals are not corrected for multiple comparisons.}
\label{tab:liwc_crt2}
\begin{tabular*}{\textwidth}{@{\extracolsep{\fill}}lrrrrrr@{}}
\toprule
LIWC category & $\beta$ & 95\% CI & SE & $t$ & $p$ & $R^{2}$ \\
\midrule
Negative emotion$^{\dagger}$ & -0.010 & [-0.076, 0.056] & 0.033 & -0.30 & .764 & 0.009 \\
Discrepancy                  & -0.178 & [-0.351, -0.005] & 0.088 & -2.02 & .221 & 0.023 \\
Differentiation              & -0.049 & [-0.208, 0.110] & 0.081 & -0.61 & .738 & 0.003 \\
Auxiliary verbs              & -0.174 & [-0.377, 0.028] & 0.103 & -1.69 & .170 & 0.008 \\
Risk words$^{\dagger}$       & -0.005 & [-0.063, 0.054] & 0.030 & -0.16 & .877 & 0.000 \\
Present-focus                & 0.140 & [-0.039, 0.319] & 0.091 & 1.54 & .125 & 0.005 \\
\bottomrule
\end{tabular*}
\end{table}

\begin{figure}[!htbp]
  \centering
  \makebox[\textwidth][c]{\includegraphics[width=170mm]{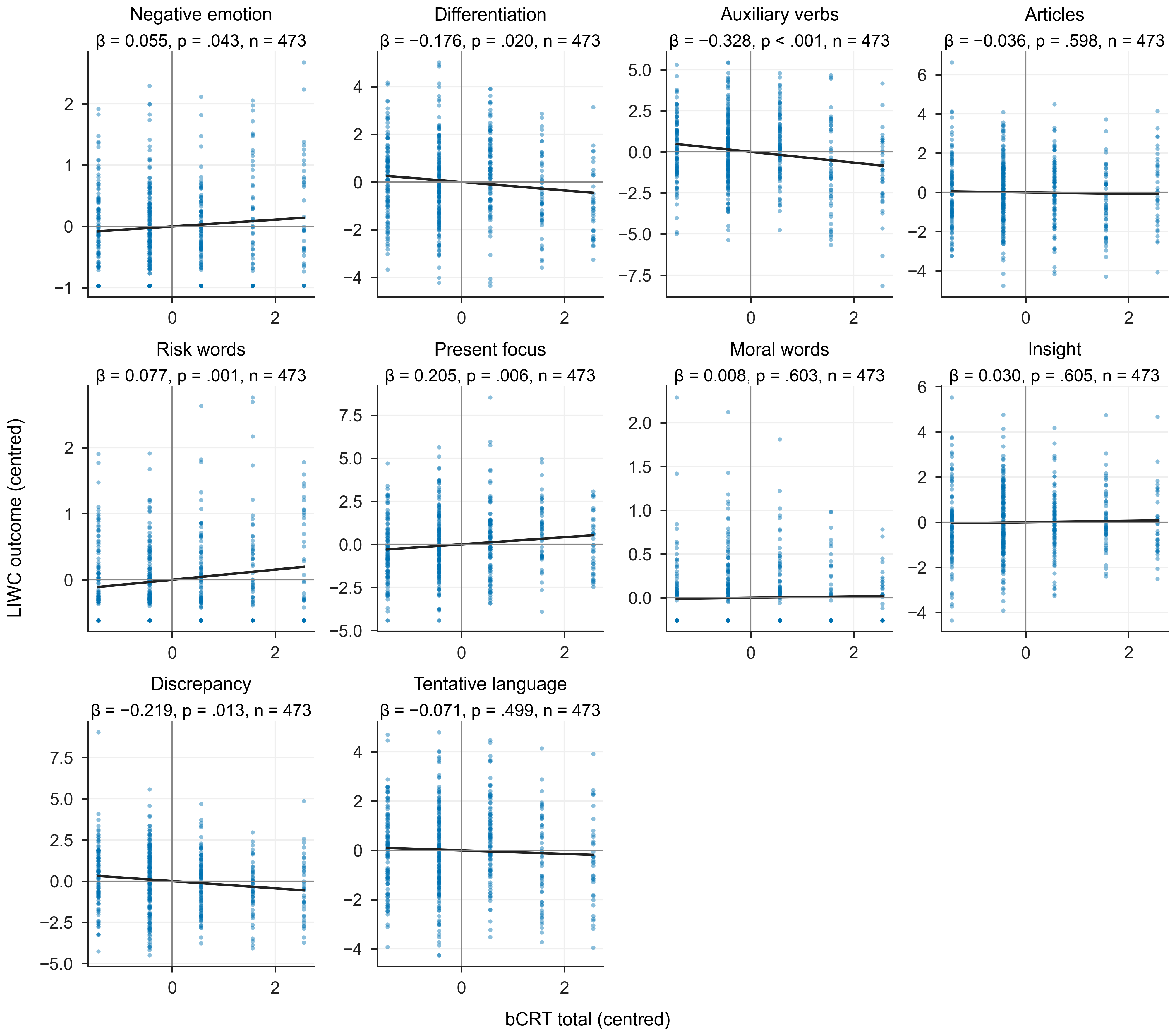}}
  \makebox[\linewidth][c]{\begin{minipage}{170mm}\captionsetup{position=below}\caption{\textbf{How does each LIWC category vary with bCRT, participant by participant?} Primary bCRT LIWC slope diagnostics for the BRT. Panels show selected LIWC categories from the primary model family, where each LIWC category is predicted by raw bCRT total; blue dots are participants, plotted as bCRT total (centred) against the LIWC outcome (centred), and the black line is the fitted slope. The main text foregrounds the psychologically interpretable effects that were either predicted or exploratorily significant.}\label{fig:liwc_grid}\end{minipage}}
\end{figure}

\begin{figure}[!htbp]
  \centering
  \includegraphics[width=\textwidth]{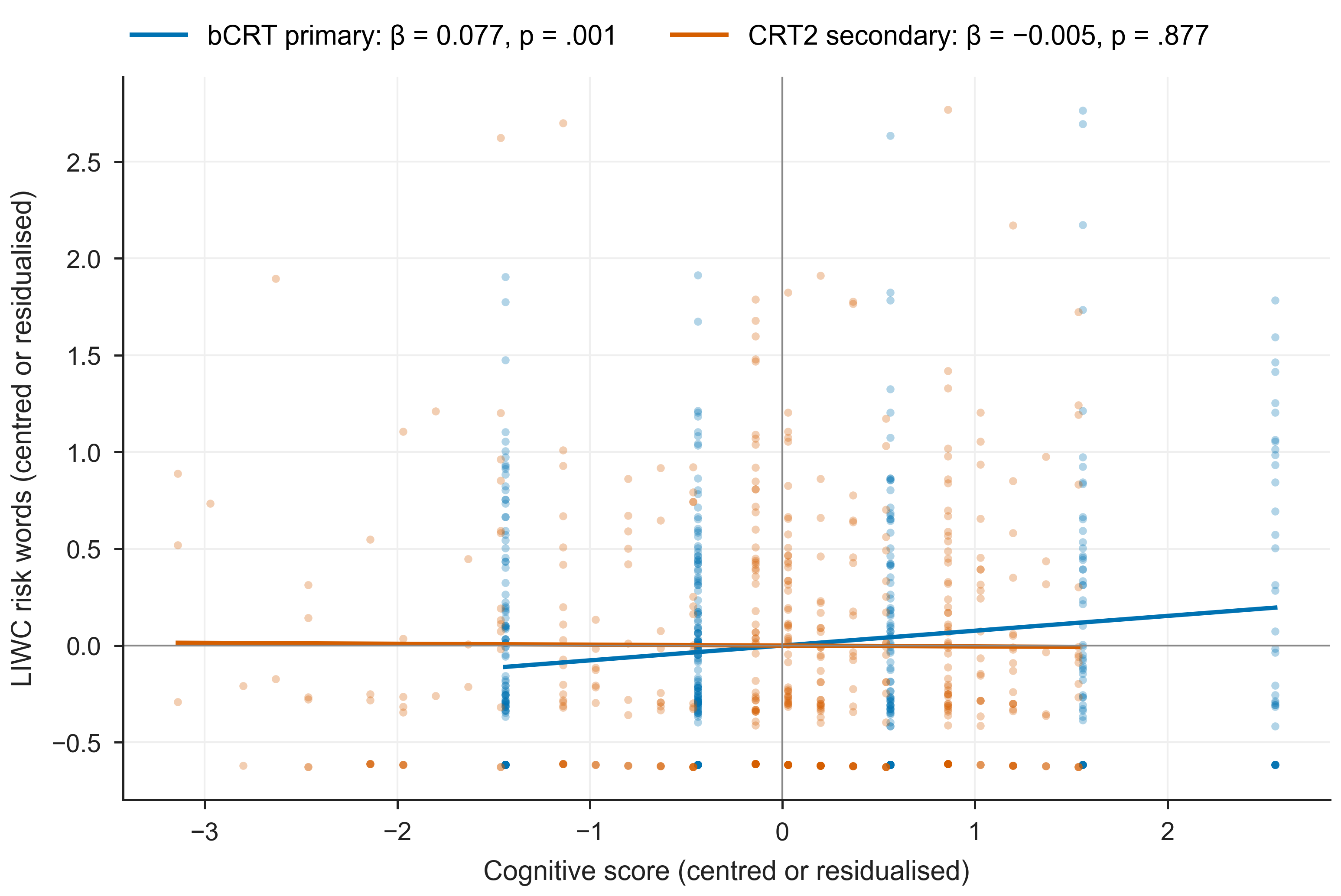}
  \caption{\textbf{How do the bCRT and CRT2 slopes for risk words compare?} Overlay of LIWC \textit{risk}-word slopes for the primary bCRT model and secondary CRT2 familiarity-adjusted model. The bCRT slope (blue) comes from \textit{risk} words $\sim$ bCRT total; the CRT2 slope (vermillion) comes from \textit{risk} words $\sim$ CRT2 total + CRT2 familiarity count. Faded points show individual participants' centred or residualised scores.}
  \label{fig:risk_slope_overlay_supp}
\end{figure}

Supplementary Figure~\ref{fig:liwc_grid} visualises the primary bCRT LIWC relationships from the raw bCRT model family. Supplementary Figure~\ref{fig:risk_slope_overlay_supp} overlays the primary bCRT slope for \textit{risk}-related language with the corresponding CRT2 familiarity-adjusted comparator slope.

\FloatBarrier
\subsection*{Supplementary Results: Predictor comparison for the BRT outcomes}
\label{app:crt2_primary_brt_sensitivity}

Supplementary Tables~\ref{tab:brt_block_crt2_raw}--\ref{tab:brt_block_joint} report every primary, exploratory and H2-sensitivity model refitted with the focal predictor replaced (Methods), and Supplementary Figure~\ref{fig:brt_predictor_comparison} sets the four predictors side by side for the primary outcomes. Entered jointly with familiarity-adjusted CRT2 and NFC, bCRT retained its associations with phoning the restaurant, consulting institutional sources, asking friends or family about the medication, and worry, but not with consulting hygiene or official records (Supplementary Table~\ref{tab:brt_block_joint}).

\phantomsection\label{app:familiarity}The familiarity adjustment uses a fixed coefficient (Methods). Re-estimated on the analysed data, the familiarity effect was 0.169 points per item seen without covariates ($n=471$), 0.176 controlling for age, with or without sex ($n=470$), and 0.204 with education added ($n=439$; education missing for 32 of the 471 participants with a CRT2 score). Because CRT2 totals and seen counts are whole numbers, any estimate below 0.25 points per item orders participants identically; only the absolute adjusted scores differ (mean 2.38--2.47).

\begin{table}[!htbp]
\centering
\footnotesize
\caption{\textbf{Does raw CRT2 predict the BRT outcomes?} Every primary, exploratory and H2-sensitivity model refitted with raw CRT2 total in place of bCRT (Prolific sample, $N=471$). OR per point of CRT2, Wald 95\% CI, likelihood-ratio $p$; $q$ re-runs the within-family Benjamini--Hochberg correction inside this block (sensitivity rows carry $p$ only); the last row is the likelihood-ratio test of the medication search-plan profile. The bCRT estimates are in Supplementary Tables~\ref{tab:brt_primary} and~\ref{tab:brt_exploratory}.}
\label{tab:brt_block_crt2_raw}
\begin{tabularx}{\textwidth}{@{}>{\raggedright\arraybackslash}Xrlrr@{}}
\toprule
Outcome & $N$ & OR [95\% CI] & $p$ & $q$ \\
\midrule
\multicolumn{5}{@{}l}{\textit{H1 Epistemic search}} \\
\quad Phones the restaurant (primary source; H1a) & 471 & 1.20 [0.89, 1.63] & .220 & .513 \\
\quad Consults regulators / journals / official guidance & 471 & 1.24 [1.01, 1.53] & .036 & .252 \\
\quad Asks friends or family (medication) & 471 & 1.03 [0.84, 1.26] & .769 & .879 \\
\quad Consults hygiene / official records & 471 & 1.27 [0.96, 1.69] & .082 & .285 \\
\quad Consults a qualified professional & 471 & 1.02 [0.84, 1.24] & .837 & .879 \\
\quad Asks friends or family (restaurant) & 471 & 1.02 [0.79, 1.32] & .879 & .879 \\
\quad Verifies the video's cited sources & 471 & 0.94 [0.79, 1.13] & .531 & .879 \\
\addlinespace
\multicolumn{5}{@{}l}{\textit{H2 Evidence-weighted decisions}} \\
\quad Keeps the booking on evidence grounds & 471 & 1.06 [0.88, 1.27] & .537 & .621 \\
\quad Accepts on institutional grounds with a verification plan & 471 & 1.13 [0.87, 1.48] & .347 & .621 \\
\quad Delays with a verification plan and a stated condition & 471 & 1.05 [0.87, 1.28] & .594 & .621 \\
\quad Acts on the claim with no doubt and no rule & 471 & 0.90 [0.71, 1.14] & .368 & .621 \\
\quad Refuses outright, worry-marked & 471 & 0.91 [0.64, 1.31] & .621 & .621 \\
\addlinespace
\multicolumn{5}{@{}l}{\textit{H3 Affective engagement}} \\
\quad Voices worry about the medication & 471 & 1.16 [0.96, 1.39] & .119 & .132 \\
\quad Hostile rebuke of the unhelpful friend & 471 & 1.19 [0.95, 1.49] & .132 & .132 \\
\addlinespace
\multicolumn{5}{@{}l}{\textit{H4 Fairness}} \\
\quad Appeals to fairness or loyalty toward Friend B & 471 & 1.09 [0.87, 1.37] & .430 & .430 \\
\bottomrule
\end{tabularx}
\end{table}

\begin{table}[!htbp]
\ContinuedFloat
\centering
\footnotesize
\caption{(continued) Exploratory outcome families, the H2 composite sensitivities and the medication search-plan profile.}
\begin{tabularx}{\textwidth}{@{}>{\raggedright\arraybackslash}Xrlrr@{}}
\toprule
Outcome & $N$ & OR [95\% CI] & $p$ & $q$ \\
\midrule
\multicolumn{5}{@{}l}{\textit{Appraises the claim's evidence}} \\
\quad Doubts the video with a stated reason & 471 & 1.22 [1.01, 1.46] & .036 & .071 \\
\quad Evaluates the plausibility, evidence or base rates of the claim & 471 & 1.08 [0.88, 1.32] & .463 & .463 \\
\addlinespace
\multicolumn{5}{@{}l}{\textit{What the decision rested on}} \\
\quad Rests on the critics' / official record & 471 & 0.96 [0.76, 1.21] & .714 & .714 \\
\quad Rests on the medical system & 471 & 1.16 [0.96, 1.41] & .126 & .252 \\
\addlinespace
\multicolumn{5}{@{}l}{\textit{Which choice they made}} \\
\quad Keeps the booking & 471 & 0.98 [0.80, 1.19] & .831 & .831 \\
\quad Takes or delays the medication & 471 & 1.12 [0.89, 1.42] & .334 & .669 \\
\addlinespace
\multicolumn{5}{@{}l}{\textit{Low-scrutiny responses}} \\
\quad Vague-only search, restaurant & 471 & 0.59 [0.34, 1.06] & .076 & .216 \\
\quad Vague-only search, medication & 471 & 0.72 [0.49, 1.06] & .108 & .216 \\
\quad Conspiratorial ideation & 471 & 1.18 [0.74, 2.04] & .515 & .616 \\
\quad Anger at the prescriber & 471 & 1.08 [0.79, 1.49] & .616 & .616 \\
\addlinespace
\multicolumn{5}{@{}l}{\textit{Miscellaneous}} \\
\quad Reads the reviews / crowd sources & 471 & 1.14 [0.95, 1.36] & .166 & .953 \\
\quad Consults lay online content & 471 & 0.97 [0.77, 1.22] & .794 & .953 \\
\quad States an if-then rule & 471 & 1.07 [0.89, 1.28] & .491 & .953 \\
\quad Wider moral reasoning & 471 & 0.97 [0.81, 1.16] & .740 & .953 \\
\quad Sunk-cost salience & 471 & 0.95 [0.79, 1.13] & .549 & .953 \\
\quad Takes the medication with any check plan & 471 & 1.00 [0.81, 1.22] & .962 & .962 \\
\addlinespace
\multicolumn{5}{@{}l}{\textit{H2 composite sensitivities}} \\
\quad Accepts on institutional grounds, exclusive-tier plan or no plan & 471 & 0.92 [0.71, 1.21] & .558 & --- \\
\quad Delays with an exclusive-tier plan and a stated condition & 471 & 1.11 [0.87, 1.41] & .408 & --- \\
\addlinespace
\multicolumn{5}{@{}l}{\textit{Medication search-plan profile}} \\
\quad Search-plan profile (likelihood-ratio test, 4 df) & 471 & $\chi^2_{4}=1.96$ & .743 & --- \\
\bottomrule
\end{tabularx}
\end{table}

\begin{table}[!htbp]
\centering
\footnotesize
\caption{\textbf{Does familiarity-adjusted CRT2 predict the BRT outcomes?} As Supplementary Table~\ref{tab:brt_block_crt2_raw}, with the familiarity-adjusted CRT2 total (Methods) in place of bCRT; OR per point of the adjusted total.}
\label{tab:brt_block_crt2_adj}
\begin{tabularx}{\textwidth}{@{}>{\raggedright\arraybackslash}Xrlrr@{}}
\toprule
Outcome & $N$ & OR [95\% CI] & $p$ & $q$ \\
\midrule
\multicolumn{5}{@{}l}{\textit{H1 Epistemic search}} \\
\quad Phones the restaurant (primary source; H1a) & 471 & 1.31 [0.95, 1.81] & .087 & .269 \\
\quad Consults regulators / journals / official guidance & 471 & 1.29 [1.04, 1.60] & .019 & .132 \\
\quad Asks friends or family (medication) & 471 & 1.09 [0.88, 1.34] & .421 & .589 \\
\quad Consults hygiene / official records & 471 & 1.25 [0.94, 1.66] & .115 & .269 \\
\quad Consults a qualified professional & 471 & 1.04 [0.85, 1.27] & .713 & .831 \\
\quad Asks friends or family (restaurant) & 471 & 1.01 [0.78, 1.31] & .951 & .951 \\
\quad Verifies the video's cited sources & 471 & 0.92 [0.76, 1.12] & .411 & .589 \\
\addlinespace
\multicolumn{5}{@{}l}{\textit{H2 Evidence-weighted decisions}} \\
\quad Keeps the booking on evidence grounds & 471 & 1.06 [0.88, 1.28] & .560 & .727 \\
\quad Accepts on institutional grounds with a verification plan & 471 & 1.12 [0.86, 1.47] & .396 & .727 \\
\quad Delays with a verification plan and a stated condition & 471 & 1.04 [0.85, 1.26] & .727 & .727 \\
\quad Acts on the claim with no doubt and no rule & 471 & 0.89 [0.70, 1.14] & .377 & .727 \\
\quad Refuses outright, worry-marked & 471 & 0.91 [0.63, 1.33] & .632 & .727 \\
\addlinespace
\multicolumn{5}{@{}l}{\textit{H3 Affective engagement}} \\
\quad Voices worry about the medication & 471 & 1.17 [0.97, 1.41] & .100 & .178 \\
\quad Hostile rebuke of the unhelpful friend & 471 & 1.17 [0.93, 1.47] & .178 & .178 \\
\addlinespace
\multicolumn{5}{@{}l}{\textit{H4 Fairness}} \\
\quad Appeals to fairness or loyalty toward Friend B & 471 & 1.16 [0.92, 1.47] & .209 & .209 \\
\bottomrule
\end{tabularx}
\end{table}

\begin{table}[!htbp]
\ContinuedFloat
\centering
\footnotesize
\caption{(continued) Exploratory outcome families, the H2 composite sensitivities and the medication search-plan profile.}
\begin{tabularx}{\textwidth}{@{}>{\raggedright\arraybackslash}Xrlrr@{}}
\toprule
Outcome & $N$ & OR [95\% CI] & $p$ & $q$ \\
\midrule
\multicolumn{5}{@{}l}{\textit{Appraises the claim's evidence}} \\
\quad Doubts the video with a stated reason & 471 & 1.22 [1.01, 1.47] & .040 & .080 \\
\quad Evaluates the plausibility, evidence or base rates of the claim & 471 & 1.09 [0.89, 1.35] & .406 & .406 \\
\addlinespace
\multicolumn{5}{@{}l}{\textit{What the decision rested on}} \\
\quad Rests on the critics' / official record & 471 & 0.95 [0.75, 1.21] & .675 & .675 \\
\quad Rests on the medical system & 471 & 1.14 [0.93, 1.38] & .202 & .404 \\
\addlinespace
\multicolumn{5}{@{}l}{\textit{Which choice they made}} \\
\quad Keeps the booking & 471 & 0.99 [0.81, 1.21] & .914 & .914 \\
\quad Takes or delays the medication & 471 & 1.13 [0.88, 1.44] & .336 & .671 \\
\addlinespace
\multicolumn{5}{@{}l}{\textit{Low-scrutiny responses}} \\
\quad Vague-only search, restaurant & 471 & 0.51 [0.29, 0.93] & .028 & .113 \\
\quad Vague-only search, medication & 471 & 0.68 [0.45, 1.01] & .062 & .123 \\
\quad Conspiratorial ideation & 471 & 1.11 [0.69, 1.91] & .679 & .722 \\
\quad Anger at the prescriber & 471 & 1.06 [0.77, 1.47] & .722 & .722 \\
\addlinespace
\multicolumn{5}{@{}l}{\textit{Miscellaneous}} \\
\quad Reads the reviews / crowd sources & 471 & 1.20 [0.99, 1.44] & .061 & .366 \\
\quad Consults lay online content & 471 & 1.01 [0.80, 1.28] & .907 & .962 \\
\quad States an if-then rule & 471 & 1.05 [0.87, 1.27] & .603 & .962 \\
\quad Wider moral reasoning & 471 & 0.96 [0.80, 1.15] & .649 & .962 \\
\quad Sunk-cost salience & 471 & 0.97 [0.81, 1.17] & .783 & .962 \\
\quad Takes the medication with any check plan & 471 & 0.99 [0.81, 1.22] & .962 & .962 \\
\addlinespace
\multicolumn{5}{@{}l}{\textit{H2 composite sensitivities}} \\
\quad Accepts on institutional grounds, exclusive-tier plan or no plan & 471 & 0.90 [0.69, 1.19] & .469 & --- \\
\quad Delays with an exclusive-tier plan and a stated condition & 471 & 1.05 [0.82, 1.35] & .674 & --- \\
\addlinespace
\multicolumn{5}{@{}l}{\textit{Medication search-plan profile}} \\
\quad Search-plan profile (likelihood-ratio test, 4 df) & 471 & $\chi^2_{4}=1.38$ & .847 & --- \\
\bottomrule
\end{tabularx}
\end{table}

\begin{table}[!htbp]
\centering
\footnotesize
\caption{\textbf{Does bCRT still predict the BRT outcomes with familiarity-adjusted CRT2 and NFC held constant?} bCRT coefficient from a joint model entering standardised bCRT, familiarity-adjusted CRT2 and NFC together (Prolific sample, $N=471$); OR per SD of bCRT; otherwise as Supplementary Table~\ref{tab:brt_block_crt2_raw}.}
\label{tab:brt_block_joint}
\begin{tabularx}{\textwidth}{@{}>{\raggedright\arraybackslash}Xrlrr@{}}
\toprule
Outcome & $N$ & OR [95\% CI] & $p$ & $q$ \\
\midrule
\multicolumn{5}{@{}l}{\textit{H1 Epistemic search}} \\
\quad Phones the restaurant (primary source; H1a) & 471 & 1.60 [1.19, 2.14] & .002 & .013 \\
\quad Consults regulators / journals / official guidance & 471 & 1.33 [1.08, 1.64] & .008 & .026 \\
\quad Asks friends or family (medication) & 471 & 1.32 [1.06, 1.63] & .011 & .026 \\
\quad Consults hygiene / official records & 471 & 1.28 [0.98, 1.67] & .071 & .122 \\
\quad Consults a qualified professional & 471 & 1.21 [0.97, 1.50] & .087 & .122 \\
\quad Asks friends or family (restaurant) & 471 & 0.96 [0.73, 1.27] & .794 & .876 \\
\quad Verifies the video's cited sources & 471 & 1.02 [0.83, 1.24] & .876 & .876 \\
\addlinespace
\multicolumn{5}{@{}l}{\textit{H2 Evidence-weighted decisions}} \\
\quad Keeps the booking on evidence grounds & 471 & 0.98 [0.81, 1.20] & .867 & .867 \\
\quad Accepts on institutional grounds with a verification plan & 471 & 1.07 [0.82, 1.40] & .631 & .867 \\
\quad Delays with a verification plan and a stated condition & 471 & 1.14 [0.93, 1.40] & .198 & .867 \\
\quad Acts on the claim with no doubt and no rule & 471 & 0.95 [0.73, 1.25] & .738 & .867 \\
\quad Refuses outright, worry-marked & 471 & 0.87 [0.57, 1.32] & .509 & .867 \\
\addlinespace
\multicolumn{5}{@{}l}{\textit{H3 Affective engagement}} \\
\quad Voices worry about the medication & 471 & 1.56 [1.26, 1.92] & $<$.001 & $<$.001 \\
\quad Hostile rebuke of the unhelpful friend & 471 & 1.07 [0.85, 1.34] & .569 & .569 \\
\addlinespace
\multicolumn{5}{@{}l}{\textit{H4 Fairness}} \\
\quad Appeals to fairness or loyalty toward Friend B & 471 & 1.16 [0.92, 1.46] & .215 & .215 \\
\bottomrule
\end{tabularx}
\end{table}

\begin{table}[!htbp]
\ContinuedFloat
\centering
\footnotesize
\caption{(continued) Exploratory outcome families, the H2 composite sensitivities and the medication search-plan profile.}
\begin{tabularx}{\textwidth}{@{}>{\raggedright\arraybackslash}Xrlrr@{}}
\toprule
Outcome & $N$ & OR [95\% CI] & $p$ & $q$ \\
\midrule
\multicolumn{5}{@{}l}{\textit{Appraises the claim's evidence}} \\
\quad Doubts the video with a stated reason & 471 & 1.09 [0.90, 1.33] & .384 & .769 \\
\quad Evaluates the plausibility, evidence or base rates of the claim & 471 & 1.01 [0.81, 1.27] & .919 & .919 \\
\addlinespace
\multicolumn{5}{@{}l}{\textit{What the decision rested on}} \\
\quad Rests on the critics' / official record & 471 & 0.91 [0.71, 1.18] & .495 & .962 \\
\quad Rests on the medical system & 471 & 1.00 [0.82, 1.23] & .962 & .962 \\
\addlinespace
\multicolumn{5}{@{}l}{\textit{Which choice they made}} \\
\quad Keeps the booking & 471 & 1.03 [0.83, 1.28] & .800 & .800 \\
\quad Takes or delays the medication & 471 & 1.10 [0.84, 1.44] & .475 & .800 \\
\addlinespace
\multicolumn{5}{@{}l}{\textit{Low-scrutiny responses}} \\
\quad Vague-only search, restaurant & 471 & 0.63 [0.21, 1.49] & .311 & .809 \\
\quad Vague-only search, medication & 471 & 1.06 [0.65, 1.74] & .809 & .809 \\
\quad Conspiratorial ideation & 471 & 0.91 [0.53, 1.48] & .708 & .809 \\
\quad Anger at the prescriber & 471 & 1.11 [0.80, 1.54] & .541 & .809 \\
\addlinespace
\multicolumn{5}{@{}l}{\textit{Miscellaneous}} \\
\quad Reads the reviews / crowd sources & 471 & 0.95 [0.78, 1.15] & .599 & .898 \\
\quad Consults lay online content & 471 & 0.84 [0.65, 1.08] & .163 & .490 \\
\quad States an if-then rule & 471 & 1.09 [0.90, 1.33] & .372 & .744 \\
\quad Wider moral reasoning & 471 & 1.01 [0.83, 1.23] & .904 & .904 \\
\quad Sunk-cost salience & 471 & 1.27 [1.04, 1.55] & .017 & .099 \\
\quad Takes the medication with any check plan & 471 & 0.98 [0.79, 1.22] & .844 & .904 \\
\addlinespace
\multicolumn{5}{@{}l}{\textit{H2 composite sensitivities}} \\
\quad Accepts on institutional grounds, exclusive-tier plan or no plan & 471 & 0.76 [0.56, 1.05] & .088 & --- \\
\quad Delays with an exclusive-tier plan and a stated condition & 471 & 1.06 [0.82, 1.36] & .664 & --- \\
\addlinespace
\multicolumn{5}{@{}l}{\textit{Medication search-plan profile}} \\
\quad Search-plan profile (likelihood-ratio test, 4 df) & 471 & $\chi^2_{4}=7.60$ & .108 & --- \\
\bottomrule
\end{tabularx}
\end{table}

\FloatBarrier
\begin{figure}[p]
  \centering
  \begin{subfigure}{\linewidth}\centering\includegraphics[width=\linewidth]{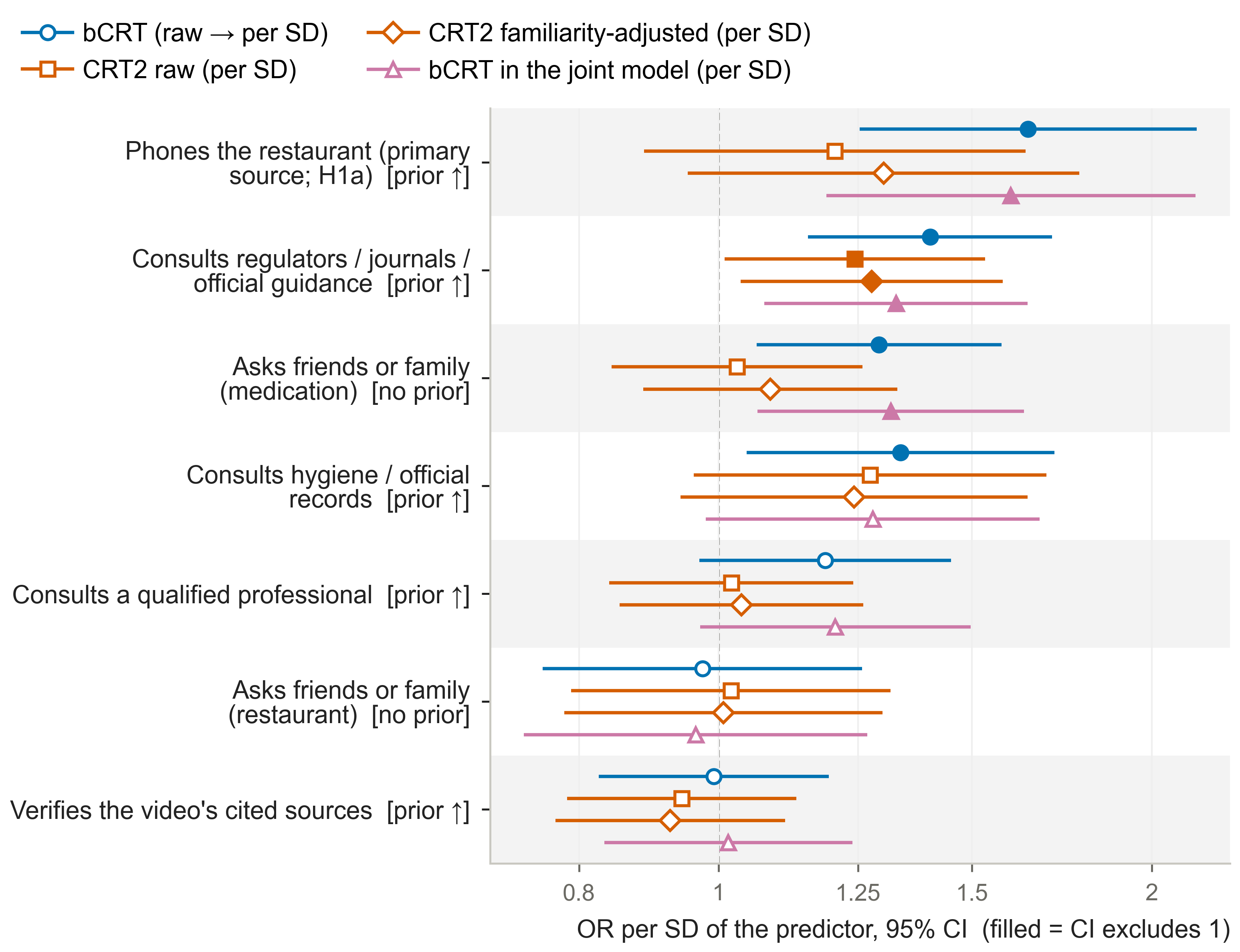}\caption{Family H1, epistemic search}\end{subfigure}
  \caption{\textbf{How do bCRT and CRT2 compare on each primary coded outcome?} Odds ratios per SD of each predictor (95\% CI) from separate logistic models for bCRT, raw CRT2 and familiarity-adjusted CRT2, and for bCRT in the joint model with familiarity-adjusted CRT2 and NFC. The dashed line marks an odds ratio of 1; the predicted direction is shown beside each outcome. bCRT alone, $n=473$; all other models, $n=471$. Estimates on each predictor's own scale in Supplementary Tables~\ref{tab:brt_primary}, \ref{tab:brt_block_crt2_raw} and~\ref{tab:brt_block_crt2_adj}, and per SD for the joint model in Supplementary Table~\ref{tab:brt_block_joint}.}\label{fig:brt_predictor_comparison}
\end{figure}

\begin{figure}[p]
  \ContinuedFloat
  \centering
  \begin{subfigure}{\linewidth}\centering\includegraphics[width=\linewidth]{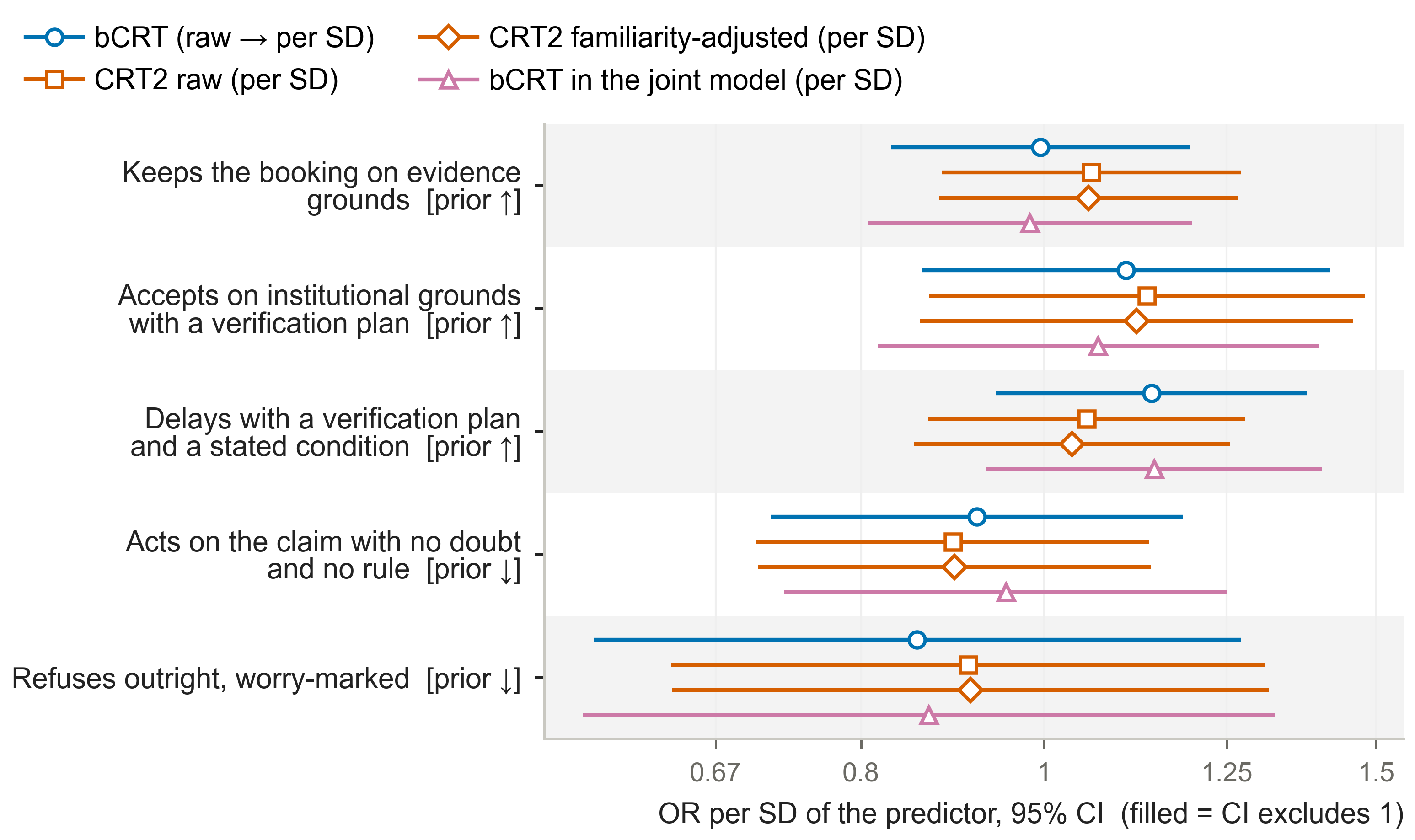}\caption{Family H2, evidence-weighted decisions}\end{subfigure}\\[4pt]
  \begin{subfigure}{\linewidth}\centering\includegraphics[width=\linewidth]{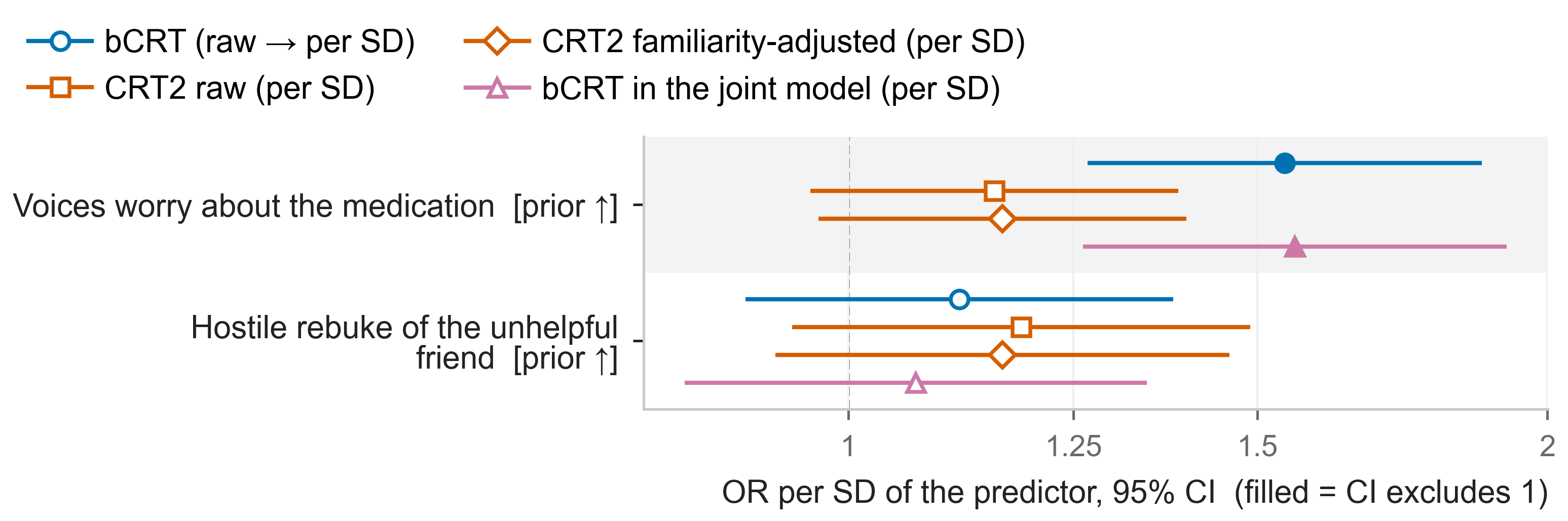}\caption{Family H3, affective engagement}\end{subfigure}\\[4pt]
  \begin{subfigure}{\linewidth}\centering\includegraphics[width=\linewidth]{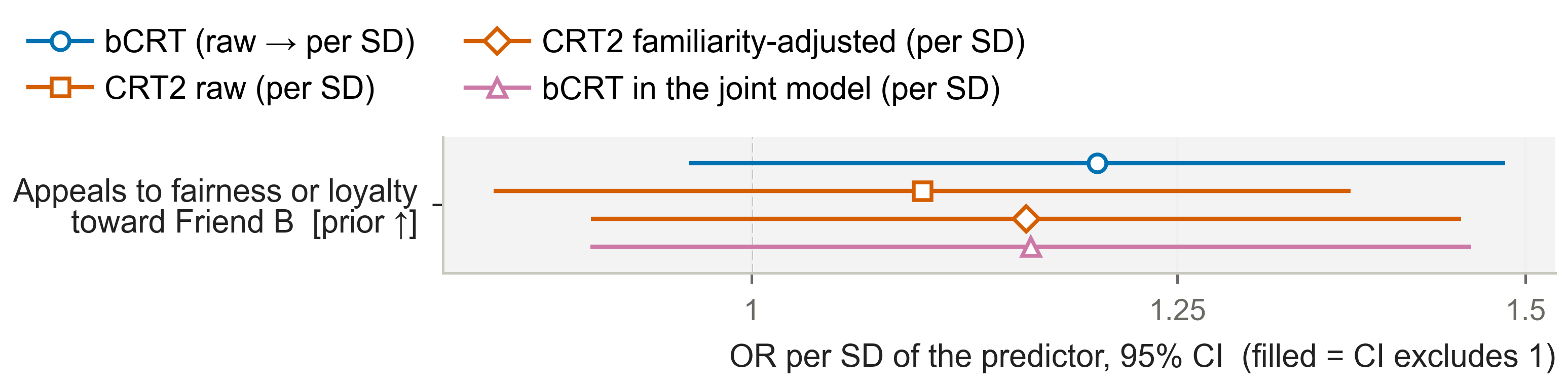}\caption{Family H4, fairness and norm-sensitive reasoning}\end{subfigure}
  \caption{(continued)}
\end{figure}

\FloatBarrier
\subsection*{Supplementary Results: Exploratory BRT outcome models}
\label{app:brt_exploratory_results}

Supplementary Table~\ref{tab:brt_exploratory} reports the five binary exploratory families and Supplementary Table~\ref{tab:brt_search_profile} the medication search-plan profile (Supplementary Methods: Exploratory and sensitivity models); Supplementary Table~\ref{tab:si_counts_by_bcrt} gives the four descriptive-only outcomes.

\begin{table}[!htbp]
\centering
\footnotesize
\caption{\textbf{Does bCRT predict the exploratory coded outcomes?} Exploratory outcome families (Prolific sample, $N=473$). Same models as the primary families: logistic regression on raw bCRT total, OR per bCRT item, Wald 95\% CI, likelihood-ratio $p$; $q$ is Benjamini--Hochberg corrected within each family. Prior, AME and Reliability as in Supplementary Table~\ref{tab:brt_primary}. $^{\dagger}$Fewer than 20 events: Firth-penalised logistic regression, profile-likelihood CI, penalised likelihood-ratio $p$.}
\label{tab:brt_exploratory}
\begin{tabularx}{\textwidth}{@{}>{\raggedright\arraybackslash}Xcrlrrrr@{}}
\toprule
Outcome & Prior & Events & OR [95\% CI] & $p$ & $q$ & AME & Reliability \\
\midrule
\multicolumn{8}{@{}l}{\textit{Appraises the claim's evidence ($m=2$; extends H2)}} \\
\quad Doubts the video with a stated reason & $\uparrow$ & 265 & 1.13 [0.97, 1.32] & .126 & .253 & +3.0 & 96.7/0.93 \\
\quad Evaluates the plausibility, evidence or base rates of the claim & $\uparrow$ & 352 & 1.03 [0.87, 1.23] & .707 & .707 & +0.6 & 90.7/0.77 \\
\addlinespace
\multicolumn{8}{@{}l}{\textit{What the decision rested on ($m=2$; extends H2)}} \\
\quad Rests on the critics' / official record & $\uparrow$ & 85 & 0.93 [0.76, 1.13] & .452 & .614 & -1.1 & 85.3/0.79 \\
\quad Rests on the medical system & $\uparrow$ & 170 & 1.04 [0.89, 1.22] & .614 & .614 & +0.9 & 79.3/0.73 \\
\addlinespace
\multicolumn{8}{@{}l}{\textit{Which choice they made ($m=2$; extends H2)}} \\
\quad Keeps the booking & $\uparrow$ & 335 & 1.01 [0.85, 1.19] & .953 & .953 & +0.1 & 98.0/0.96 \\
\quad Takes or delays the medication & $\uparrow$ & 395 & 1.11 [0.90, 1.37] & .326 & .651 & +1.4 & 90.0/0.85 \\
\addlinespace
\multicolumn{8}{@{}l}{\textit{Low-scrutiny responses ($m=4$; extends H1, H2)}} \\
\quad Vague-only search, restaurant$^{\dagger}$ & $\downarrow$ & 8 & 0.56 [0.23, 1.11] & .101 & .405 & -1.1 & 98.7/0.74 \\
\quad Vague-only search, medication & $\downarrow$ & 21 & 0.92 [0.63, 1.35] & .670 & .692 & -0.3 & 99.3/0.95 \\
\quad Conspiratorial ideation$^{\dagger}$ & $\downarrow$ & 19 & 0.92 [0.61, 1.35] & .692 & .692 & -0.3 & 100.0/1.00 \\
\quad Anger at the prescriber & $\downarrow$ & 44 & 1.10 [0.85, 1.42] & .454 & .692 & +0.8 & 97.3/0.83 \\
\addlinespace
\multicolumn{8}{@{}l}{\textit{Miscellaneous ($m=6$; extends ---)}} \\
\quad Reads the reviews / crowd sources & --- & 258 & 1.02 [0.88, 1.19] & .782 & .961 & +0.5 & 96.7/0.93 \\
\quad Consults lay online content & --- & 92 & 0.88 [0.72, 1.08] & .217 & .645 & -1.9 & 96.0/0.89 \\
\quad States an if-then rule & --- & 219 & 1.08 [0.93, 1.26] & .323 & .645 & +1.9 & 92.7/0.84 \\
\quad Wider moral reasoning & --- & 226 & 1.00 [0.86, 1.17] & .962 & .962 & +0.1 & 77.3/0.54 \\
\quad Sunk-cost salience & --- & 246 & 1.18 [1.01, 1.38] & .036 & .219 & +4.1 & 99.3/0.99 \\
\quad Takes the medication with any check plan & --- & 134 & 0.98 [0.83, 1.16] & .801 & .961 & -0.4 & 89.3/0.77 \\
\bottomrule
\end{tabularx}
\end{table}

\begin{table}[!htbp]
\centering
\footnotesize
\caption{\textbf{Does bCRT predict which kind of verification plan participants make for the medication?} One multinomial logistic regression of the five unordered plan types on raw bCRT total (Prolific sample, $N=473$; extends H1). Reference: no evidence-grade or professional plan ($n=80$); relative-risk ratio (RRR) per bCRT item with 95\% CI. Family test: likelihood-ratio $\chi^2(4)=7.47$, $p=.113$.}
\label{tab:brt_search_profile}
\begin{tabular*}{\textwidth}{@{\extracolsep{\fill}}lrrr@{}}
\toprule
Plan type & $n$ & RRR & 95\% CI \\
\midrule
Evidence-grade checks, no lay checks & 134 & 1.26 & [0.99, 1.62] \\
Evidence-grade checks with lay checks & 124 & 1.37 & [1.07, 1.76] \\
Professional consultation only (no lay checks) & 78 & 1.14 & [0.86, 1.51] \\
Professional consultation with lay checks & 57 & 1.31 & [0.97, 1.76] \\
\bottomrule
\end{tabular*}
\end{table}

\begin{sidewaystable}
\centering
\footnotesize
\caption{\textbf{How often do the four descriptive-only outcomes occur at each bCRT score?} Count/$n$, then percentage with Clopper--Pearson 95\% interval; no model is fitted. $N=473$.}
\label{tab:si_counts_by_bcrt}
\begin{tabular}{@{}lrrrrrr@{}}
\toprule
Outcome & \multicolumn{5}{c}{bCRT total} & All \\ \cmidrule(lr){2-6} & 0 & 1 & 2 & 3 & 4 & \\
\midrule
$n$ at each bCRT score & 109 & 174 & 100 & 53 & 37 & 473 \\
\addlinespace
Credulous switching, restaurant & \makecell[r]{4/109\\ 3.7\% [1.0, 9.1]} & \makecell[r]{1/174\\ 0.6\% [0.0, 3.2]} & \makecell[r]{3/100\\ 3.0\% [0.6, 8.5]} & \makecell[r]{1/53\\ 1.9\% [0.0, 10.1]} & \makecell[r]{0/37\\ 0.0\% [0.0, 9.5]} & \makecell[r]{9/473\\ 1.9\% [0.9, 3.6]} \\[3pt]
Groundless scepticism, medication & \makecell[r]{2/109\\ 1.8\% [0.2, 6.5]} & \makecell[r]{3/174\\ 1.7\% [0.4, 5.0]} & \makecell[r]{1/100\\ 1.0\% [0.0, 5.4]} & \makecell[r]{0/53\\ 0.0\% [0.0, 6.7]} & \makecell[r]{0/37\\ 0.0\% [0.0, 9.5]} & \makecell[r]{6/473\\ 1.3\% [0.5, 2.7]} \\[3pt]
Misread the scenario, restaurant & \makecell[r]{5/109\\ 4.6\% [1.5, 10.4]} & \makecell[r]{5/174\\ 2.9\% [0.9, 6.6]} & \makecell[r]{4/100\\ 4.0\% [1.1, 9.9]} & \makecell[r]{3/53\\ 5.7\% [1.2, 15.7]} & \makecell[r]{2/37\\ 5.4\% [0.7, 18.2]} & \makecell[r]{19/473\\ 4.0\% [2.4, 6.2]} \\[3pt]
Misread the scenario, medication & \makecell[r]{1/109\\ 0.9\% [0.0, 5.0]} & \makecell[r]{5/174\\ 2.9\% [0.9, 6.6]} & \makecell[r]{4/100\\ 4.0\% [1.1, 9.9]} & \makecell[r]{2/53\\ 3.8\% [0.5, 13.0]} & \makecell[r]{2/37\\ 5.4\% [0.7, 18.2]} & \makecell[r]{14/473\\ 3.0\% [1.6, 4.9]} \\[3pt]
\bottomrule
\end{tabular}
\end{sidewaystable}

\FloatBarrier
\subsection*{Supplementary Results: Sensitivity analyses and response length}\phantomsection\label{app:sensitivity}

Supplementary Tables~\ref{tab:brt_cov_nosplit} and~\ref{tab:brt_ordered_factor} report the sensitivity analyses (Supplementary Methods: Exploratory and sensitivity models); Supplementary Table~\ref{tab:brt_response_length} reports the association between response length and the search outcomes.

With the eight attention-check failures dropped ($n=465$; Supplementary Methods: Data Cleaning and Participant Flow), negative emotion, a predicted category, was no longer below .05 ($p=.052$; full sample, $p=.043$); every other category, for bCRT and for the CRT2 comparator, fell on the same side of .05 as in the full sample (Supplementary Table~\ref{tab:si_liwc_attncheck465}).

\begin{table}[!htbp]
\centering
\footnotesize
\caption{\textbf{Do the primary estimates hold with demographic covariates, and without split-coded responses?} The 15 primary-family members (per-test $p$ only; OR per bCRT item). \textit{Covariates}: bCRT total with age, gender and education entered alongside ($N=470$; missing gender or education entered as its own category); $^{\dagger}$Firth-penalised fit (profile-likelihood CI) where a sparse covariate level produced separation. \textit{No-split refit}: the primary model refitted after dropping respondents for whom the three models produced no majority on any field the outcome is built from.}
\label{tab:brt_cov_nosplit}
\begin{tabular*}{\textwidth}{@{\extracolsep{\fill}}>{\raggedright\arraybackslash}p{38mm}rlr|rlr@{}}
\toprule
 & \multicolumn{3}{c|}{Covariates} & \multicolumn{3}{c}{No-split refit} \\
Outcome & $N$ & OR [95\% CI] & $p$ & $N$ & OR [95\% CI] & $p$ \\
\midrule
Phones the restaurant (primary source; H1a) & 470 & 1.47 [1.14, 2.07]$^{\dagger}$ & .003 & 473 & 1.52 [1.21, 1.91] & $<$.001 \\
Consults regulators / journals / official guidance & 470 & 1.28 [1.09, 1.52]$^{\dagger}$ & .003 & 473 & 1.33 [1.13, 1.57] & $<$.001 \\
Asks friends or family (medication) & 470 & 1.21 [1.02, 1.43] & .031 & 473 & 1.24 [1.05, 1.47] & .010 \\
Consults hygiene / official records & 470 & 1.27 [1.02, 1.57] & .034 & 473 & 1.28 [1.04, 1.57] & .022 \\
Consults a qualified professional & 470 & 1.16 [0.97, 1.38] & .092 & 473 & 1.16 [0.97, 1.37] & .093 \\
Asks friends or family (restaurant) & 470 & 0.96 [0.77, 1.20] & .705 & 473 & 0.98 [0.79, 1.21] & .845 \\
Verifies the video's cited sources & 470 & 0.98 [0.85, 1.14]$^{\dagger}$ & .758 & 473 & 0.99 [0.85, 1.16] & .937 \\
Keeps the booking on evidence grounds & 470 & 1.00 [0.85, 1.17] & .989 & 420 & 1.00 [0.85, 1.18] & .971 \\
Accepts on institutional grounds with a verification plan & 470 & 1.07 [0.86, 1.33]$^{\dagger}$ & .547 & 433 & 1.08 [0.87, 1.34] & .477 \\
Delays with a verification plan and a stated condition & 470 & 1.10 [0.93, 1.30] & .262 & 462 & 1.12 [0.95, 1.32] & .171 \\
Acts on the claim with no doubt and no rule & 470 & 0.94 [0.75, 1.16]$^{\dagger}$ & .558 & 462 & 0.93 [0.75, 1.16] & .523 \\
Refuses outright, worry-marked & 470 & 0.92 [0.64, 1.27]$^{\dagger}$ & .606 & 462 & 0.88 [0.63, 1.23] & .435 \\
Voices worry about the medication & 470 & 1.42 [1.20, 1.68]$^{\dagger}$ & $<$.001 & 473 & 1.44 [1.22, 1.70] & $<$.001 \\
Hostile rebuke of the unhelpful friend & 470 & 1.11 [0.93, 1.34] & .253 & 473 & 1.10 [0.92, 1.31] & .312 \\
Appeals to fairness or loyalty toward Friend B & 470 & 1.13 [0.94, 1.36] & .191 & 473 & 1.17 [0.97, 1.40] & .099 \\
\bottomrule
\end{tabular*}
\end{table}

\begin{table}[!htbp]
\centering
\footnotesize
\caption{\textbf{Does the bCRT association depend on treating the score as linear?} bCRT entered as an ordered factor (four indicators, reference: 0 items correct), each primary-family member ($N=473$). OR [95\% CI] for each bCRT score relative to 0.}
\label{tab:brt_ordered_factor}
\begin{tabular*}{\textwidth}{@{\extracolsep{\fill}}>{\raggedright\arraybackslash}p{55mm}cccc@{}}
\toprule
Outcome & bCRT $=1$ & bCRT $=2$ & bCRT $=3$ & bCRT $=4$ \\
\midrule
Phones the restaurant (primary source; H1a) & \makecell{1.74\\{}[0.66, 4.59]} & \makecell{2.79\\{}[1.03, 7.58]} & \makecell{3.05\\{}[1.00, 9.31]} & \makecell{6.36\\{}[2.12, 19.05]} \\ \addlinespace[2pt]
Consults regulators / journals / official guidance & \makecell{1.51\\{}[0.86, 2.65]} & \makecell{1.53\\{}[0.81, 2.87]} & \makecell{2.87\\{}[1.41, 5.84]} & \makecell{3.18\\{}[1.44, 7.03]} \\ \addlinespace[2pt]
Asks friends or family (medication) & \makecell{0.89\\{}[0.51, 1.55]} & \makecell{1.49\\{}[0.82, 2.71]} & \makecell{1.75\\{}[0.87, 3.54]} & \makecell{1.97\\{}[0.90, 4.32]} \\ \addlinespace[2pt]
Consults hygiene / official records & \makecell{2.22\\{}[1.01, 4.90]} & \makecell{1.10\\{}[0.42, 2.89]} & \makecell{3.25\\{}[1.27, 8.31]} & \makecell{3.57\\{}[1.29, 9.85]} \\ \addlinespace[2pt]
Consults a qualified professional & \makecell{0.82\\{}[0.49, 1.38]} & \makecell{0.81\\{}[0.45, 1.44]} & \makecell{2.12\\{}[0.93, 4.85]} & \makecell{1.57\\{}[0.65, 3.81]} \\ \addlinespace[2pt]
Asks friends or family (restaurant) & \makecell{1.15\\{}[0.58, 2.28]} & \makecell{1.19\\{}[0.56, 2.56]} & \makecell{1.11\\{}[0.44, 2.82]} & \makecell{0.76\\{}[0.24, 2.45]} \\ \addlinespace[2pt]
Verifies the video's cited sources & \makecell{1.20\\{}[0.73, 1.96]} & \makecell{1.25\\{}[0.72, 2.18]} & \makecell{1.09\\{}[0.56, 2.13]} & \makecell{0.90\\{}[0.41, 1.96]} \\ \addlinespace[2pt]
Keeps the booking on evidence grounds & \makecell{0.81\\{}[0.50, 1.32]} & \makecell{0.68\\{}[0.39, 1.18]} & \makecell{0.88\\{}[0.45, 1.72]} & \makecell{1.16\\{}[0.53, 2.52]} \\ \addlinespace[2pt]
Accepts on institutional grounds with a verification plan & \makecell{1.30\\{}[0.65, 2.60]} & \makecell{0.84\\{}[0.36, 1.94]} & \makecell{1.58\\{}[0.65, 3.83]} & \makecell{1.58\\{}[0.58, 4.29]} \\ \addlinespace[2pt]
Delays with a verification plan and a stated condition & \makecell{1.04\\{}[0.62, 1.74]} & \makecell{1.35\\{}[0.76, 2.41]} & \makecell{1.40\\{}[0.70, 2.78]} & \makecell{1.40\\{}[0.64, 3.06]} \\ \addlinespace[2pt]
Acts on the claim with no doubt and no rule & \makecell{0.99\\{}[0.52, 1.86]} & \makecell{0.65\\{}[0.30, 1.41]} & \makecell{0.84\\{}[0.34, 2.07]} & \makecell{0.92\\{}[0.34, 2.50]} \\ \addlinespace[2pt]
Refuses outright, worry-marked & \makecell{1.18\\{}[0.45, 3.05]} & \makecell{0.77\\{}[0.24, 2.50]} & \makecell{0.57\\{}[0.11, 2.85]} & \makecell{0.83\\{}[0.17, 4.20]} \\ \addlinespace[2pt]
Voices worry about the medication & \makecell{1.75\\{}[1.08, 2.84]} & \makecell{2.05\\{}[1.18, 3.55]} & \makecell{3.01\\{}[1.51, 6.01]} & \makecell{5.16\\{}[2.16, 12.31]} \\ \addlinespace[2pt]
Hostile rebuke of the unhelpful friend & \makecell{1.42\\{}[0.78, 2.57]} & \makecell{1.33\\{}[0.68, 2.60]} & \makecell{1.17\\{}[0.51, 2.65]} & \makecell{1.88\\{}[0.80, 4.43]} \\ \addlinespace[2pt]
Appeals to fairness or loyalty toward Friend B & \makecell{1.12\\{}[0.61, 2.06]} & \makecell{1.33\\{}[0.68, 2.60]} & \makecell{1.76\\{}[0.81, 3.79]} & \makecell{1.65\\{}[0.69, 3.94]} \\ \addlinespace[2pt]
\bottomrule
\end{tabular*}
\end{table}

\begin{table}[!htbp]
\centering
\footnotesize
\caption{\textbf{Do longer responses name more sources?} Separate logistic regressions of each H1 epistemic-search outcome on log word count (Prolific sample, $N=473$); OR per log-word. Length is the LIWC word count over both scenarios. Reported as a property of the data; word count is not a covariate in any model.}
\label{tab:brt_response_length}
\begin{tabular*}{\textwidth}{@{\extracolsep{\fill}}lrrrr@{}}
\toprule
H1 outcome & Events & OR & 95\% CI & $p$ \\
\midrule
Phones the restaurant (primary source; H1a) & 54 & 5.28 & [2.18, 12.75] & $<$.001 \\
Consults regulators / journals / official guidance & 142 & 3.89 & [2.12, 7.13] & $<$.001 \\
Asks friends or family (medication) & 138 & 4.16 & [2.24, 7.70] & $<$.001 \\
Consults hygiene / official records & 68 & 6.16 & [2.69, 14.12] & $<$.001 \\
Consults a qualified professional & 328 & 2.80 & [1.65, 4.74] & $<$.001 \\
Asks friends or family (restaurant) & 70 & 2.49 & [1.20, 5.18] & .012 \\
Verifies the video's cited sources & 191 & 8.67 & [4.56, 16.48] & $<$.001 \\
\bottomrule
\end{tabular*}
\end{table}

\begin{table}[!htbp]
\centering
\footnotesize
\caption{\textbf{Do the LIWC associations change when the eight attention-check failures are dropped?} Each row a separate OLS model, as in Table~\ref{tab:liwc_bcrt} and Supplementary Tables~\ref{tab:liwc_bcrt_full} and~\ref{tab:liwc_crt2}; $\dagger$ predicted category (per-test $p$); other rows Benjamini--Hochberg within family, except a family's single exploratory category (raw $p$); intervals are not corrected for multiple comparisons. $n$: bCRT 473 (all) and 465 (subset); CRT2 471 and 463.}
\label{tab:si_liwc_attncheck465}
\begin{tabular*}{\textwidth}{@{\extracolsep{\fill}}llrrrr@{}}
\toprule
& & \multicolumn{2}{c}{All participants} & \multicolumn{2}{c}{Attention-check subset} \\ \cmidrule(lr){3-4}\cmidrule(lr){5-6} Family & Category & $\beta$ [95\% CI] & $p$ & $\beta$ [95\% CI] & $p$ \\
\midrule
\multicolumn{6}{@{}l}{\textbf{a. bCRT}} \\
QTY    & Word count & 15.797 [7.938, 23.657] & $<$.001 & 15.506 [7.570, 23.441] & $<$.001 \\
       & Words per sentence & 1.068 [-0.079, 2.214] & .068 & 1.091 [-0.070, 2.251] & .065 \\
\addlinespace
FUNC   & Articles & -0.036 [-0.165, 0.093] & .598 & -0.029 [-0.160, 0.101] & .657 \\
       & Prepositions & -0.162 [-0.319, -0.004] & .103 & -0.163 [-0.322, -0.004] & .139 \\
       & First-person singular & 0.050 [-0.107, 0.207] & .598 & 0.060 [-0.098, 0.217] & .642 \\
       & Other pronouns & -0.116 [-0.252, 0.019] & .161 & -0.107 [-0.244, 0.030] & .218 \\
       & Auxiliary verbs & -0.328 [-0.492, -0.164] & $<$.001 & -0.323 [-0.488, -0.158] & $<$.001 \\
       & Conjunctions & 0.038 [-0.103, 0.178] & .598 & 0.032 [-0.110, 0.174] & .657 \\
       & Adjectives & 0.145 [0.016, 0.273] & .096 & 0.122 [-0.005, 0.250] & .139 \\
\addlinespace
COG    & Insight$^{\dagger}$ & 0.030 [-0.085, 0.145] & .605 & 0.029 [-0.087, 0.145] & .626 \\
       & Causation & 0.026 [-0.050, 0.102] & .530 & 0.030 [-0.046, 0.106] & .548 \\
       & Discrepancy & -0.219 [-0.361, -0.077] & .013 & -0.212 [-0.355, -0.070] & .018 \\
       & Tentativeness & -0.071 [-0.204, 0.063] & .499 & -0.073 [-0.207, 0.061] & .473 \\
       & Differentiation & -0.176 [-0.306, -0.046] & .020 & -0.171 [-0.301, -0.041] & .026 \\
       & Certainty & 0.015 [-0.032, 0.061] & .530 & 0.009 [-0.037, 0.056] & .691 \\
\addlinespace
AFFECT & Negative emotion$^{\dagger}$ & 0.055 [0.002, 0.109] & .043 & 0.053 [-0.001, 0.107] & .052 \\
\addlinespace
TIME   & Present-focus & 0.205 [0.060, 0.351] & .006 & 0.199 [0.052, 0.346] & .008 \\
\addlinespace
MOT    & Risk words$^{\dagger}$ & 0.077 [0.030, 0.124] & .001 & 0.072 [0.024, 0.119] & .003 \\
\addlinespace
MOS    & Moral language$^{\dagger}$ & 0.008 [-0.021, 0.037] & .603 & 0.008 [-0.021, 0.037] & .584 \\
\bottomrule
\end{tabular*}
\end{table}

\begin{table}[!htbp]
\ContinuedFloat
\centering
\footnotesize
\caption{(continued)}
\begin{tabular*}{\textwidth}{@{\extracolsep{\fill}}llrrrr@{}}
\toprule
& & \multicolumn{2}{c}{All participants} & \multicolumn{2}{c}{Attention-check subset} \\ \cmidrule(lr){3-4}\cmidrule(lr){5-6} Family & Category & $\beta$ [95\% CI] & $p$ & $\beta$ [95\% CI] & $p$ \\
\midrule
\multicolumn{6}{@{}l}{\textbf{b. CRT2, items seen held constant}} \\
QTY    & Word count & 19.641 [10.034, 29.248] & $<$.001 & 19.183 [9.398, 28.967] & $<$.001 \\
       & Words per sentence & 1.650 [0.250, 3.050] & .021 & 1.648 [0.219, 3.077] & .024 \\
\addlinespace
FUNC   & Articles & 0.138 [-0.019, 0.295] & .170 & 0.125 [-0.035, 0.285] & .219 \\
       & Prepositions & -0.080 [-0.273, 0.113] & .484 & -0.090 [-0.287, 0.106] & .513 \\
       & First-person singular & -0.234 [-0.425, -0.044] & .113 & -0.207 [-0.401, -0.014] & .179 \\
       & Other pronouns & -0.054 [-0.220, 0.111] & .518 & -0.053 [-0.221, 0.116] & .539 \\
       & Auxiliary verbs & -0.174 [-0.377, 0.028] & .170 & -0.205 [-0.411, 0.001] & .179 \\
       & Conjunctions & 0.145 [-0.026, 0.316] & .170 & 0.148 [-0.026, 0.322] & .219 \\
       & Adjectives & 0.072 [-0.085, 0.230] & .484 & 0.055 [-0.102, 0.212] & .539 \\
\addlinespace
COG    & Insight$^{\dagger}$ & -0.017 [-0.157, 0.123] & .811 & -0.023 [-0.166, 0.119] & .751 \\
       & Causation & -0.047 [-0.140, 0.046] & .738 & -0.042 [-0.136, 0.052] & .614 \\
       & Discrepancy & -0.178 [-0.351, -0.005] & .221 & -0.179 [-0.354, -0.003] & .229 \\
       & Tentativeness & 0.006 [-0.157, 0.169] & .943 & 0.012 [-0.153, 0.177] & .888 \\
       & Differentiation & -0.049 [-0.208, 0.110] & .738 & -0.072 [-0.234, 0.089] & .614 \\
       & Certainty & 0.016 [-0.041, 0.072] & .738 & 0.020 [-0.037, 0.078] & .614 \\
\addlinespace
AFFECT & Negative emotion$^{\dagger}$ & -0.010 [-0.076, 0.056] & .764 & -0.012 [-0.079, 0.054] & .713 \\
\addlinespace
TIME   & Present-focus & 0.140 [-0.039, 0.319] & .125 & 0.155 [-0.027, 0.337] & .095 \\
\addlinespace
MOT    & Risk words$^{\dagger}$ & -0.005 [-0.063, 0.054] & .877 & -0.004 [-0.063, 0.055] & .900 \\
\addlinespace
MOS    & Moral language$^{\dagger}$ & 0.028 [-0.007, 0.063] & .121 & 0.027 [-0.008, 0.063] & .133 \\
\bottomrule
\end{tabular*}
\end{table}

\FloatBarrier
\subsection*{Supplementary Results: Completion-time distributions}

\begin{figure}[htbp]
    \centering
    \includegraphics[width=\textwidth]{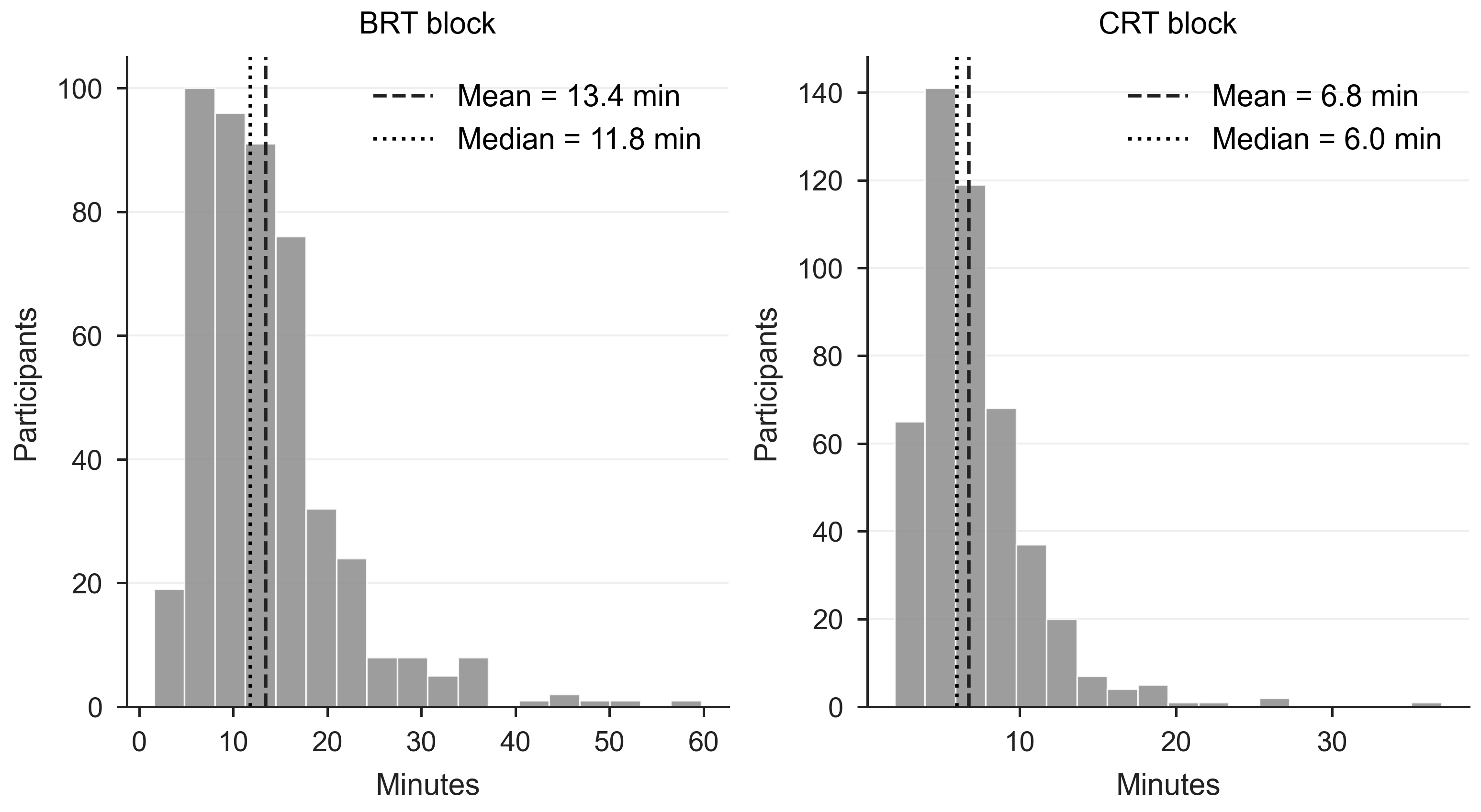}
    \caption{\textbf{How long did participants take on the BRT and CRT reasoning blocks?} Completion-time distributions in the Prolific validation sample. Dashed line: mean; dotted line: median.}\label{fig:block_time_distributions}
\end{figure}
\end{document}